\documentclass[10pt,preprint2]{aastex}

\usepackage{bm}
\usepackage{textcomp}
\usepackage{graphicx}
\usepackage{amssymb}
\usepackage{gensymb}
\usepackage[titletoc]{appendix}

\begin{document}
\date{}
\title{POLARIZED SCATTERING OF LIGHT FOR ARBITRARY MAGNETIC FIELDS WITH
LEVEL-CROSSINGS FROM THE COMBINATION OF HYPERFINE AND FINE STRUCTURE SPLITTINGS}
\author{K. SOWMYA$^{1}$, K. N. NAGENDRA$^{1}$, M. SAMPOORNA$^{1}$ AND
J. O. STENFLO$^{2,3}$}
\affil{$^1$Indian Institute of Astrophysics, Koramangala, Bengaluru, India}
\affil{$^2$Institute of Astronomy,
ETH Zurich, CH-8093 Zurich, Switzerland }
\affil{$^3$Istituto Ricerche Solari Locarno, Via Patocchi,
6605 Locarno-Monti, Switzerland}
\email{ksowmya@iiap.res.in; knn@iiap.res.in; sampoorna@iiap.res.in;
stenflo@astro.phys.ethz.ch}

\begin{abstract}
Interference between magnetic substates of the hyperfine structure states belonging
to different fine structure states of the same term influences the polarization for some
of the  diagnostically important lines of the Sun's spectrum, like the sodium and lithium
doublets. The polarization signatures of this combined interference contain information
on the properties of the solar magnetic fields. Motivated by this,
in the present paper, we study the problem of polarized scattering on a two-term
atom with hyperfine structure by accounting for the partial redistribution in the
photon frequencies arising due to the Doppler motions of the atoms. We consider the
scattering atoms to be under the influence of a magnetic field of arbitrary strength
and develop a formalism based on the Kramers--Heisenberg approach to calculate the
scattering cross section for this process. We explore the rich polarization effects
that arise from various level-crossings in the Paschen--Back regime in a single
scattering case using the lithium atomic system as a concrete example that is
relevant to the Sun.
\end{abstract}


\section{INTRODUCTION}
\label{intro}
In the present paper, we address the problem of quantum interference between
the magnetic substates of the hyperfine structure ($F$) states pertaining to
different fine structure ($J$) states of a given term, in the presence of
magnetic fields of arbitrary strength covering the Hanle, Zeeman, and
Paschen--Back (PB) effect regimes. We will refer to this as ``combined
interference'' or the ``$F+J$ state interference''. We develop the
necessary theory including the effects of partial frequency redistribution (PRD) in
the absence of collisions, assuming the lower levels to be unpolarized and infinitely
sharp. We refer to this theory as the ``combined theory'' throughout the paper.

\begin{figure}
\begin{center}
\includegraphics[scale=0.43]{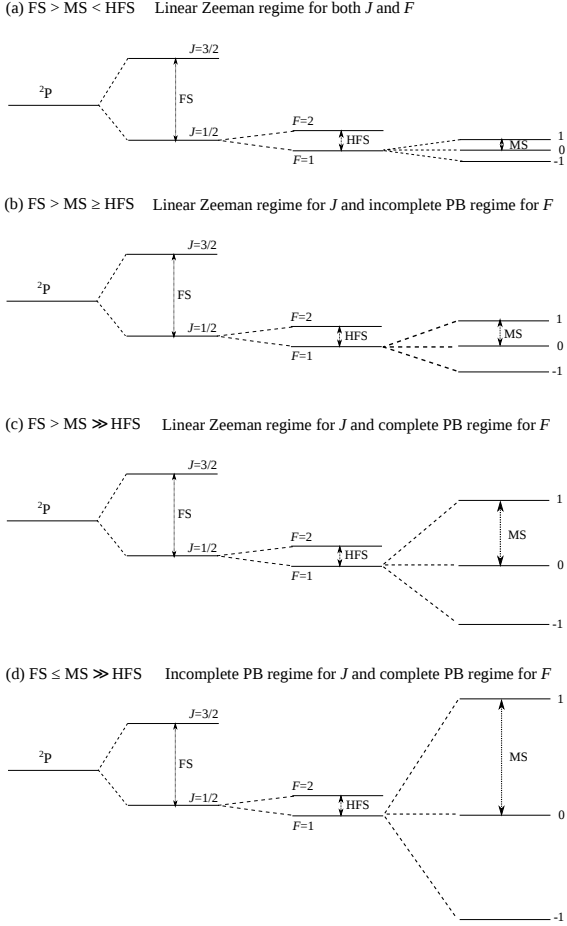}
\end{center}
\caption{Illustration of the magnetic field strength regimes in the combined theory.
For illustration purpose, a $^2$P term with nuclear spin 3/2 is considered.
The various splittings indicated are not to scale. Panels (a)--(d)
show the first four regimes of the field strength. When MS is much greater than FS,
we have a complete PB regime for both $J$ and $F$, which we call the fifth regime
(not illustrated in the figure).
\label{lev-spl}}
\end{figure}

We consider a two-term atom with hyperfine structure under the assumption that
the lower term is unpolarized. In the absence of a magnetic field, the atomic
transitions in a two-term atom take place between the degenerate magnetic substates
belonging to the $F$ states. An applied magnetic field lifts the degeneracies and
modifies the energies of these magnetic substates. The amount of splitting (or the
energy change) produced by the magnetic field defines the regimes in which Zeeman
and PB effects act. Depending on the relative magnitudes of the fine structure
splitting (FS), the hyperfine structure splitting (HFS), and the magnetic splitting
(MS), we characterize the magnetic field strength into five regimes. These regimes
are illustrated schematically in Figure~\ref{lev-spl}. In the approach presented
in this paper, we account for the interferences between the magnetic substates pertaining
to the same $F$ state, the magnetic substates belonging to different $F$ states of
the same $J$ state, and the magnetic substates belonging to different $F$ states
pertaining to different $J$ states. Although all three types of interference are
always present, depending on the field strength one or two of them would dominate as
depicted in the different panels of Figure~\ref{lev-spl}.

Within the framework of non-relativistic quantum electrodynamics, \citet{cm05}
formulated a theory for polarized scattering on a multi-term atom with hyperfine
structure in the presence of an arbitrary strength magnetic field under the
approximation of complete frequency redistribution (CRD). In the present paper, we
restrict our treatment to a two-term atom with HFS and consider the limit of coherent
scattering in the atomic frame with Doppler frequency redistribution in the observer's
frame. We base our formalism on the Kramers--Heisenberg coherency matrix
approach of \citet{s94}. In our combined theory, we do not account for the coherences
among the states in the lower term. In a recent paper, \citet{s15} indicated how
they may be included by extending the coherency matrix approach to the multi-level case.

Based on the concept of ``metalevels'', \citet{landi97} formulated a theory
that is able to treat coherent scattering in the atomic rest frame for a two-term
atom with hyperfine structure. Recently, \citet{casinietal14} presented a generalized
frequency redistribution function for the polarized two-term atom in arbitrary fields,
based on a new formulation of the quantum scattering theory. Our approach
is an alternative approach to the same problem and is conceptually more transparent,
although limited to infinitely sharp and unpolarized lower levels. 

\citet{bel09} studied the linear polarization produced due to scattering on the D
lines of neutral lithium isotopes. They employed the density matrix formalism of
\citet[][hereafter LL04]{ll04}, together with the approximation of CRD, to treat the
quantum interference between the fine and hyperfine structure states. They restricted
their study to the non-magnetic case. However, they explored the sensitivity of the
Stokes profiles to the microturbulent magnetic fields. For our study in the
present paper, we consider the same D lines of lithium isotopes and present in
detail the effects of a deterministic magnetic field of arbitrary strength. For
this atomic line system, the PB effect in both the fine and the hyperfine structure
states occurs for the magnetic field strengths encountered on the Sun. We restrict
our treatment to the single scattering case, since our aim here is to explore the
basic physical effects of the combined theory.

\section{THE ATOMIC MODEL}
\label{atmod}
In this section, we describe the structure of the model atom considered 
for our studies and its interaction with an external magnetic field. We 
consider a two-term atom, each state of which is designated by the quantum 
numbers $L$ (orbital), $S$ (electron spin), $J$ ($=L+S$), $I_s$ (nuclear spin), 
$F$ ($=J+I_s$), and $\mu$ (projection of $F$ onto the quantization axis). 

\subsection{The Atomic Hamiltonian}
\label{ath}
Under the $L-S$ coupling scheme, the atomic Hamiltonian for a two-term atom
with hyperfine structure is given by
\begin{eqnarray}
&& \!\!\!\!\!\!\!\!\!\!\!\mathcal{H}_{A}=\zeta(LS){\bm L}\cdot{\bm S}
 \nonumber \\ &&
\!\!\!\!\!\!\!\!\!\!\!+\mathcal{A}_J {\bm I_s}\cdot{\bm J}+
\frac{\mathcal{B}_J}{2I_s(2I_s-1)J(2J-1)}
 \nonumber \\ &&
\!\!\!\!\!\!\!\!\!\!\!\times\bigg\{3({\bm I_s}\cdot{\bm J})^2
+\frac{3}{2}({\bm I_s}\cdot{\bm J})-I_s(I_s+1)J(J+1)\bigg\}\ ,
 \nonumber \\ &&
\label{atom-ham}
\end{eqnarray}
where $\zeta(LS)$ is a constant having the dimensions of energy, and $\mathcal{A}_J$
and $\mathcal{B}_J$ are the magnetic dipole and electric quadrupole hyperfine
structure constants, respectively. The first term in the above equation is a
measure of the FS while the second and the third terms provide a measure of
the HFS. The eigenvalues of the atomic Hamiltonian represent the energies of
the $F$ states, calculated with respect to the energy of the corresponding term.

\subsection{The Magnetic and the Total Hamiltonians}
\label{mah}
An external magnetic field lifts the degeneracies of the magnetic substates of
the $F$ states and changes their energies by an amount given by the eigenvalues of
the magnetic Hamiltonian
\begin{eqnarray}
 && \mathcal{H}_B=\mu_0 ({\bm J}+{\bm S})\cdot {\bm B}\ .
\label{mag-ham}
\end{eqnarray}
Assuming the quantization axis to be along the magnetic field ({\it z}-axis of
the reference system), the matrix elements of the total Hamiltonian,
$\mathcal{H}_T=\mathcal{H}_A+\mathcal{H}_B$, can be written as
\begin{eqnarray}
 && \!\!\!\!\!\!\!\!\!\!\!\!\!
\langle LSJI_sF\mu|\mathcal{H}_T|LSJ^\prime I_sF^\prime\mu\rangle=
\delta_{JJ^\prime}\delta_{FF^\prime}\nonumber \\ &&
\!\!\!\!\!\!\!\!\!\!\!\!\!\times\bigg[\frac{1}{2}\zeta(LS)\{J(J+1)-L(L+1)-S(S+1)\}
\nonumber \\ &&
\!\!\!\!\!\!\!\!\!\!\!\!\!
+\frac{1}{2}\mathcal{A}_J \mathcal{K} +
\frac{\mathcal{B}_J}{8I_s(2I_s-1)J(2J-1)}\nonumber \\ &&
\!\!\!\!\!\!\!\!\!\!\!\!\!\times\{3\mathcal{K}(\mathcal{K}+1)-4J(J+1)I_s(I_s+1)\}\bigg]
\nonumber \\ &&
\!\!\!\!\!\!\!\!\!\!\!\!\!+\mu_0 B (-1)^{L+S+J+J^\prime+I_s-\mu+1}
\nonumber \\ &&
\!\!\!\!\!\!\!\!\!\!\!\!\!\times\sqrt{(2J+1)(2J^\prime+1)(2F+1)(2F^\prime+1)}
\nonumber \\ &&
\!\!\!\!\!\!\!\!\!\!\!\!\!\times\left (
\begin{array}{ccc}
F & F^\prime & 1\\
-\mu & \mu & 0 \\
\end{array}
\right )
\left\lbrace
\begin{array}{ccc}
J & J^\prime & 1\\
F^\prime & F & I_s \\
\end{array}
\right\rbrace
\nonumber \\ &&
\!\!\!\!\!\!\!\!\!\!\!\!\!\times\bigg[\delta_{JJ^\prime}(-1)^{L+S+J+1}
\frac{\sqrt{J(J+1)}}{\sqrt{2J+1}}
\nonumber \\ &&
\!\!\!\!\!\!\!\!\!\!\!\!\!+(-1)^{J-J^\prime}\sqrt{S(S+1)(2S+1)}
\left\lbrace
\begin{array}{ccc}
J & J^\prime & 1\\
S & S & L \\
\end{array}
\right\rbrace\bigg]\ ,
\nonumber \\ &&
\label{tot-ham}
\end{eqnarray}
where $\mathcal{K}=F(F+1)-I_s(I_s+1)-J(J+1)$, and $\mu_0$ is the Bohr magneton.
The total Hamiltonian matrix in the combined theory is no longer a symmetric
tridiagonal matrix, unlike the case of the PB effect in fine or hyperfine structure
states. Instead, it is a full symmetric matrix and we diagonalize it using the
Givens--Householder method described in \citet{ort68}. We test the diagonalization
code written for the problem at hand using the principle of spectroscopic
stability presented in Appendix~\ref{a-b}.

\subsection{Eigenvalues and Eigenvectors}
\label{ee}
The diagonalization of the total Hamiltonian gives the energy eigenvectors
in terms of the linear Zeeman effect regime basis $|LSJI_sF\mu\rangle$
through the expansion coefficients $C^{k}_{JF}$ as
\begin{equation}
|LSI_s,k\mu\rangle= 
\sum_{JF}C^{k}_{JF}(LSI_s,\mu)
\ |LSJI_sF\mu\rangle\ .
\label{basis-good}
\end{equation}
The symbol $k$ labels different states corresponding to the given values
of $(L,S,I_s,\mu)$ and its dimension is given by 
\begin{eqnarray}
 && \!\!\!\!\!\!\!\!\!\!\!\!\!\!\!N_k=\sum^{L+S}_{d=|L-S|} 1+d+I_s-{\rm max}
(|\mu|,|d-I_s|)\ .
\label{kdim}
\end{eqnarray}
We assume the $C$-coefficients appearing in Equation~(\ref{basis-good}) 
to be real because the total Hamiltonian is real. We obtain the $C$-coefficients
and the corresponding eigenvalues denoted here as $E_k(LSI_s,\mu)$ after
diagonalizing the atomic and magnetic Hamiltonians presented in Sections
\ref{ath} and \ref{mah}.

\section{THE REDISTRIBUTION MATRIX FOR THE COMBINED $J$ AND $F$ STATE INTERFERENCES}
The methodology followed to derive the PRD matrix (RM) for the combined case of $J$ and
$F$ state interferences in the presence of a magnetic field is similar to that
presented in \citet{sow14b} for $F$ state interference alone. For the sake 
of clarity, we only present the important equations involved in the derivation.

In a single scattering event, the scattered radiation is related to the 
incident radiation through the Mueller matrix given by
\begin{equation}
{\bf M}={\bf TWT^{-1}}\ .
\label{muel}
\end{equation}
Here, ${\bf T}$ and ${\bf T^{-1}}$ are the purely mathematical transformation matrices
and ${\bf W}$ is the coherency matrix for a transition $a\rightarrow b\rightarrow f$
defined by
\begin{equation}
{\bf W} = \sum_{a}\sum_{f} {\bm w}\otimes {\bm w}^*\ . 
\label{w-mat}
\end{equation}
Note that the summations over the initial ($a$) and final ($f$) states
are incoherent, and therefore do not allow the lower levels to interfere. ${\bm w}$ in
the above equation is the Jones matrix and its elements are given by
the Kramers--Heisenberg formula, which gives the complex probability
amplitudes for the scattering $a\rightarrow b\rightarrow f$ as
\begin{equation}
w_{\alpha\beta} \sim \sum_b
\frac{\langle f|{\bf r}\cdot{\bf e}_\alpha|b\rangle
\langle b|{\bf r}\cdot{\bf e}_\beta|a\rangle}{\omega_{bf}-\omega-{\rm i}\gamma/2}\ .
\label{jones}
\end{equation}
Here, $\omega=2\pi\xi$ is the circular frequency of the scattered radiation.
$\hbar\omega_{bf}$ is the energy difference between the excited and final
levels and $\gamma$ is the damping constant.

Using Equation~(\ref{basis-good}) in the Kramers--Heisenberg formula, and noting
that $L_f=L_a$ and using the Wigner--Eckart theorem (see Equations (2.96) and
(2.108) of LL04), we arrive at
\begin{eqnarray}
&& \!\!\!\!\!\!\!w_{\alpha\beta} \sim (2L_a+1)\sum_{k_b\mu_b}
\sum_{J_aJ_fJ_bJ_{b^{\prime\prime}}F_aF_fF_bF_{b^{\prime\prime}}} \nonumber \\ && 
\!\!\!\!\!\!\! \times\sum_{qq\prime\prime}(-1)^{q-q^{\prime\prime}}  
 (-1)^{J_f+J_a+J_b+J_{b^{\prime\prime}}} \nonumber \\ && 
\!\!\!\!\!\!\! \times C^{k_f}_{J_fF_f}(L_aSI_s,\mu_f)C^{k_a}_{J_aF_a}(L_aSI_s,\mu_a)
\nonumber \\ &&\!\!\!\!\!\!\! \times C^{k_b}_{J_bF_b}(L_bSI_s,\mu_b)
C^{k_b}_{J_{b^{\prime\prime}}F_{b^{\prime\prime}}}(L_bSI_s,\mu_b)
\nonumber \\ && 
\!\!\!\!\!\!\! \times\sqrt{(2F_a+1)(2F_f+1)(2F_b+1)(2F_{b^{\prime\prime}}+1)}
\nonumber \\ && 
\!\!\!\!\!\!\! \times\sqrt{(2J_a+1)(2J_f+1)(2J_b+1)(2J_{b^{\prime\prime}}+1)}
\nonumber \\ && 
\!\!\!\!\!\!\! \times\left (
\begin{array}{ccc}
F_b & F_f & 1\\
-\mu_b & \mu_f & -q \\
\end{array}
\right )
\left (
\begin{array}{ccc}
F_{b^{\prime\prime}} & F_a & 1\\
-\mu_b & \mu_a & -q^{\prime\prime} \\
\end{array}
\right )
\nonumber \\ && 
\!\!\!\!\!\!\! \times\left\lbrace
\begin{array}{ccc}
J_f & J_b & 1\\
F_b & F_f & I_s \\
\end{array}
\right\rbrace
\left\lbrace
\begin{array}{ccc}
J_a & J_{b^{\prime\prime}} & 1\\
F_{b^{\prime\prime}} & F_a & I_s \\
\end{array}
\right\rbrace
\nonumber \\ && 
\!\!\!\!\!\!\! \times\left\lbrace
\begin{array}{ccc}
L_a & L_b & 1\\
J_b & J_f & S \\
\end{array}
\right\rbrace
\left\lbrace
\begin{array}{ccc}
L_a & L_b & 1\\
J_{b^{\prime\prime}} & J_a & S \\
\end{array}
\right\rbrace
\nonumber \\ && 
\!\!\!\!\!\!\!\times \ 
\varepsilon^{\alpha^*}_{q} \varepsilon^{\beta}_{q^{\prime\prime}}
\ \Phi_{\gamma}(\nu_{k_b\mu_bk_f\mu_f}-\xi)\ .
\label{coh-ele}
\end{eqnarray}
Here, $\varepsilon$ are the spherical vector components of the polarization
unit vectors (${\bf e}_\alpha$ and ${\bf e}_\beta$) with $\alpha$ and 
$\beta$ referring to the scattered and incident rays,
respectively. $\Phi_{\gamma}(\nu_{k_b\mu_bk_f\mu_f}-\xi)$
is the frequency-normalized profile function defined as
\begin{equation}
\Phi_{\gamma}(\nu_{k_b\mu_bk_f\mu_f}-\xi)=\frac{1/\pi {\rm i}}
{\nu_{k_b\mu_bk_f\mu_f}-\xi-{\rm i}\gamma/4\pi}\ ,
\label{norm-prof}
\end{equation}
where we have used an abbreviation 
\begin{eqnarray}
&& \!\!\!\!\!\!\!\!\!\!\nu_{k_b\mu_bk_f\mu_f}=\nu_{L_bSI_sk_b\mu_b,L_aSI_sk_f\mu_f}
\nonumber \\ &&
\!\!\!\!\!\!\!\!\!\!=\nu_{L_bL_a} + \frac{E_{k_b}(L_bSI_s,\mu_b)
-E_{k_f}(L_aSI_s,\mu_f)}{h}\ ,
\nonumber \\ &&
\label{freq}
\end{eqnarray}
with $h$ being the Planck constant.

Inserting Equation~(\ref{coh-ele}) into Equation~(\ref{w-mat}), and after elaborate
algebra \citep[see for example][]{sow14b}, we obtain the normalized RM,
${\bf R}^{\rm II}_{ij}$, for type-II scattering in the laboratory frame as
\begin{eqnarray}
&&\!\!\!\!\!\!\!\!\!\!{\bf R}_{ij}^{\rm II}
(x,{\bm n},x^\prime,{\bm n}^\prime;{\bm B})=
\frac{3(2L_b+1)}{(2S+1)(2I_s+1)}
\nonumber \\ && \!\!\!\!\!\!\!\!\!\! \times
\sum_{KK^\prime Q}
\sum_{k_a\mu_ak_f\mu_fk_b\mu_bk_{b^\prime}\mu_{b^\prime}}
\sum_{qq^\prime q^{\prime\prime}
q^{\prime\prime\prime}}
(-1)^{q-q^{\prime\prime\prime}+Q}
\nonumber \\ && \!\!\!\!\!\!\!\!\!\! \times
\sqrt{(2K+1)(2K^\prime+1)}
\cos\beta_{k_{b^\prime}\mu_{b^\prime}k_b\mu_b}
{\rm e}^{{\rm i}\beta_{k_{b^\prime}\mu_{b^\prime}k_b\mu_b}}
\nonumber \\ && \!\!\!\!\!\!\!\!\!\! \times
[(h^{\rm II}_{k_b\mu_b,k_{b^\prime}\mu_{b^\prime}})_{k_a\mu_ak_f\mu_f}+
{\rm i}(f^{\rm II}_{k_b\mu_b,k_{b^\prime}\mu_{b^\prime}})_{k_a\mu_ak_f\mu_f}]
\nonumber \\ && \!\!\!\!\!\!\!\!\!\! \times
\sum_{J_aJ_{a^\prime}J_fJ_{f^\prime}J_bJ_{b^\prime}
J_{b^{\prime\prime}}J_{b^{\prime\prime\prime}}}
\sum_{F_aF_{a^\prime}F_fF_{f^\prime}F_bF_{b^\prime}
F_{b^{\prime\prime}}F_{b^{\prime\prime\prime}}}
\nonumber \\ && \!\!\!\!\!\!\!\!\!\! \times 
C^{k_f}_{J_fF_f}(L_aSI_s,\mu_f) C^{k_a}_{J_aF_a}(L_aSI_s,\mu_a) 
\nonumber \\ && \!\!\!\!\!\!\!\!\!\! \times
C^{k_b}_{J_bF_b}(L_bSI_s,\mu_b)
C^{k_b}_{J_{b^{\prime\prime}}F_{b^{\prime\prime}}}(L_bSI_s,\mu_b)
\nonumber \\ && \!\!\!\!\!\!\!\!\!\! \times
C^{k_f}_{J_{f^\prime}F_{f^\prime}}(L_aSI_s,\mu_f)
C^{k_a}_{J_{a^\prime}F_{a^\prime}}(L_aSI_s,\mu_a)
\nonumber \\ && \!\!\!\!\!\!\!\!\!\! \times
C^{k_{b^\prime}}_{J_{b^\prime}F_{b^\prime}}(L_bSI_s,\mu_{b^\prime})
C^{k_{b^\prime}}_{J_{b^{\prime\prime\prime}}F_{b^{\prime\prime\prime}}}
(L_bSI_s,\mu_{b^\prime})
\nonumber \\ && \!\!\!\!\!\!\!\!\!\! \times
(-1)^{J_a+J_{a^\prime}+J_f+J_{f^\prime}+J_b+J_{b^\prime}+J_{b^{\prime\prime}}
+J_{b^{\prime\prime\prime}}}
\nonumber \\ && \!\!\!\!\!\!\!\!\!\! \times
\sqrt{(2J_a+1)(2J_f+1)(2J_{a^\prime}+1)(2J_{f^\prime}+1)}
\nonumber \\ && \!\!\!\!\!\!\!\!\!\! \times
\sqrt{(2J_b+1)(2J_{b^\prime}+1)(2J_{b^{\prime\prime}}+1)
(2J_{b^{\prime\prime\prime}}+1)}
\nonumber \\ && \!\!\!\!\!\!\!\!\!\! \times
\sqrt{(2F_a+1)(2F_f+1)(2F_{a^\prime}+1)(2F_{f^\prime}+1)}
\nonumber \\ && \!\!\!\!\!\!\!\!\!\! \times
\sqrt{(2F_b+1)(2F_{b^\prime}+1)(2F_{b^{\prime\prime}}+1)
(2F_{b^{\prime\prime\prime}}+1)}
\nonumber \\ && \!\!\!\!\!\!\!\!\!\! \times
\left (
\begin{array}{ccc}
F_b & F_f & 1\\
-\mu_b & \mu_f & -q \\
\end{array}
\right )
\left (
\begin{array}{ccc}
F_{b^\prime} & F_{f^\prime} & 1\\
-\mu_{b^\prime} & \mu_f & -q^{\prime} \\
\end{array}
\right )
\nonumber \\ && \!\!\!\!\!\!\!\!\!\! \times
\left (
\begin{array}{ccc}
F_{b^{\prime\prime}} & F_a & 1\\
-\mu_b & \mu_a & -q^{\prime\prime} \\
\end{array}
\right )
\left (
\begin{array}{ccc}
F_{b^{\prime\prime\prime}} & F_{a^\prime} & 1\\
-\mu_{b^\prime} & \mu_a & -q^{\prime\prime\prime} \\
\end{array}
\right )
\nonumber \\ && \!\!\!\!\!\!\!\!\!\! \times
\left (
\begin{array}{ccc}
1 & 1 & K\\
q & -q^{\prime} & -Q \\
\end{array}
\right )
\left (
\begin{array}{ccc}
1 & 1 & K^\prime\\
q^{\prime\prime\prime} & -q^{\prime\prime} & Q\\
\end{array}
\right )
\nonumber \\ && \!\!\!\!\!\!\!\!\!\! \times
\left\lbrace
\begin{array}{ccc}
J_f & J_b & 1\\
F_b & F_f & I_s \\
\end{array}
\right\rbrace
\left\lbrace
\begin{array}{ccc}
J_{f^\prime} & J_{b^\prime} & 1\\
F_{b^\prime} & F_{f^\prime} & I_s \\
\end{array}
\right\rbrace
\nonumber \\ && \!\!\!\!\!\!\!\!\!\! \times
\left\lbrace
\begin{array}{ccc}
J_a & J_{b^{\prime\prime}} & 1\\
F_{b^{\prime\prime}} & F_a & I_s \\
\end{array}
\right\rbrace
\left\lbrace
\begin{array}{ccc}
J_{a^\prime} & J_{b^{\prime\prime\prime}} & 1\\
F_{b^{\prime\prime\prime}} & F_{a^\prime} & I_s \\
\end{array}
\right\rbrace
\nonumber \\ && \!\!\!\!\!\!\!\!\!\! \times
\left\lbrace
\begin{array}{ccc}
L_a & L_b & 1\\
J_b & J_f & S \\
\end{array}
\right\rbrace
\left\lbrace
\begin{array}{ccc}
L_a & L_b & 1\\
J_{b^\prime} & J_{f^\prime} & S \\
\end{array}
\right\rbrace
\nonumber \\ && \!\!\!\!\!\!\!\!\!\! \times
\left\lbrace
\begin{array}{ccc}
L_a & L_b & 1\\
J_{b^{\prime\prime}} & J_a & S \\
\end{array}
\right\rbrace
\left\lbrace
\begin{array}{ccc}
L_a & L_b & 1\\
J_{b^{\prime\prime\prime}} & J_{a^\prime} & S \\
\end{array}
\right\rbrace
\nonumber \\ && \!\!\!\!\!\!\!\!\!\! \times
(-1)^Q \mathcal{T}^K_{-Q}(i,{\bm n})
\mathcal{T}^{K^\prime}_Q(j,{\bm n}^\prime)\ .
\label{final-rm}
\end{eqnarray}
Here, $\mathcal{T}^K_Q(i,\bm n)$ are the irreducible spherical tensors for
polarimetry \citep{landi84} with $i=0,1,2,3$ referring to the Stokes parameters, the
multipolar index $K=0,1,2,$ and $Q\in[-K,K]$. $\bm n^\prime$ and $\bm n$ represent the
directions of the incident and scattered rays, respectively, and $\bm B$ the
vector magnetic field. $x^\prime$ and $x$ are the non-dimensional frequencies in Doppler
width units (see Appendix~\ref{a-a}). $\beta_{k_{b^\prime}\mu_{b^\prime}k_b\mu_b}$ is
the Hanle angle given by
\begin{equation}
 \tan\beta_{k_{b^\prime}\mu_{b^\prime}k_b\mu_b}=
\frac{\nu_{k_{b^\prime}\mu_{b^\prime}k_a\mu_a}-\nu_{k_b\mu_bk_a\mu_a}}{\gamma/2\pi}\ .
\label{hanle-beta}
\end{equation}
The explicit forms of the auxiliary functions $h^{\rm II}$ and $f^{\rm II}$ appearing
in Equation~(\ref{final-rm}) are given in Appendix~\ref{a-a}. When $I_s=0$,
Equation~(\ref{final-rm}) reduces to the PRD matrix for $J$ state interference
alone \citep[see Equation~(11) of][]{sow14a}. When FS is neglected,
Equation~(\ref{final-rm}) reduces to the expression of RM for pure $F$ state
interference \citep[see Equation~(16) of][]{sow14b}. When we neglect both FS and HFS,
we recover RM for $L_a\rightarrow L_b\rightarrow L_a$ transition (analogous to a
two-level atom case) in the presence of a magnetic field.

\section{RESULTS}
\label{res}
In this section, we present the results obtained from the combined theory
for the case of the single scattering of an unpolarized, spectrally flat incident
radiation beam by an atom with both non-zero
electron and nuclear spins. Considering the relevance to solar applications, we
choose the D line system at 6708 \AA{} from neutral $^6$Li and $^7$Li isotopes as an
example to test the formalism developed. We take the values of the atomic parameters
and isotope abundances for this system from Table 1 of \citet{bel09}.

\subsection{Level-crossings and Avoided Crossings}
\label{sec-1}

\begin{figure*}
\begin{center}
\includegraphics[scale=0.48]{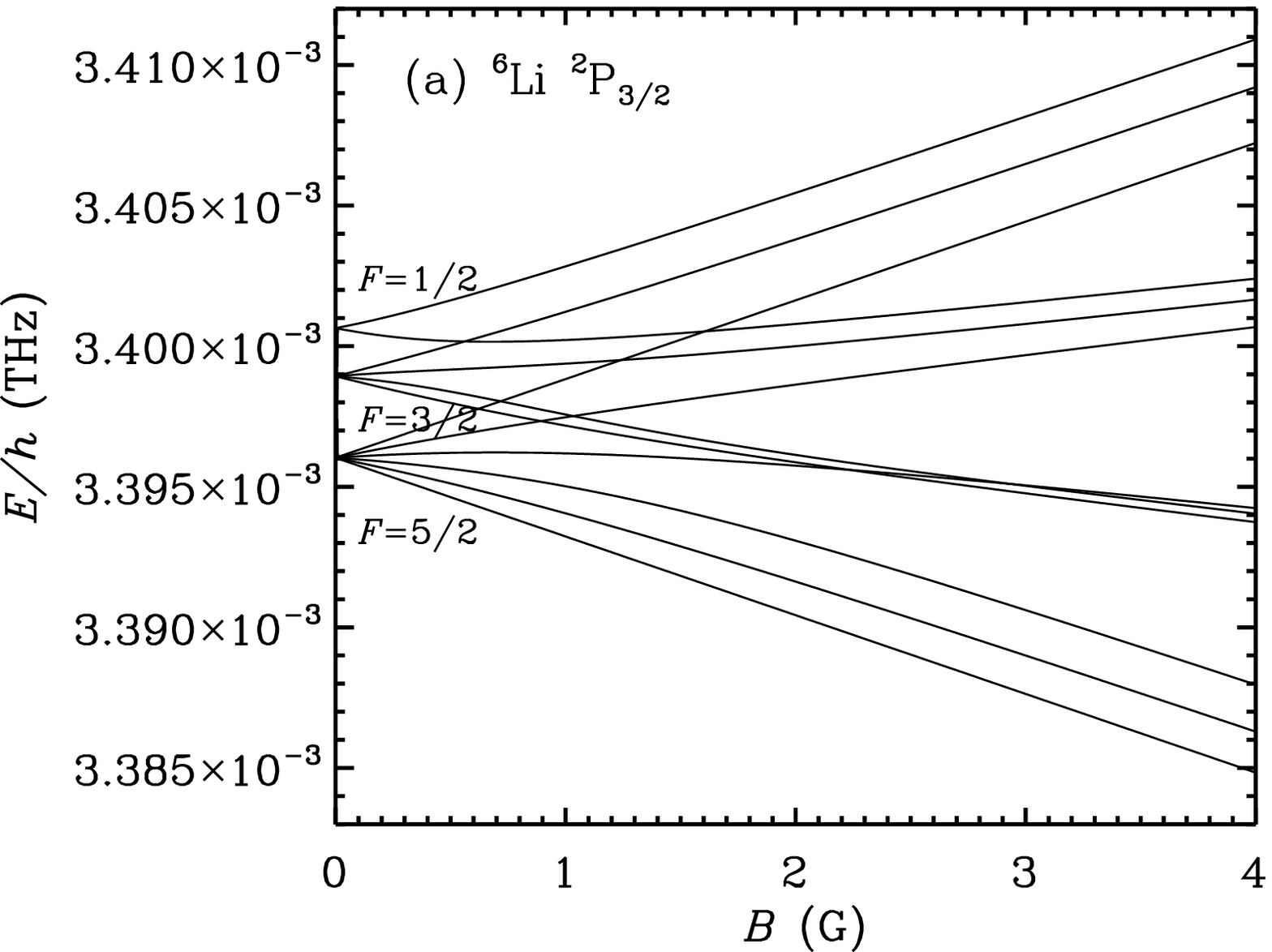}
\includegraphics[scale=0.48]{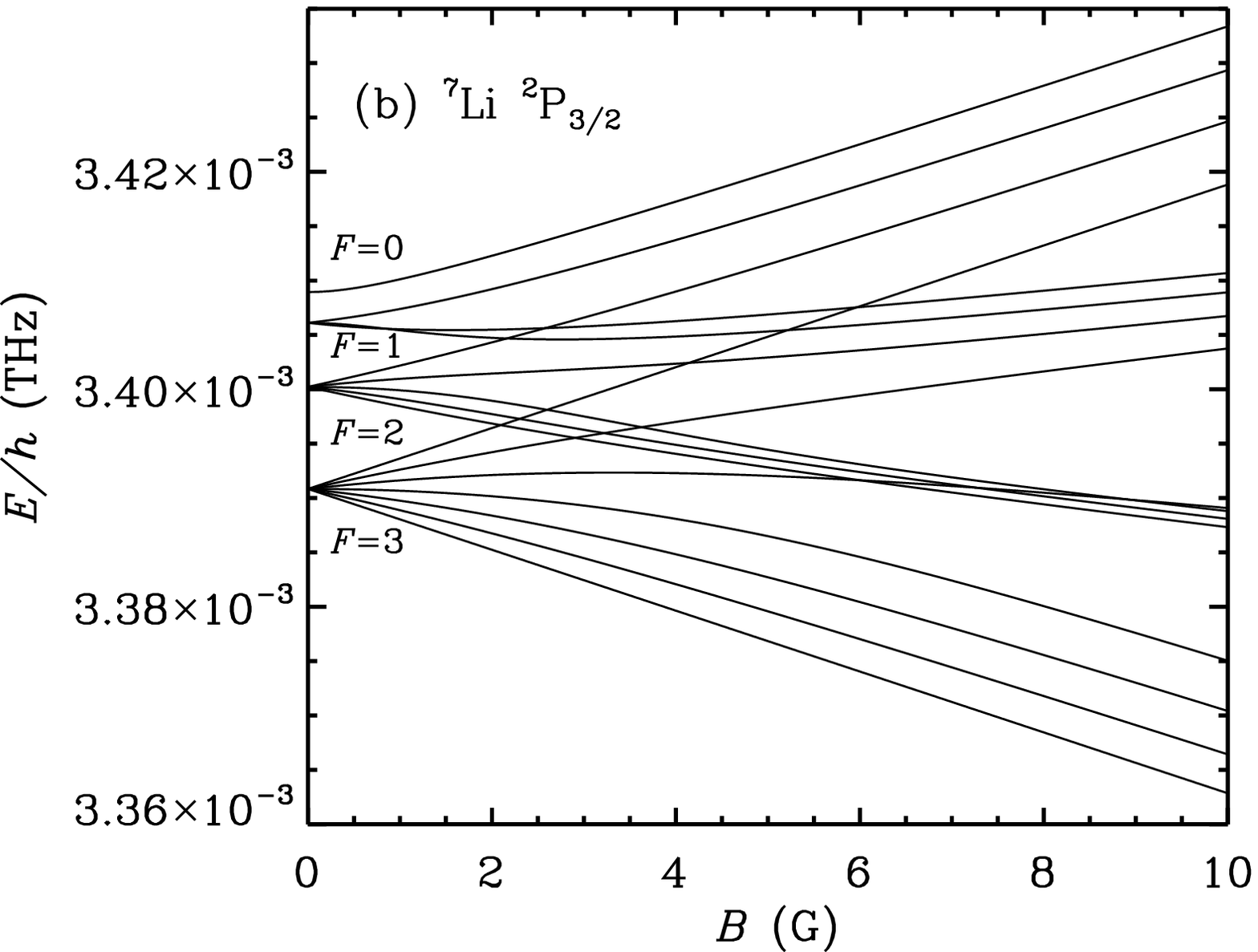}
\includegraphics[scale=0.48]{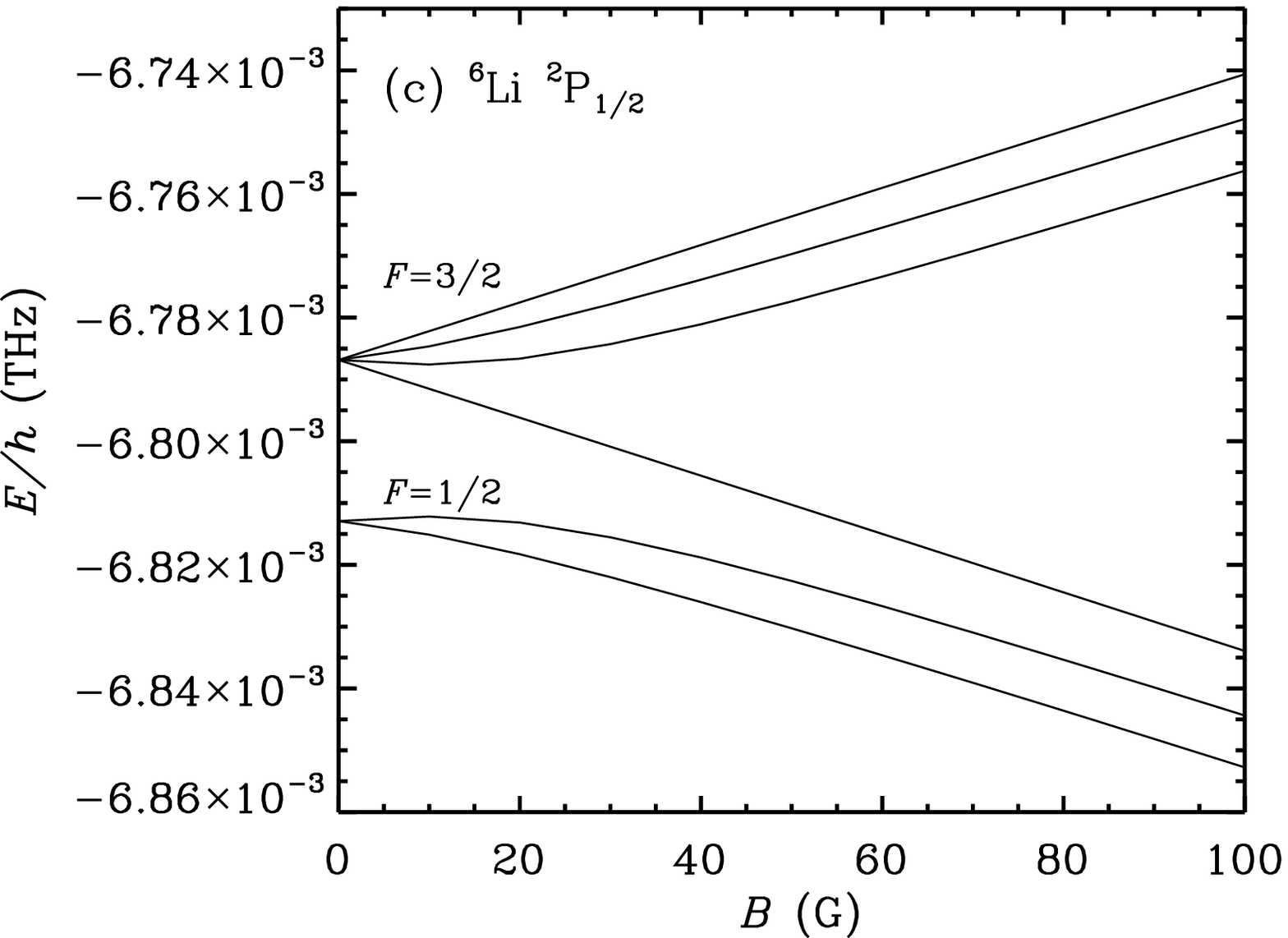}
\includegraphics[scale=0.48]{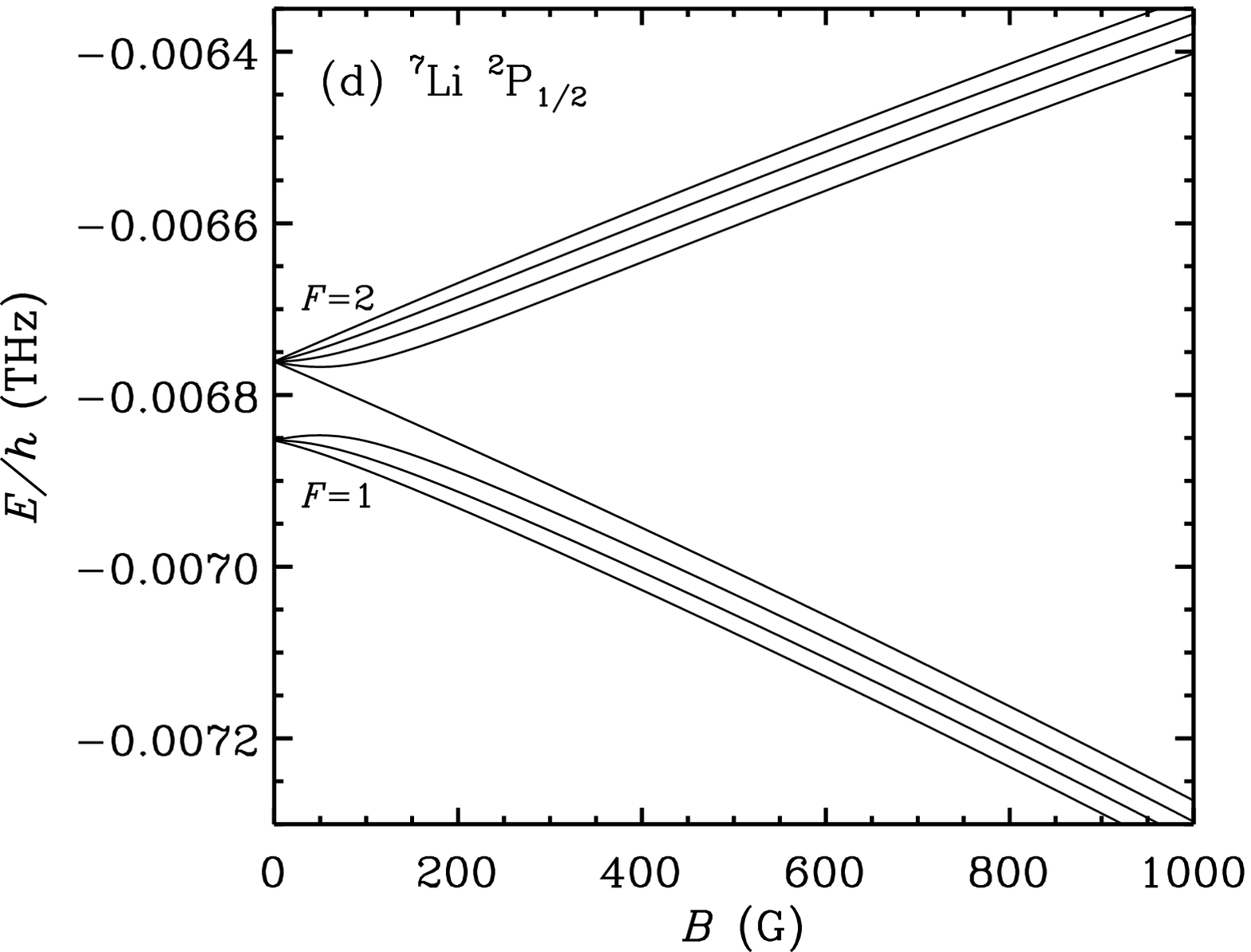}
\caption{Energies of the HFS magnetic substates as a function of the magnetic
field strength for $^6$Li (left column) and $^7$Li (right column). Panels (a) and (b)
correspond, respectively, to the $^2$P$_{3/2}$ levels of $^6$Li and $^7$Li, while panels
(c) and (d) correspond to the $^2$P$_{1/2}$ levels of $^6$Li and $^7$Li, respectively.
The nuclear spins of $^6$Li and $^7$Li are 1 and 3/2, respectively.
\label{level-fig-1}
}
\end{center}
\end{figure*}
\begin{figure*}
\begin{center}
\includegraphics[scale=0.48]{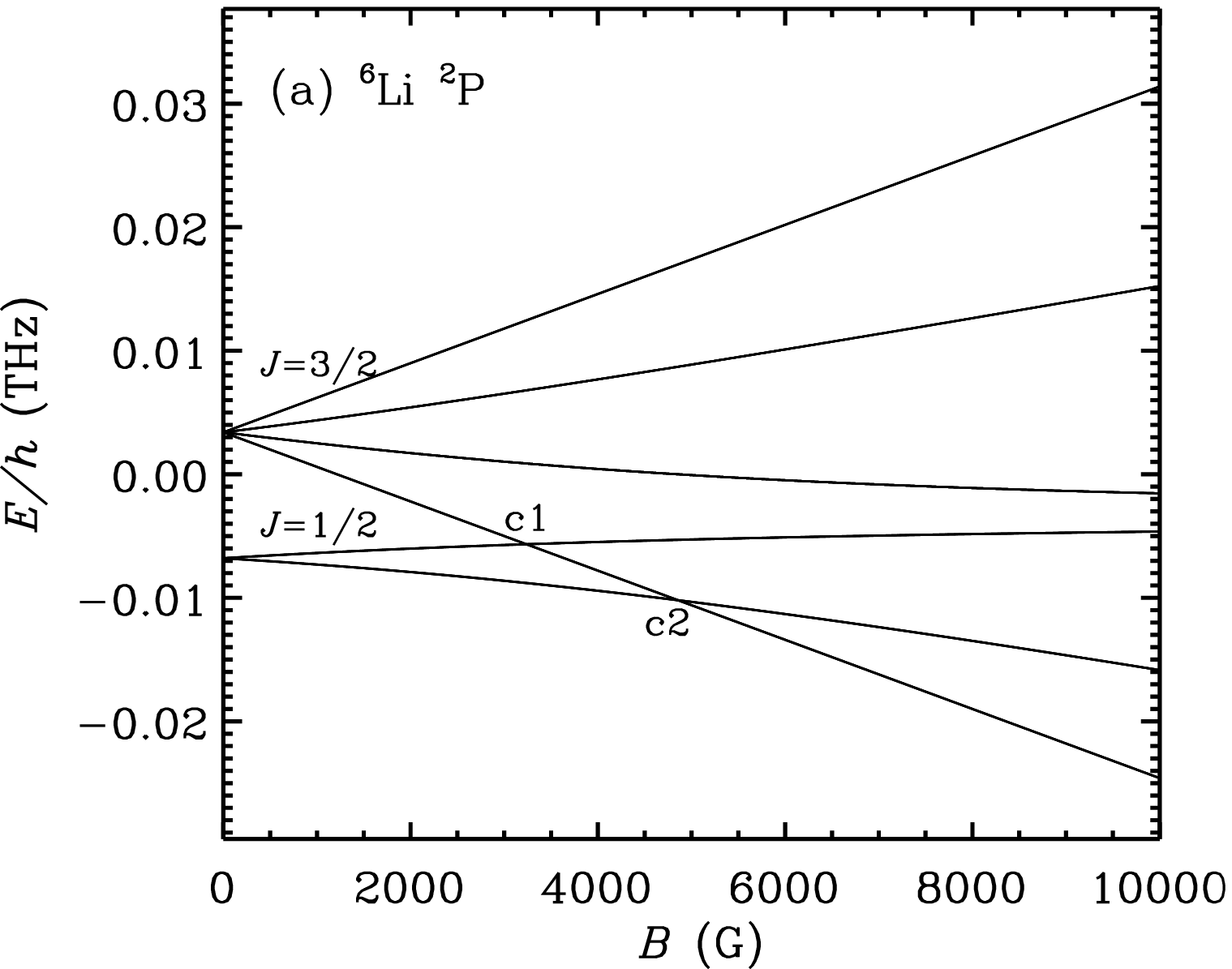}
\includegraphics[scale=0.48]{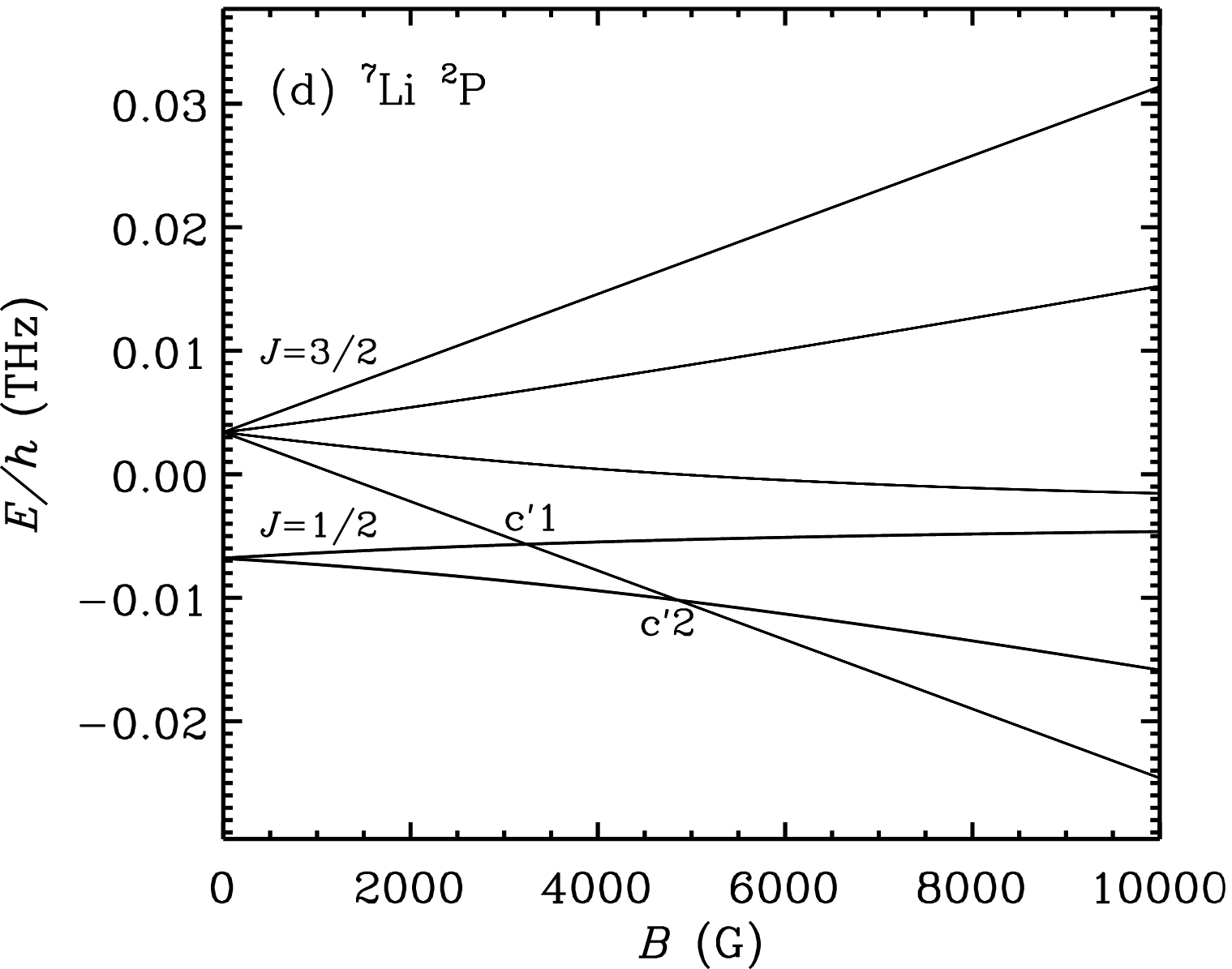}
\includegraphics[scale=0.48]{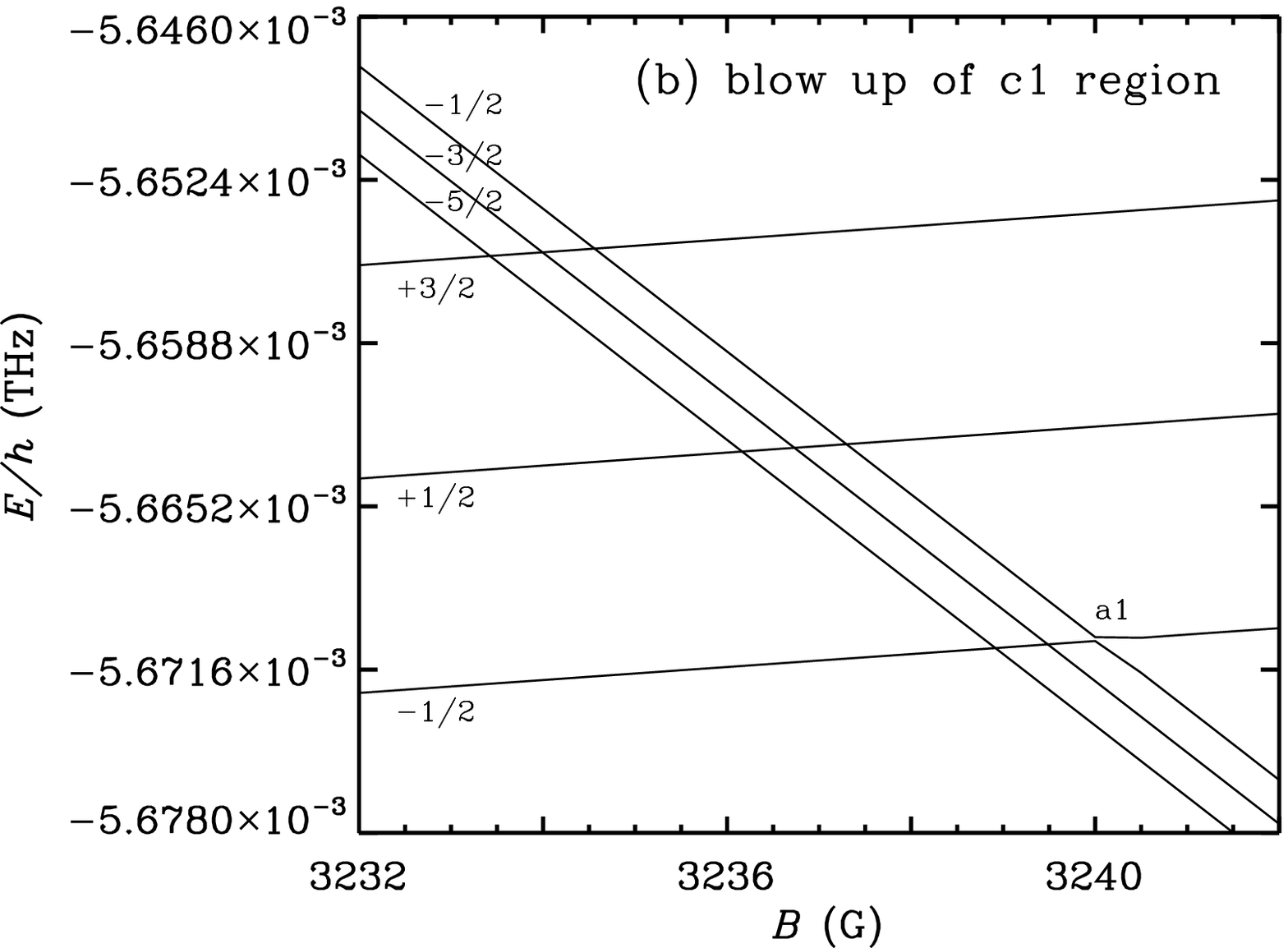}
\includegraphics[scale=0.48]{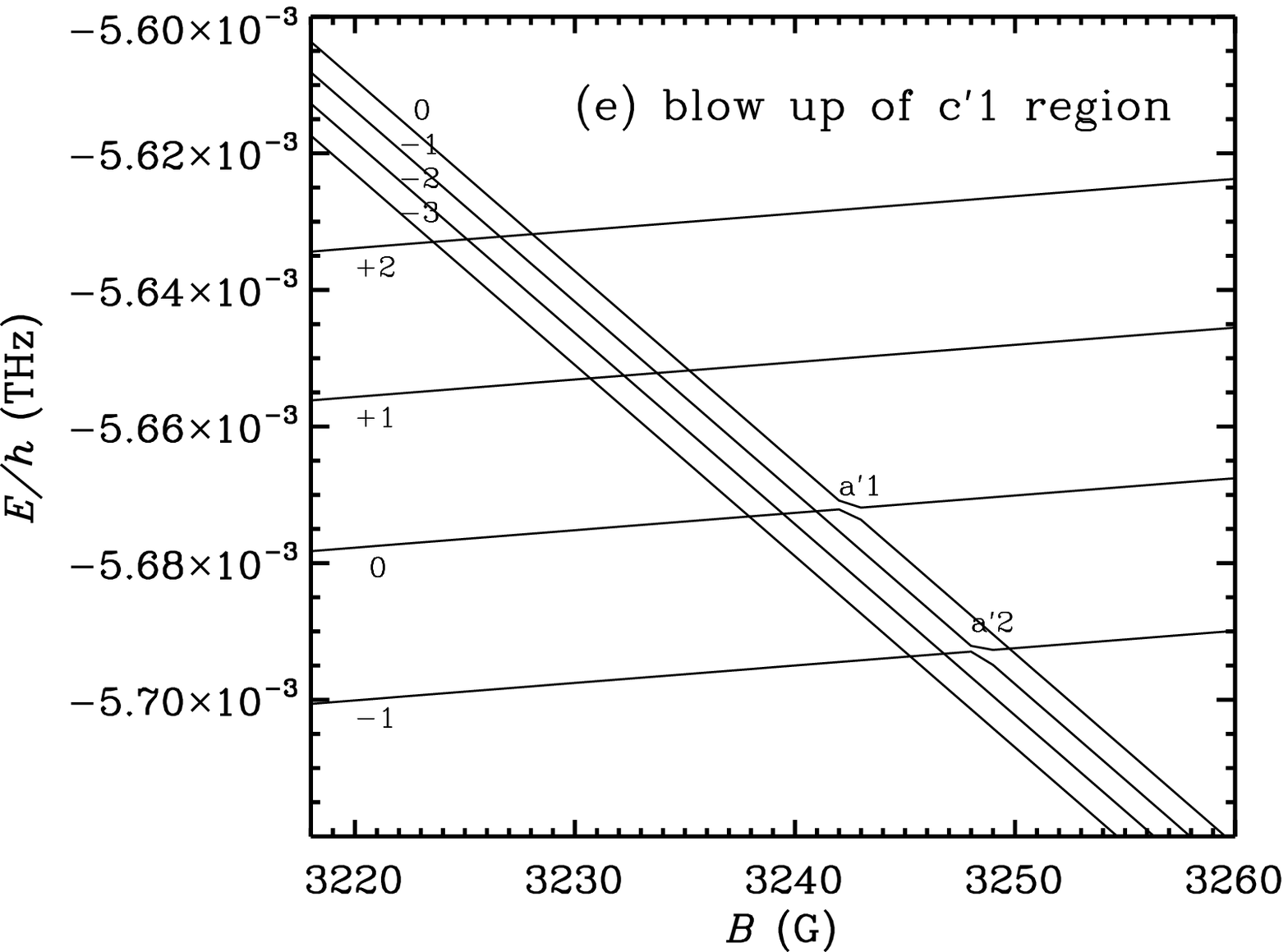}
\includegraphics[scale=0.47]{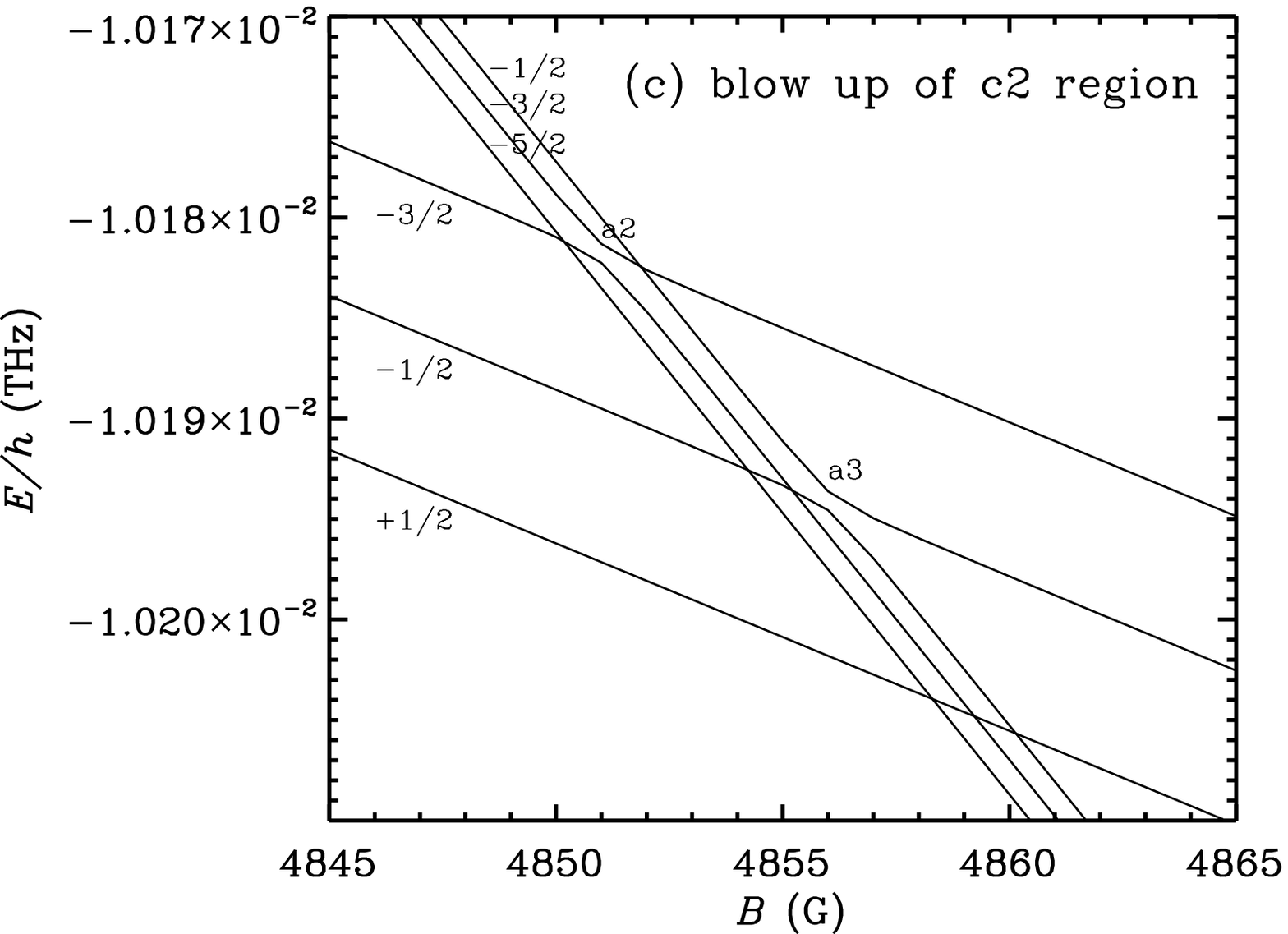}
\includegraphics[scale=0.47]{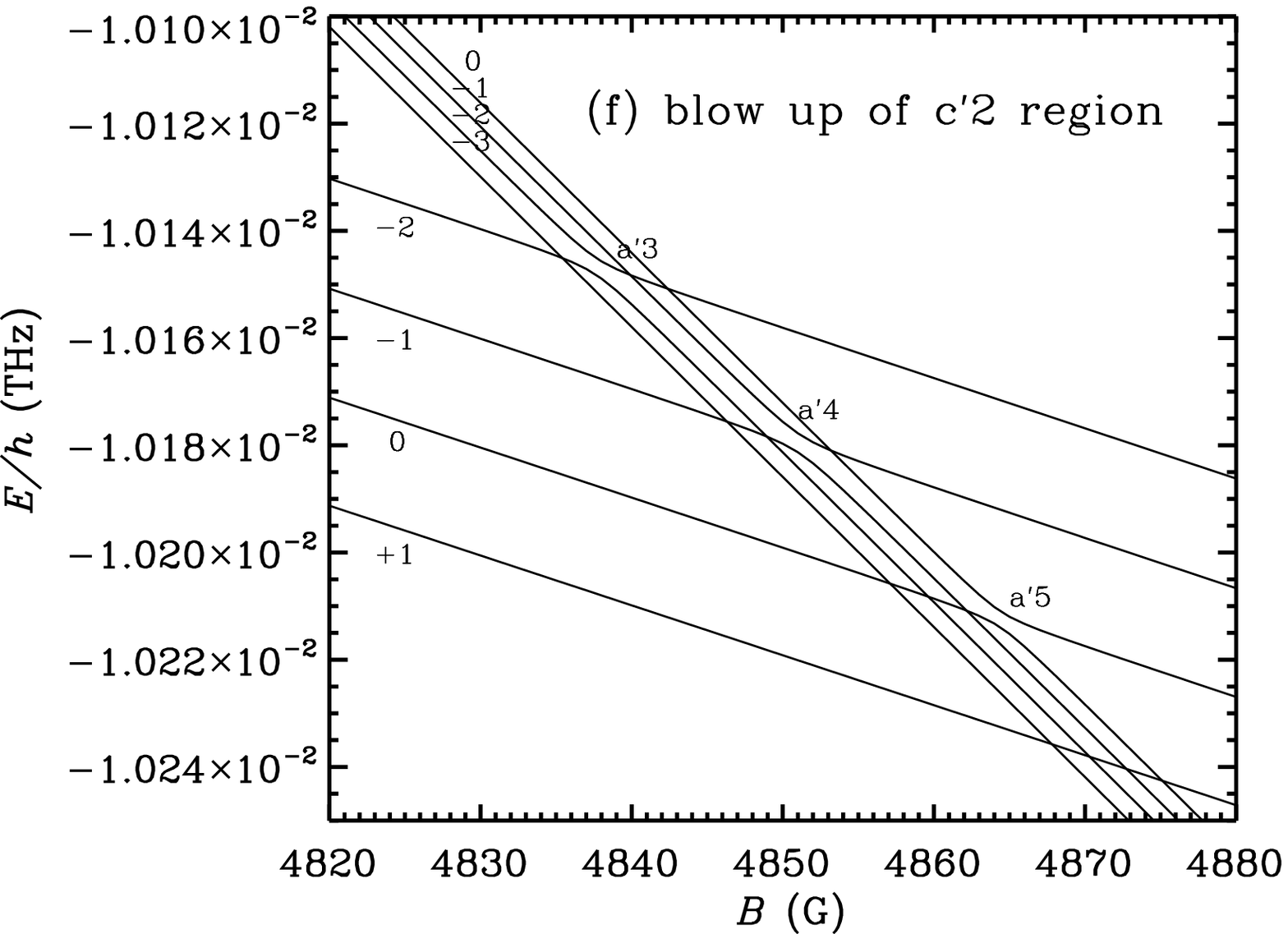}
\caption{Energies of the magnetic substates belonging to the $^2$P terms as a function
of the magnetic field strength for $^6$Li (a) and $^7$Li (d). Blow up of the crossing
regions c1 (b) and c2 (c) in $^6$Li and c$^\prime$1 (e) and c$^\prime$2 (f) in $^7$Li.
In the panels (b), (c), (e), and (f) the levels are identified by their magnetic
quantum number values $\mu$.
\label{level-fig}
}
\end{center}
\end{figure*}

In Figures~\ref{level-fig-1} and  \ref{level-fig}, we show the dependence
of the energies of the levels in the $^2$P terms of the $^6$Li and $^7$Li isotopes
on the magnetic field strength. Such figures provide us with the information on
the field strength regimes in which processes like the Zeeman effect, incomplete PB
effect, and complete PB effect operate. They help us to choose the magnetic field
strength values for studying the effects of level-crossing on the Stokes profiles.
We choose different scales for the {\it x}-axes in different panels to
bring out the level-crossings which occur at different field strengths due to the
difference in the magnitudes of FS and HFS. The {\it y}-axes in all of the panels in both
figures denote the energy shift of the levels from the parent $L=1$ level. 

In panels (a) and (c) of Figure~\ref{level-fig-1}, we plot the energies
of the magnetic substates of the $F$ states belonging to the $^2$P$_{3/2}$
and $^2$P$_{1/2}$ levels of $^6$Li, respectively, as a function of the field
strength. Since the nuclear spin of $^6$Li is 1, we have half-integer values
for $F$. In these panels, we see that the magnetic substates of the $F$ states
of $^2$P$_{3/2}$ cross at nine points while those of $^2$P$_{1/2}$ do not cross.
We note a similar behavior in the case of the $F$ states belonging to the
$^2$P$_{3/2}$ and $^2$P$_{1/2}$ levels of $^7$Li (see panels (b) and (d),
respectively). The magnetic substates of the $F$ states of $^2$P$_{1/2}$ do
not cross while those of $^2$P$_{3/2}$ cross at 14 points. In the weak field regime
(e.g., $0-60$ G), we see PB effect for the $F$ states,
and in the strong field regime (for kG
fields) we see PB effect for the $J$ states. In Tables~\ref{tab-2} and \ref{tab-3}, we
list the quantum numbers of the levels which cross along with their corresponding field
strengths for the weak field regime. The numbers indicated in boldface in these tables
correspond to those crossings which satisfy $\Delta\mu=\mu_{b^\prime}-\mu_b=\pm2$. We
discuss the effects of these level-crossings on the polarization in later sections.

In panels (a) and (d) of Figure~\ref{level-fig}, we plot the energies of the
magnetic substates of the $^2$P terms of $^6$Li and $^7$Li as a function of the magnetic
field strength. In these panels, the points where the levels cross are denoted as c1
and c2 for $^6$Li and as c$^\prime$1 and c$^\prime$2 for $^7$Li. When we zoom into these
crossing points, we see other interesting phenomena (see panels (b), (c), (e), and (f)).
For example, at c1, we see a crossing of the bunch of lowermost three levels going
downward in Figure~\ref{level-fig-1}(a) with the three levels going upward in
Figure~\ref{level-fig-1}(c). Although the magnetic substates of the
$F$ states appear to be degenerate in Figure~\ref{level-fig}(a), they are not fully
degenerate, as can be seen in Figure~\ref{level-fig}(b). Similar behavior can be seen
in Figures~\ref{level-fig}(c), (e), and (f), and the levels correspond to the magnetic
substates of the $F$ states shown in Figure~\ref{level-fig-1}.

In addition to the usual level-crossings, we see several avoided crossings in
Figures~\ref{level-fig}(b), (c), (e), and (f). For example, in panel (b), we see one
avoided crossing marked a1, two in panel (c) marked a2 and a3, two in panel (e)
marked a$^\prime$1 and a$^\prime$2, and three in panel (f) marked a$^\prime$3,
a$^\prime$4, and a$^\prime$5. As we can see from the figure, these avoided crossings
take place between the magnetic substates with the same $\mu$ values ($-1/2$ in panel (b),
$-3/2$ and $-1/2$ in panel (c), 0 and $-1$ in panel (e), and $-2$, $-1$, and
0 in panel (f)). The levels with the same $\mu$ cannot
cross owing to the small interaction that takes
place between them. This interaction is determined by the off-diagonal elements of the
magnetic hyperfine interaction Hamiltonian which couple the states with different
$J$ values \citep{bro67,we67,ari77}. A rapid transformation in the eigenvector basis
takes place around the region of avoided crossing. This is described in \citet{bom80}
and in LL04 \citep[see also][]{sow14a,sow14b}.

\begin{table*}[ht]
 \begin{centering}
\begin{tabular}{cccccc}
\hline
\hline
$F_b\diagdown F_{b^\prime}$ & & 1/2 & 3/2 & 3/2 & 3/2 \\
\hline
\ \ \ & $\mu_b\diagdown \mu_{b^\prime}$ & 1/2 & $-1/2$ & 1/2 & 3/2 \\
\hline
3/2 & $-3/2$ & {\bf 0.57} & ... & ... & ...\\
\hline
5/2 & $-5/2$ & 1.61 & {\bf 1.26} & 0.72 & 0.63 \\

5/2 & $-3/2$ & ... & ... & {\bf 1.3} & 0.9 \\

5/2 & $-1/2$ & ... & ... & 2.93 & {\bf 2.25} \\
\hline
\end{tabular}
\caption{Magnetic field strengths (approximate values in G) for which the magnetic
substates of the $F$ states cross in the $^6$Li isotope. For instance, the crossing
between $(\mu_b=-3/2,F_b=3/2)$ and $(\mu_{b^\prime}=1/2,F_{b^\prime}=1/2)$ occurs
at $B\sim0.57$ G. The numbers highlighted in boldface represent the field strength
values for which the level-crossings corresponding to
$\Delta\mu=\mu_{b^\prime}-\mu_b=\pm2$ occur.
\label{tab-2}
}
\end{centering}
\end{table*}

\begin{table*}[ht]
 \begin{centering}
\begin{tabular}{cccccccc}
\hline
\hline
$F_b\diagdown F_{b^\prime}$ & & 1 & 1 & 2 & 2 & 2 & 2 \\
\hline
\ \ \ & $\mu_b\diagdown \mu_{b^\prime}$ & 0 & +1 & $-1$ & 0 & +1 & +2 \\
\hline
2 & $-2$ & {\bf 2.2} & 2.6 & ... & ... & ... & ...\\
\hline
3 & $-3$ & 5.2 & 5.95 & {\bf 4.15} & 2.65 & 2.35 & 2.1 \\

3 & $-2$ & ... & ... & ... & {\bf 3.7} & 3.25 & 2.95 \\

3 & $-1$ & ... & ... & ... & 8.8 & {\bf 7.25} & 6.0 \\
\hline
\end{tabular}
\caption{Magnetic field strengths (approximate values in G) for which the magnetic
substates of the $F$ states cross in the $^7$Li isotope. For instance, the crossing
between $(\mu_b=-2,F_b=2)$ and $(\mu_{b^\prime}=0,F_{b^\prime}=1)$ occurs at
$B\sim2.2$ G. The numbers highlighted in boldface represent the field strength
values for which the level-crossings corresponding to
$\Delta\mu=\mu_{b^\prime}-\mu_b=\pm2$ occur.
\label{tab-3}
}
\end{centering}
\end{table*}

\subsection{Line Splitting Diagrams}
\label{sec-2}
\begin{figure*}
\begin{center}
\includegraphics[scale=0.4]{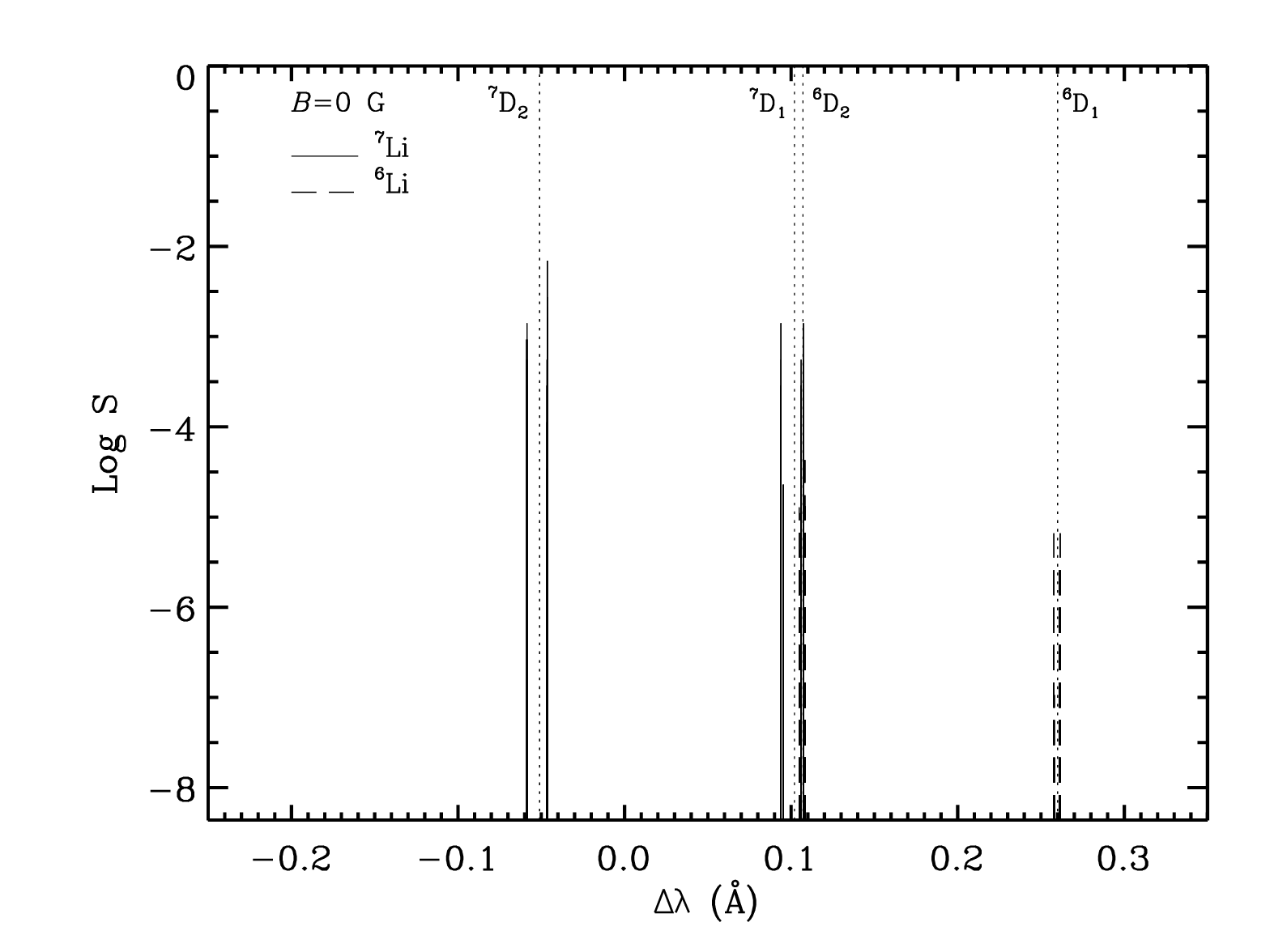}
\includegraphics[scale=0.4]{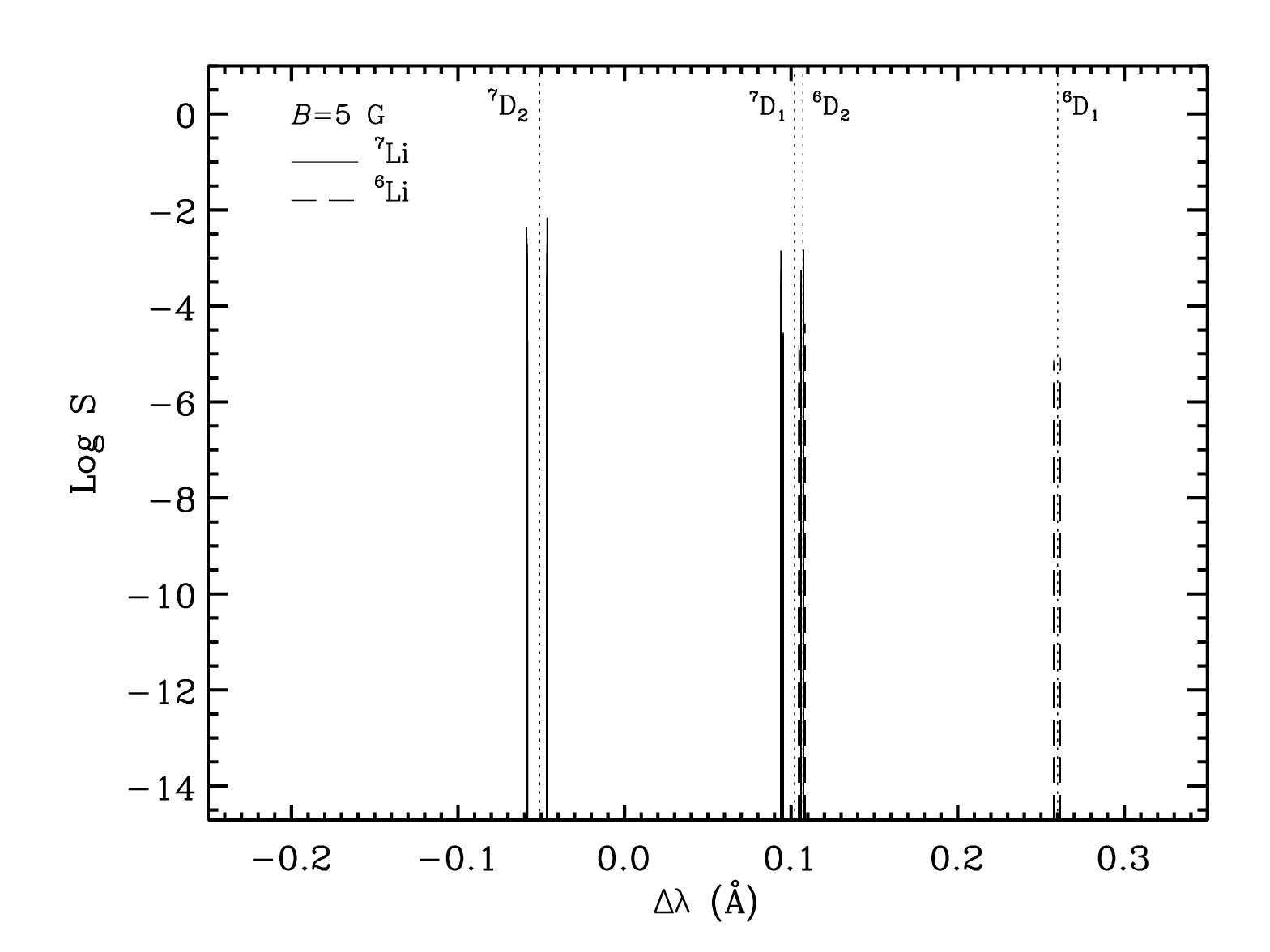}
\includegraphics[scale=0.4]{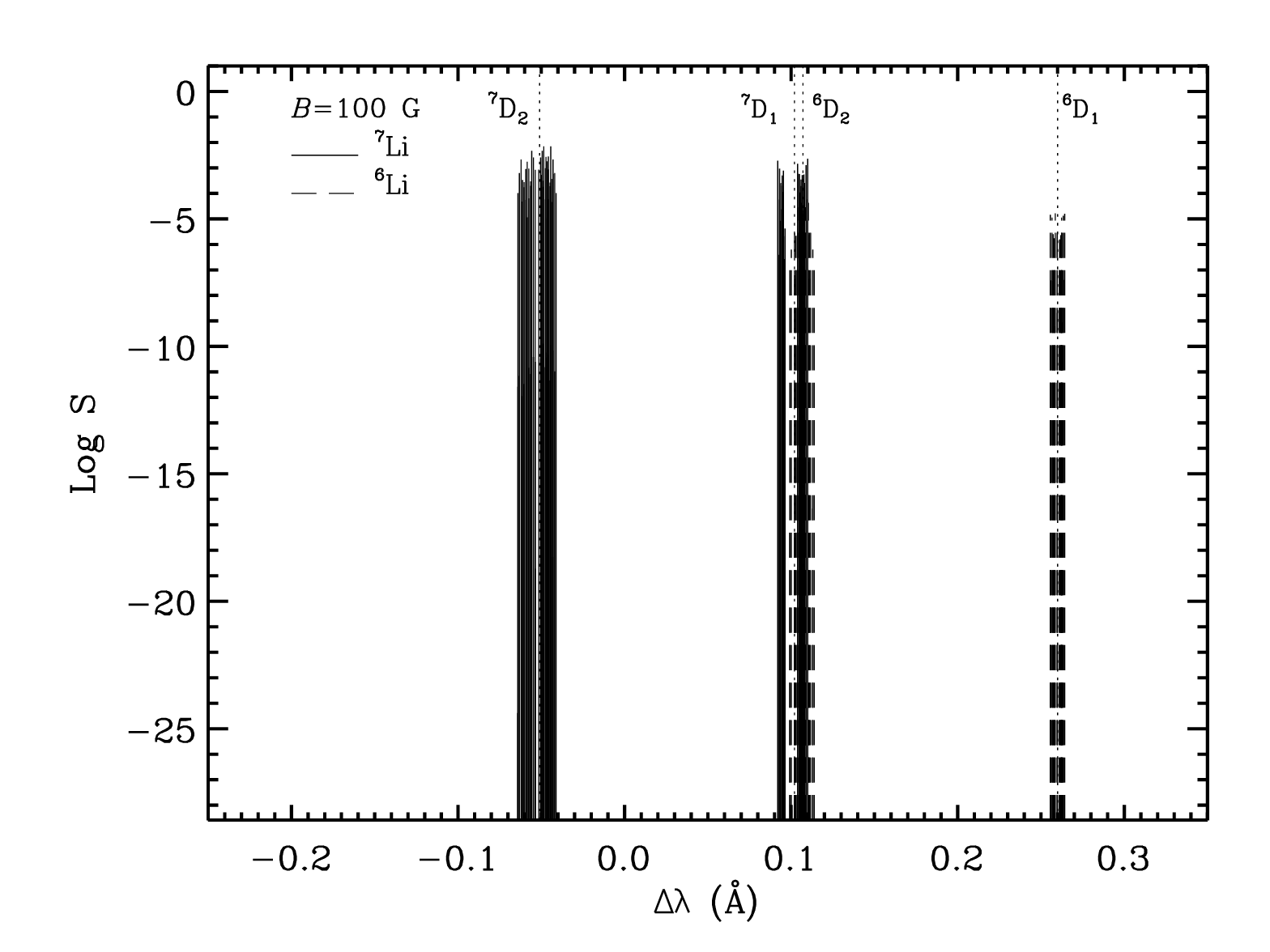}
\includegraphics[scale=0.4]{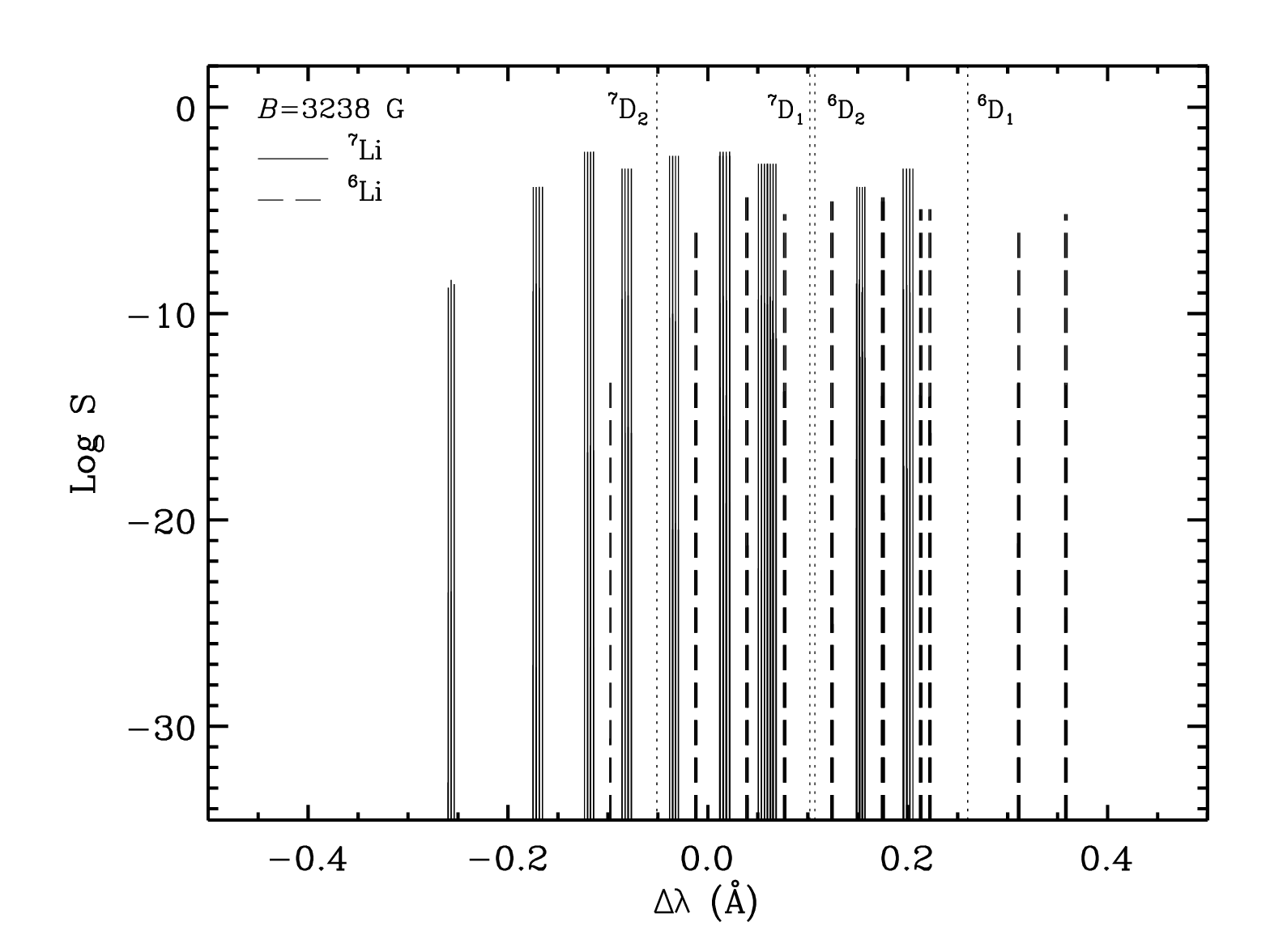}
\includegraphics[scale=0.4]{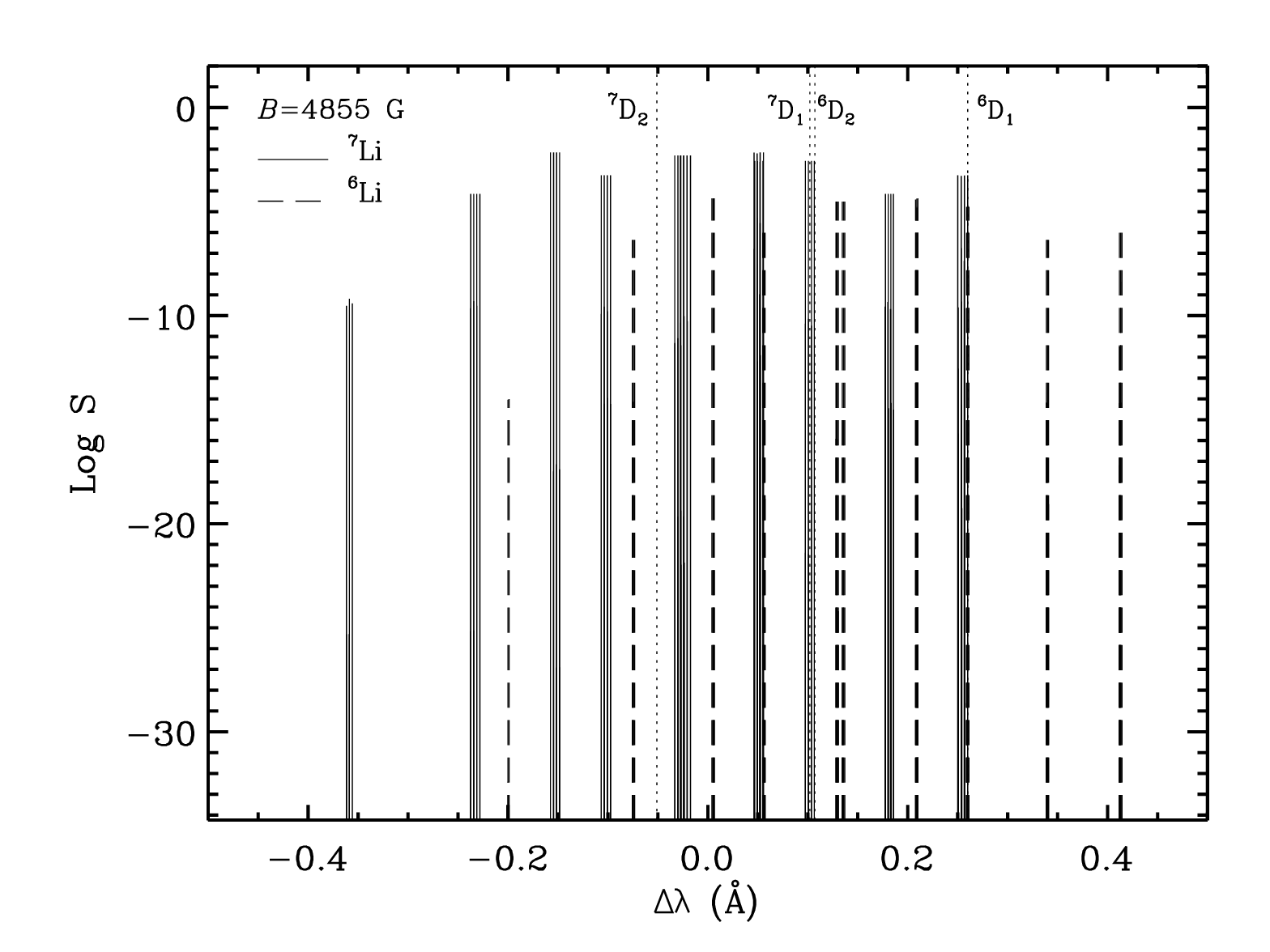}
\includegraphics[scale=0.4]{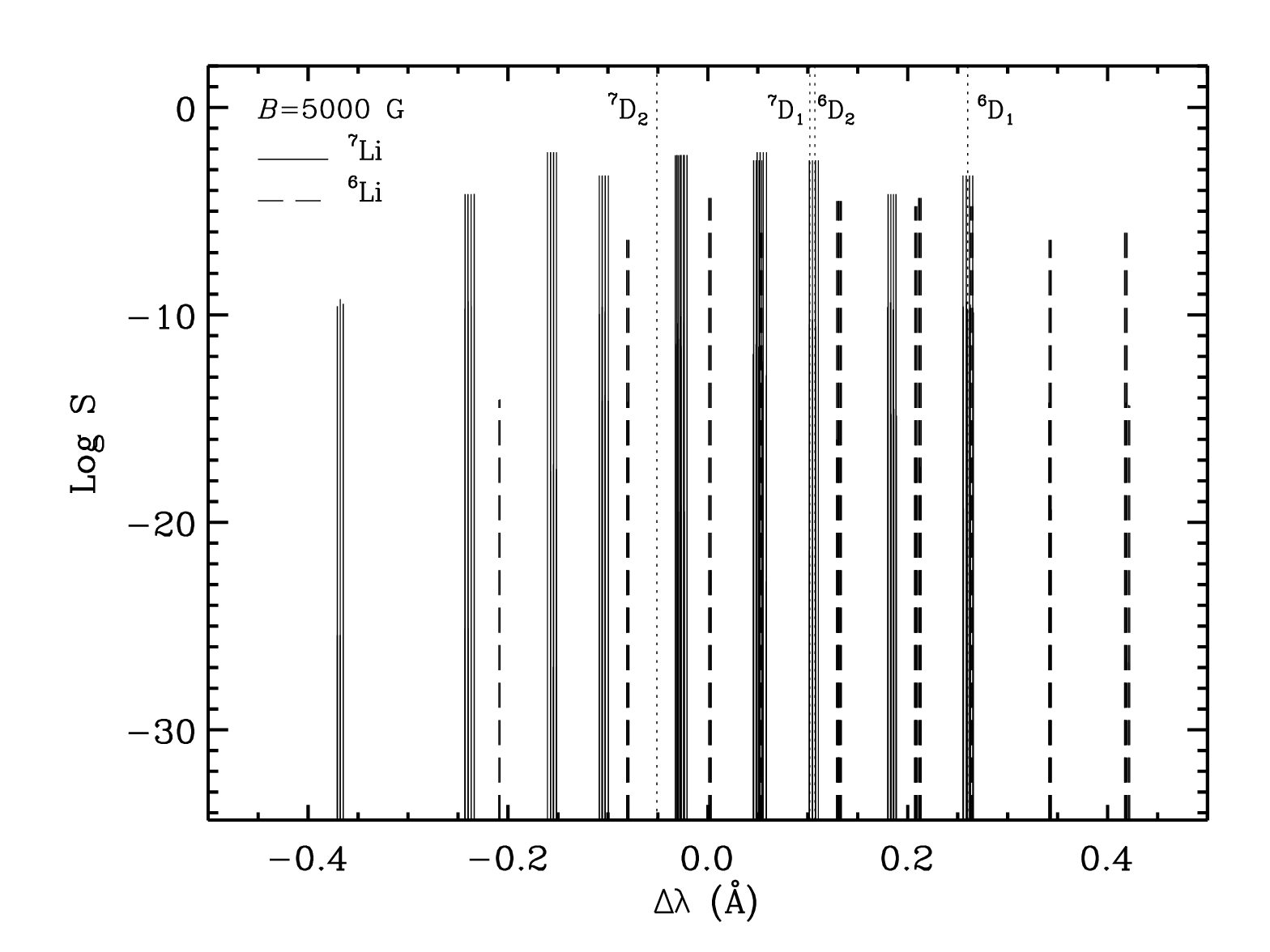}
\caption{Line splitting diagrams for the two lithium isotopes for the field strengths
indicated. The solid lines represent the magnetic components of $^7$Li while the dashed
lines represent those of $^6$Li. Vertical dotted lines mark the positions of the
D lines of the two isotopes. $\Delta\lambda=0$ corresponds to the line center
wavelength of $L=0\rightarrow1\rightarrow0$ transition in $^7$Li.
\label{spl-1}
}
\end{center}
\end{figure*}

The line splitting diagram shows the displacement of the magnetic components from the
line center (corresponding to the wavelength of the $L=0\rightarrow1\rightarrow0$
transition in the reference isotope $^7$Li) and the strengths of these components
for a given field strength. In Figure~\ref{spl-1}, we show the line splitting diagrams
for different $B$ values. We take into account the isotope shift and the solar
abundances of the two isotopes while computing the strengths and magnetic shifts. As
mentioned earlier, the components arising for $B=0$ correspond to the transitions
between the unperturbed $F$ states. We see that the hyperfine structure components of
the D lines are well separated when $B=0$ due to the relatively large FS. When the
magnetic field is applied, the degeneracy of the magnetic substates is lifted. As a
result, 70 allowed transitions take place in $^6$Li and 106 in $^7$Li. This explains
why the diagrams become crowded as the field strength increases. We see that the
magnetic displacements increase with an increase in $B$ as expected. In the diagrams
shown, we note that the MS is nonlinear and is a characteristic of the
incomplete PB regime.
\begin{figure}
\begin{center}
\includegraphics[scale=0.9]{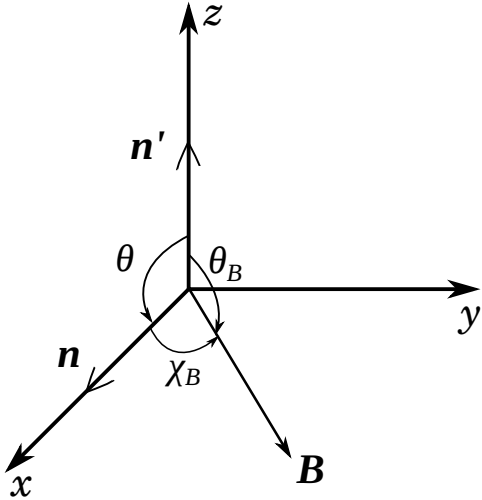}
\caption{Scattering geometry considered for the results presented in
Section~\ref{sec-3}.
\label{geom}}
\end{center}
\end{figure}
\begin{figure}
\begin{center}
\includegraphics[scale=0.45]{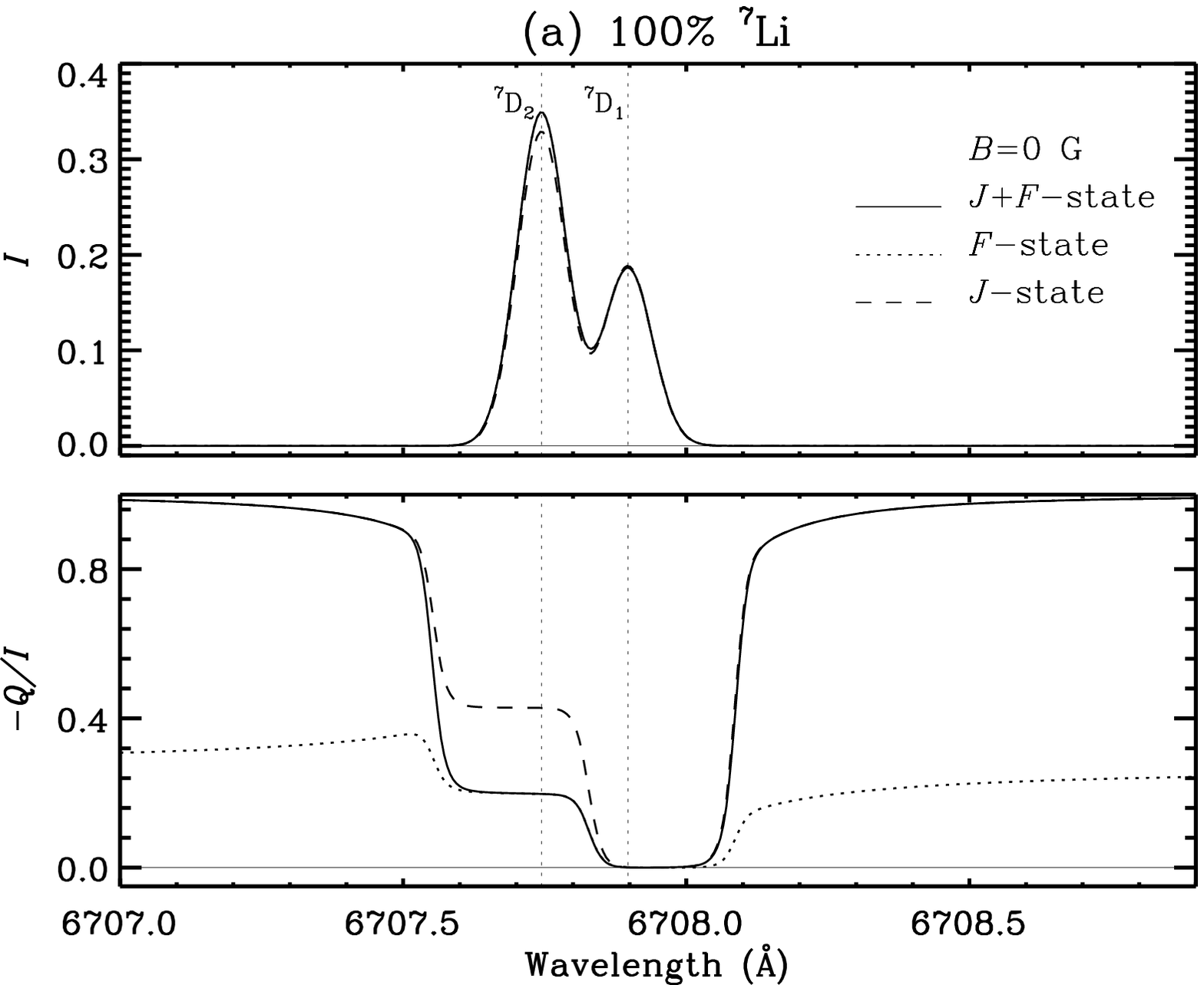}
\includegraphics[scale=0.45]{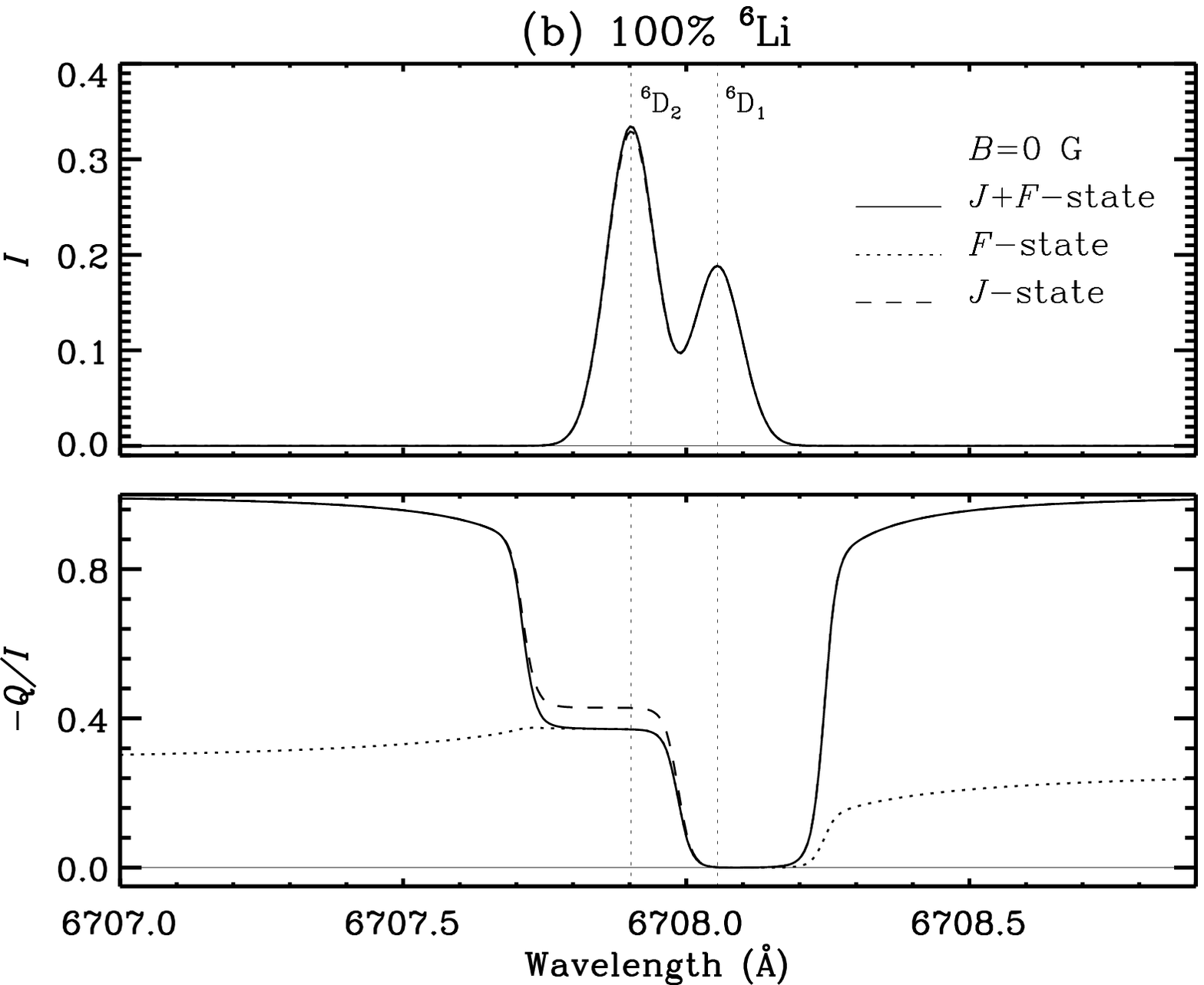}
\includegraphics[scale=0.45]{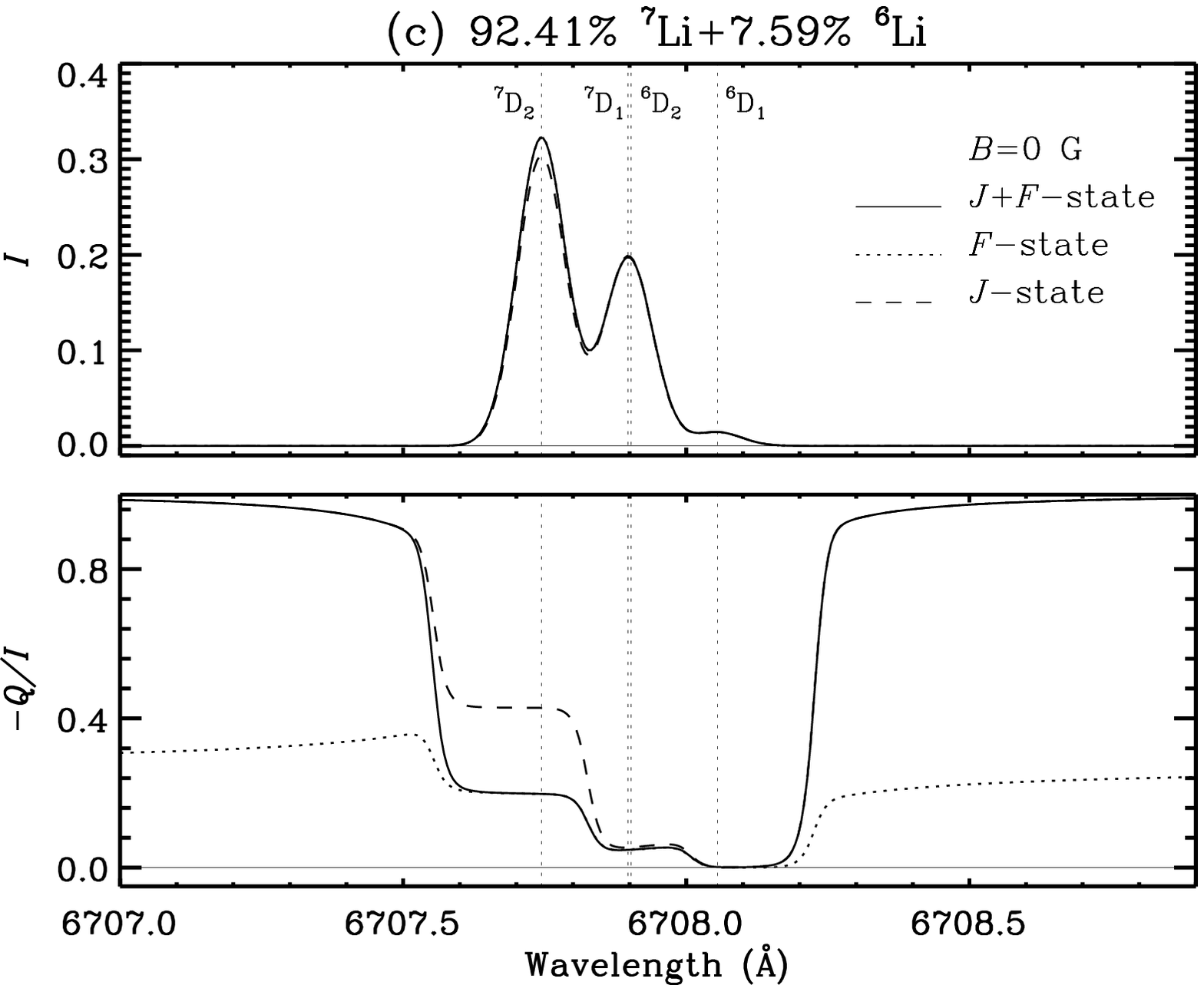}
\caption{Single scattered Stokes profiles for the lithium D line system in the
absence of a magnetic field: (a) 100\% $^7$Li, (b) 100\% $^6$Li, and (c) $^7$Li and $^6$Li
combined according to their percentage abundance. The line types are indicated in
the intensity panels. The geometry considered for scattering is $\mu=0$, $\mu^\prime=1$,
$\chi=0\degree$, and $\chi^\prime=0\degree$. The vertical dotted lines represent the
line center wavelength positions of the $^7$Li D$_2$, $^7$Li D$_1$,
$^6$Li D$_2$, and $^6$Li D$_1$ lines in the absence of magnetic fields.
\label{st-1}
}
\end{center}
\end{figure}

\subsection{Single Scattered Stokes Profiles}
\label{sec-3}
\begin{figure*}
\begin{center}
\includegraphics[scale=0.45]{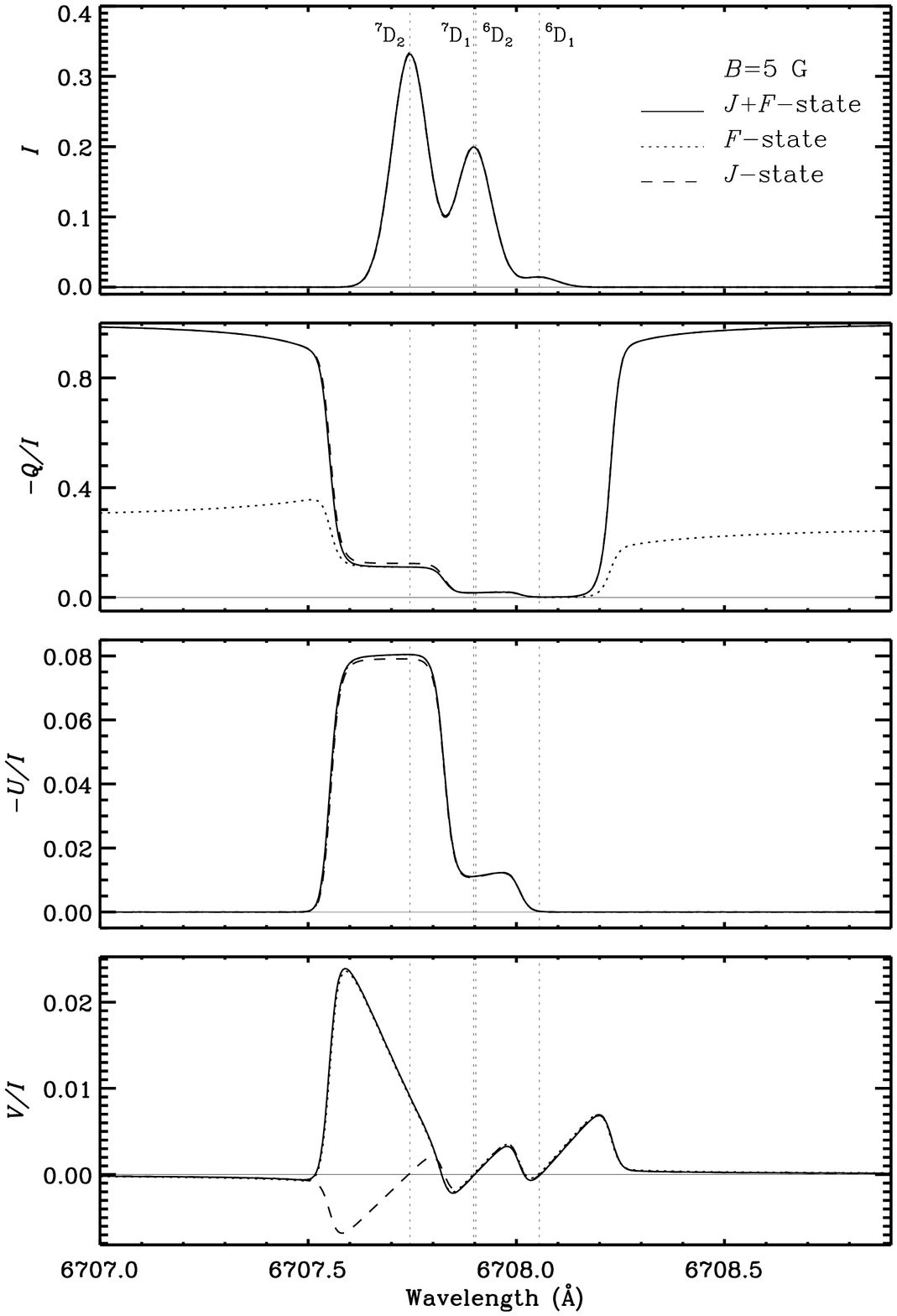}
\includegraphics[scale=0.45]{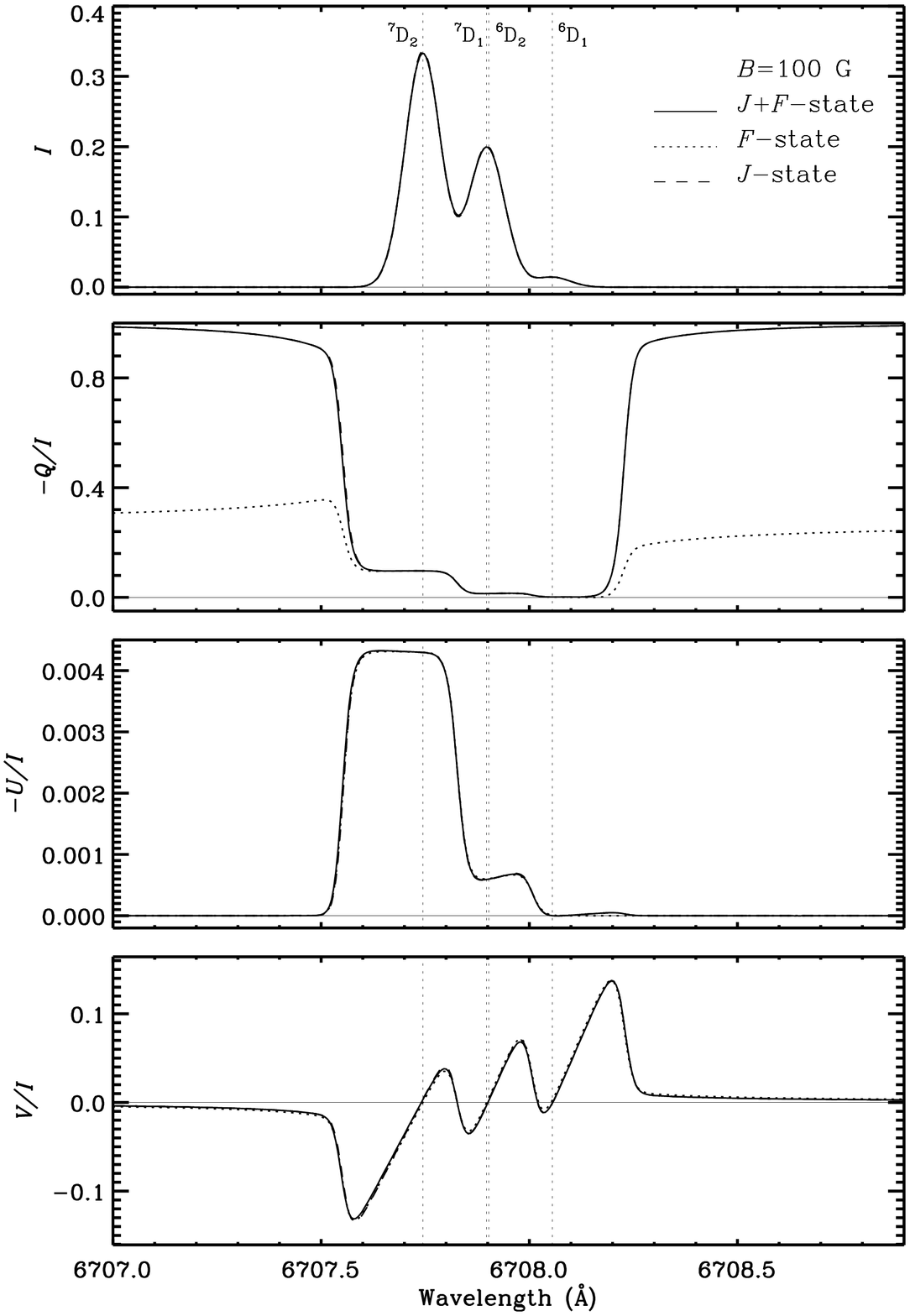}
\caption{Same as Figure~\ref{st-1} but in the presence of a magnetic field. The left and
the right panels correspond to different field strength values. The field orientation
($\theta_B=90\degree$, $\chi_B=45\degree$) is the same in both the panels. Refer to
Section~\ref{sec-3} for the scattering geometry. 
\label{st-2}
}
\end{center}
\end{figure*}

\begin{figure*}
\begin{center}
\includegraphics[scale=0.32]{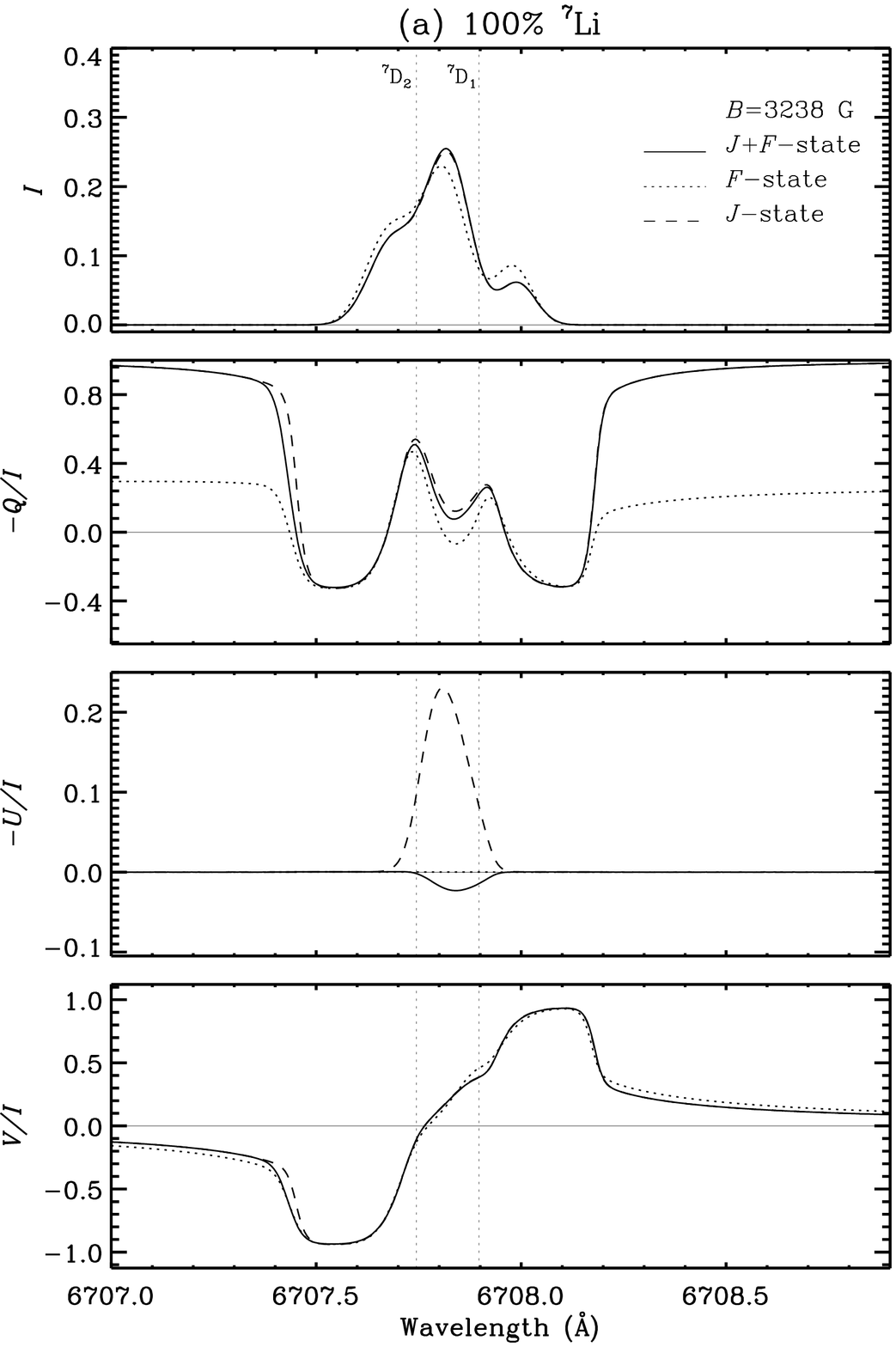}
\includegraphics[scale=0.32]{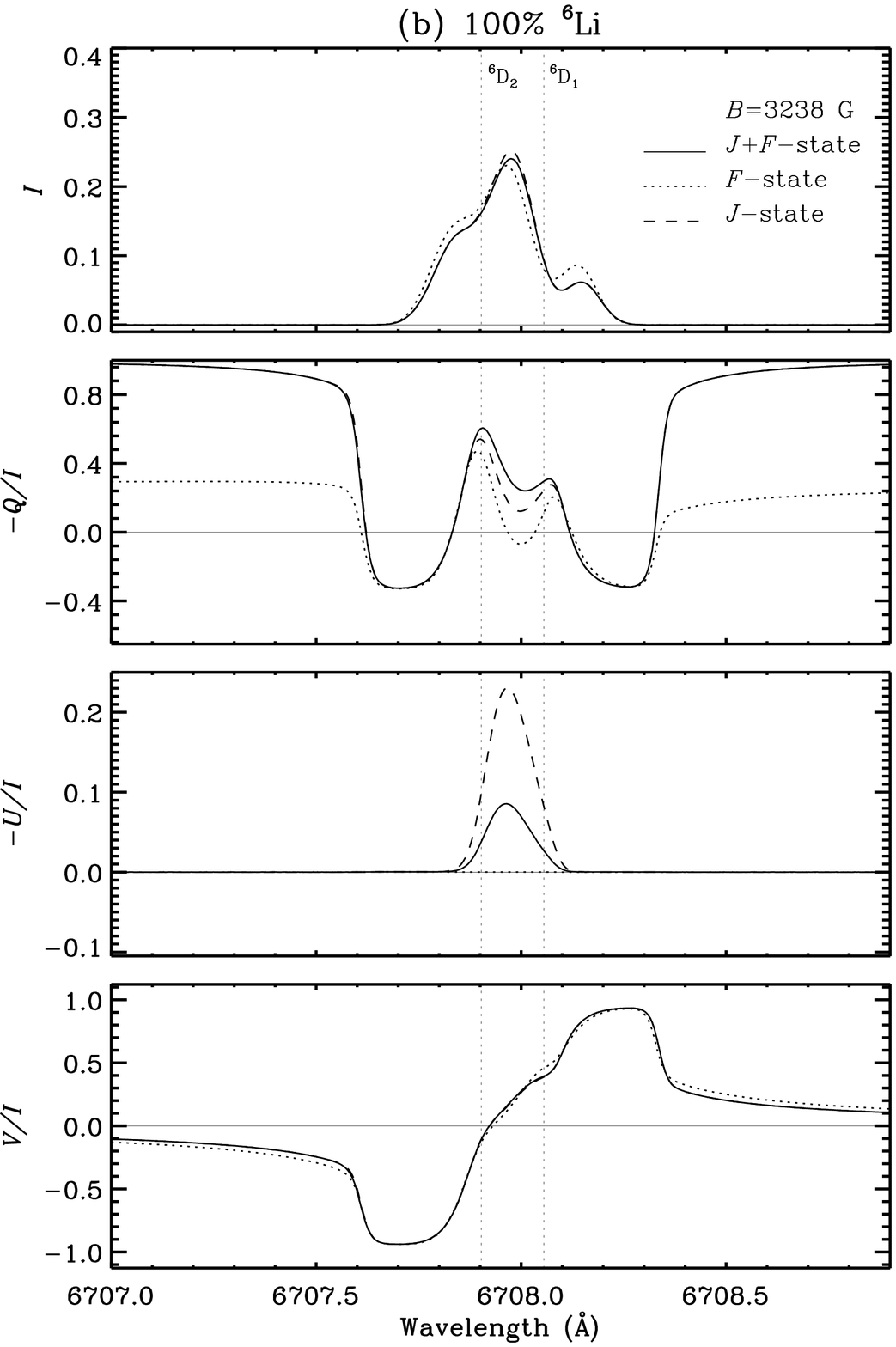}
\includegraphics[scale=0.32]{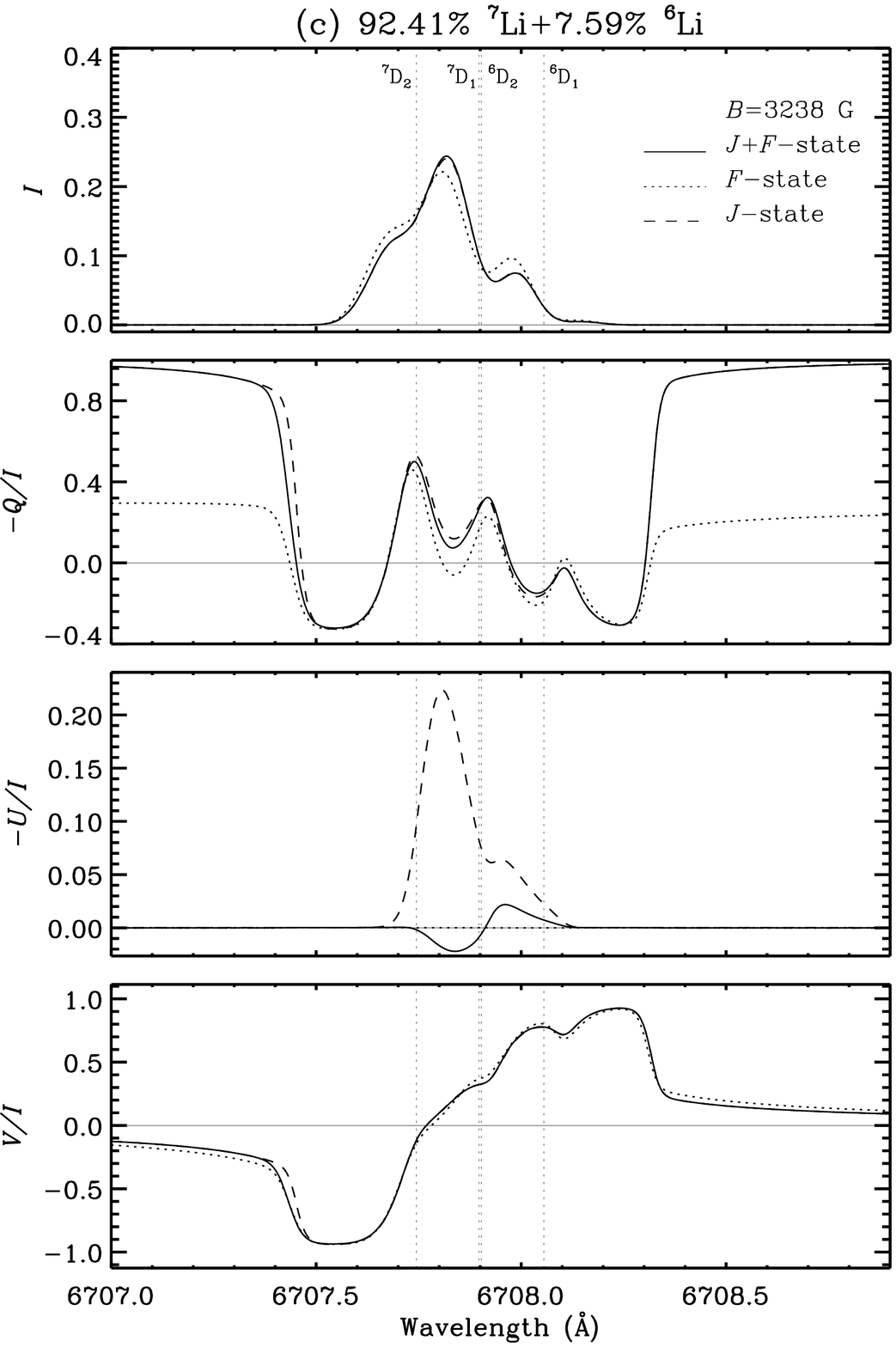}
\caption{Stokes profiles obtained for $B=3238$ G: (a) 100\% $^7$Li,
(b) 100\% $^6$Li, and (c) $^7$Li and $^6$Li combined according to their percentage
abundance. Refer to Section~\ref{sec-3} for the scattering geometry. When $B=3238$ G,
the $U/I$ values are so small for the dotted line case that they become indistinguishable 
from the zero line.
\label{st-3}
}
\end{center}
\end{figure*}

\begin{figure*}
\begin{center}
\includegraphics[scale=0.45]{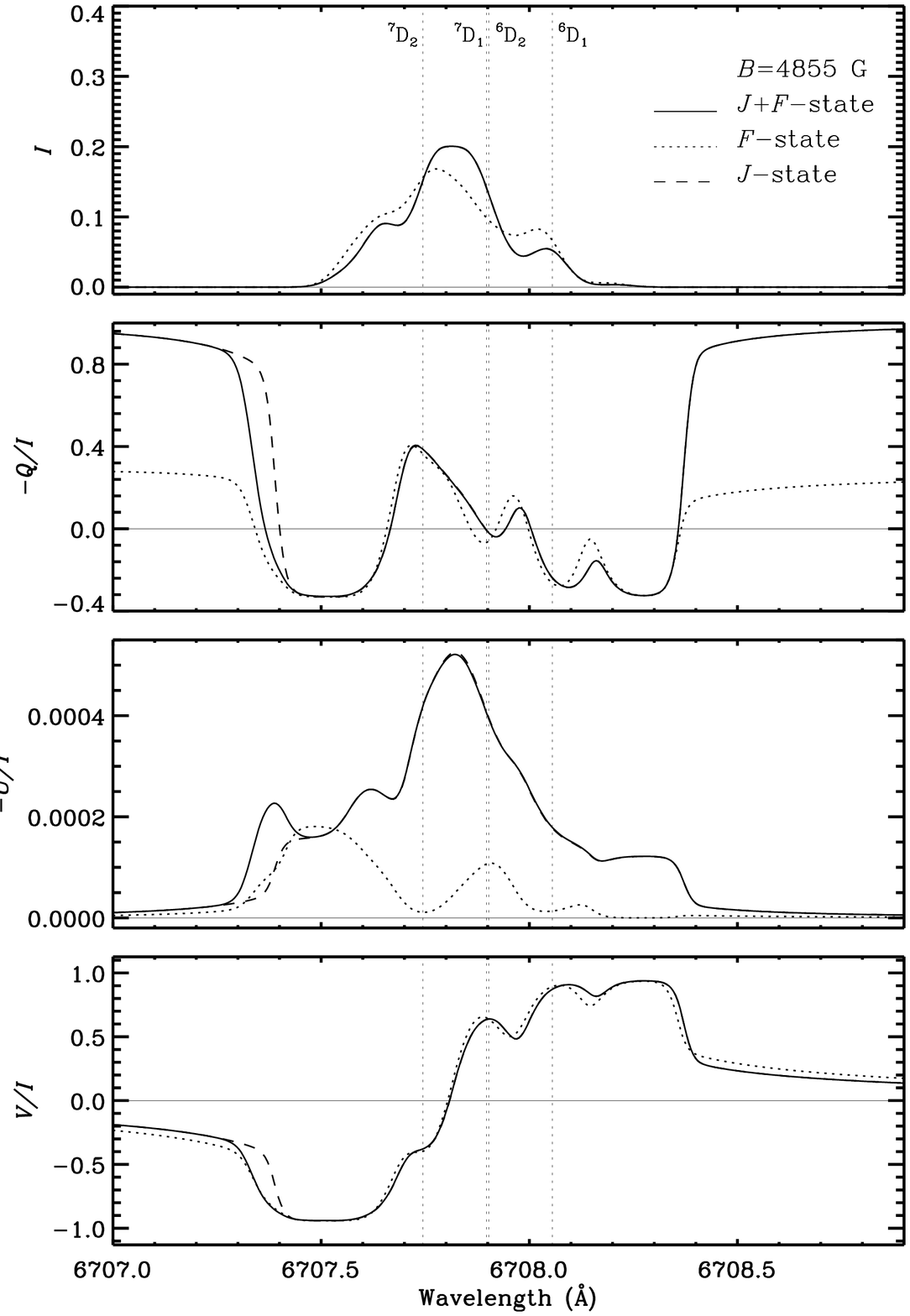}
\includegraphics[scale=0.45]{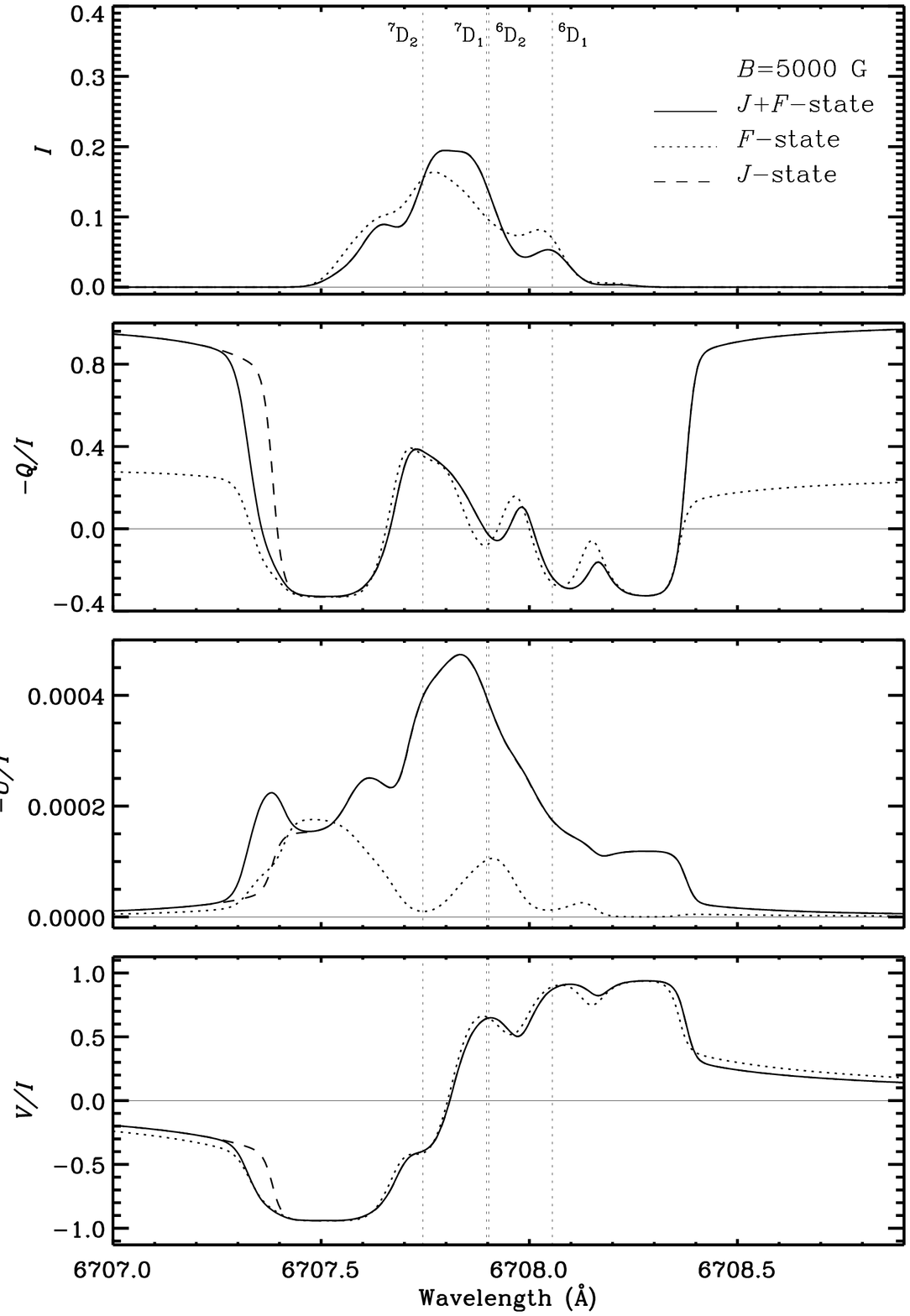}
\caption{Stokes profiles obtained for $B=4855$ G and $B=5000$ G. Refer to
Section~\ref{sec-3} for the scattering geometry.
\label{st-4}
}
\end{center}
\end{figure*}

In this section, we present the Stokes profiles for various $B$ values computed using
the combined theory for the single scattering case. We choose a coordinate system
(see Figure~\ref{geom}) in which the magnetic field lies in the horizontal ({\it xy})
plane making angles $\theta_B=90\degree$ and $\chi_B=45\degree$.
We make this choice following \citet{s98} in order to bring out clearly the effects
of the magnetic field. We assume the unpolarized incident ray to be along the vertical
({\it z}-axis) and the scattered ray (or the line of sight) to lie in the horizontal
plane along the {\it x}-axis. Thus, the angles for the incident and the scattered
rays become $\mu^\prime=1$, $\chi^\prime=0\degree$, $\mu=0$, and $\chi=0\degree$.
We use the fact that the lithium lines are optically thin and only
single scattering is considered here to add the Stokes profiles computed for the
individual isotopes after weighting them by their respective abundances. In
Figures~\ref{st-1} -- \ref{st-4}, we compare the single scattered Stokes profiles for
three cases: the cases of pure $F$ state interference (dotted lines) represented by
a two-level atom with hyperfine structure, pure $J$ state interference (dashed lines)
represented by a two-term atom without hyperfine structure, and the combined theory
(solid lines) represented by a two-term atom with hyperfine structure. We choose a
Doppler width of 60 m\AA{} for all of the components of the multiplet when computing
the Stokes profiles. For this particular value of the Doppler width, the theoretical
$Q/I$ profile closely resembles the observed $Q/I$ profile \citep[see][]{bel09}.
We use the Einstein $A$ coefficient of $3.689\times10^7$ s$^{-1}$ for all of
the components. 

In Figure~\ref{st-1}, we show the Stokes profiles computed in the absence of magnetic
fields for 100\% $^7$Li in panel (a), for 100\% $^6$Li in panel (b), and for both the
isotopes combined according to their percentage abundance in panel (c). In panels (a)
and (b), we see two peaks corresponding to the D lines of the two isotopes in intensity.
The intensities of the D lines in both the isotopes are of similar magnitude since we
have assumed 100\% abundance for the two isotopes. We also note that the wavelength
positions of the D lines of $^6$Li are different from those of $^7$Li owing to the
isotope shift. In panel (c), we see two distinct peaks in intensity. The first peak
to the left is due to the $^7$Li D$_2$ line. The second peak falls at the line center
positions of $^7$Li D$_1$ and $^6$Li D$_2$. However, the dominant contribution
comes from the $^7$Li D$_1$ due to its relatively larger abundance. A small bump to
the right of the second peak is due to the $^6$Li D$_1$ line. A small difference in the
intensity at the $^7$Li D$_2$ peak between the dashed lines and the other two cases
is seen in panels (a) and (c). It is clear from the figure that this discrepancy is
caused by $^7$Li. Comparing the solid, dotted, and dashed profiles, we come to the
conclusion that the HFS is at the origin of this discrepancy. This is because the
solid and dotted lines computed by including HFS perfectly match and only
the dashed lines computed without HFS differ from the other two cases. The discrepancy
is very small in the case of $^6$Li because of smaller HFS in $^6$Li compared to that
in $^7$Li. The reason for this discrepancy is due to the asymmetric splitting of the
HFS components about the given $J$ state and also due to finite widths of the components.
This difference decreases (graphically indistinguishable) when a magnetic field is
applied (for example, when $B=5$ G as seen in Figure~\ref{st-2}) because of the
superposition of a large number of magnetic components. In contrast, this difference
is about an order of magnitude larger in the non-magnetic case. As we increase the field
strength, the intensity profiles broaden due to an increased separation between the
magnetic components. 

When $B=0$, the $Q/I$ profiles exhibit a multi-step behavior around the line center
positions of the D$_1$ and D$_2$ lines of both isotopes. We see the effects of
quantum interference clearly in $Q/I$. In the $^7$Li D$_2$ core, significant
depolarization is caused by the HFS compared to the case where this splitting
is neglected (compare the solid and dashed lines in panels (a) and (c)). A similar
depolarization is also exhibited by the core of the $^6$Li D$_2$ line (see panels (b)
and (c)). However, in the scale adopted, the solid and dashed lines appear to merge
around the core of $^6$Li D$_2$ in panels (c), as the $Q/I$ values of $^6$Li D$_2$ are
an order of magnitude smaller than those of $^7$Li D$_2$ because of their relative
abundances. The D$_1$ lines remain upolarized. As expected, the solid lines merge with
the dotted line in the cores of lithium lines while they coincide with the dashed lines
in the wings. When a magnetic field is applied, we see a depolarization in $Q/I$ and a
generation of $U/I$ signal in the cores of the lithium lines due to the Hanle effect.
We note that the combined theory results match more closely the pure $J$ state
interference results for fields of the order of 100 G. This behavior continues until the
level-crossing field strength of $B=3238$ G for fine structure is reached.

For kG fields, we are by far in the complete PB regime for the $F$ states. In this
regime, the $J$ and $I_s$ couple strongly to the magnetic field and the interaction
between $J$ and $I_s$ becomes negligible. Therefore, one would expect the HFS magnetic
substates to be fully degenerate, and therefore the solid and dashed lines
should match closely for fields of the order of kG. However, for the
level-crossing field strengths, we see considerable differences between the solid
and the dashed lines, especially in $U/I$. In order to understand this, we compare
the Stokes profiles for $^7$Li and $^6$Li separately in panel (a) and (b) of
Figure~\ref{st-3} with the combined profiles in panel (c). We do this to check
whether a particular isotope is giving rise to this difference. We note that this
difference between the solid and dashed lines prevails in all three panels
(i.e., in both isotopes). We attribute this difference in the shape and amplitude
between the solid and the dashed lines to HFS, the level-crossings, and avoided crossings
between the HFS magnetic substates. When we look at Figure~\ref{level-fig}, we find
that the HFS magnetic substates have finite energy differences and are not fully
degenerate in the complete PB regime for the $F$ states.
We see several crossings as well as
a few avoided crossings. These level-crossings and avoided crossings between
the non-degenerate HFS magnetic substates lead to a modification of the coherence
and significant Hanle rotation, thereby affecting the shape and amplitude of the $U/I$
profiles. The HFS effects show more prominently in the polarization
diagrams which will be discussed in Section~\ref{sec-6}. For the geometry under
consideration, this effect is significantly seen for $B=3238$ G. For a level-crossing
field strength of $4855$ G, the Stokes profiles show somewhat different behavior.

We also note that for fields of the order of kG, differences between the solid and
dashed lines remain only in the far left wing (see Figures~\ref{st-3} and \ref{st-4}).
From Figure~\ref{st-3} it is clear that this difference in the far blue wings is
only due to the $^7$Li isotope (compare panels (a)--(c)). This can be
understood with the help of the line splitting diagrams for level-crossing fields in
Figure~\ref{spl-1} in comparison with the corresponding diagrams in Figure 3 of
\citet[][a direct comparison of the displacements can be made as the zero points in the
two figures are the same]{sow14a}. In a two-term atom without HFS, when a magnetic
field is applied, the various FS magnetic components are either blue or redshifted
from the line center depending on their energies. When HFS is included, the HFS magnetic
components are distributed around the positions of the FS magnetic components in the
absence of HFS. We find that the positions of the HFS magnetic components in
Figure~\ref{spl-1} correspond well with the wavelength positions of the FS magnetic
components in Figure 3 of \citet{sow14a}, except for the bunch of magnetic components to
the extreme left represented by solid lines. The magnetic field leads to a large
blue shift of this bunch, which consists of three $\sigma_b$ ($\Delta\mu=\mu_b-\mu_a=+1$),
two $\pi$ ($\Delta\mu=0$) and one $\sigma_r$ ($\Delta\mu=-1$) components. These
components (otherwise not present at this wavelength position when
HFS is neglected) give rise to the systematic difference in
$Q/I$, $U/I$, and $V/I$ in the far blue wing of the D$_2$ line of $^7$Li. However, they
do not affect the intensity.

The $V/I$ profiles remain somewhat indistinguishable between the three cases considered,
except for very weak fields like 5 G as in Figure~\ref{st-2}. $F$ state interference
significantly changes the $V/I$ profile at the $^7$Li D$_2$ wavelength position. This is
a signature of the alignment-to-orientation (A-O) conversion mechanism \citep[see][and
LL04]{landi82} acting in the incomplete PB regime for the $F$ states. As described in
LL04, this occurs because of the double summation over $K$ and $K^\prime$ appearing in
Equation~(\ref{final-rm}) and because the spherical tensor $\mathcal{T}^K_{Q}(3,{\bm n})$
is non-zero only when $K=1$ (see Equation~(\ref{tkq3}) of Appendix~\ref{a-c}). This
means that circular polarization can be generated by resonance scattering
even if the atom is not exposed to circularly polarized light. The alignment present
in the radiation field is converted to the orientation in the upper term. This orientation
in the upper $F$ states gives rise to circularly polarized light. As discussed earlier,
small differences appear in the far blue wings for fields equal to or larger than the
level-crossing field strengths.

Finally, we remark that the discussion presented above concerning the comparison of the
single scattered Stokes profiles between the three cases (namely, the pure $J$ state, pure
$F$ state, and combined $J$ and $F$ interference) also remains valid for other scattering
geometries. 


\begin{figure}
\begin{center}
\includegraphics[scale=0.45]{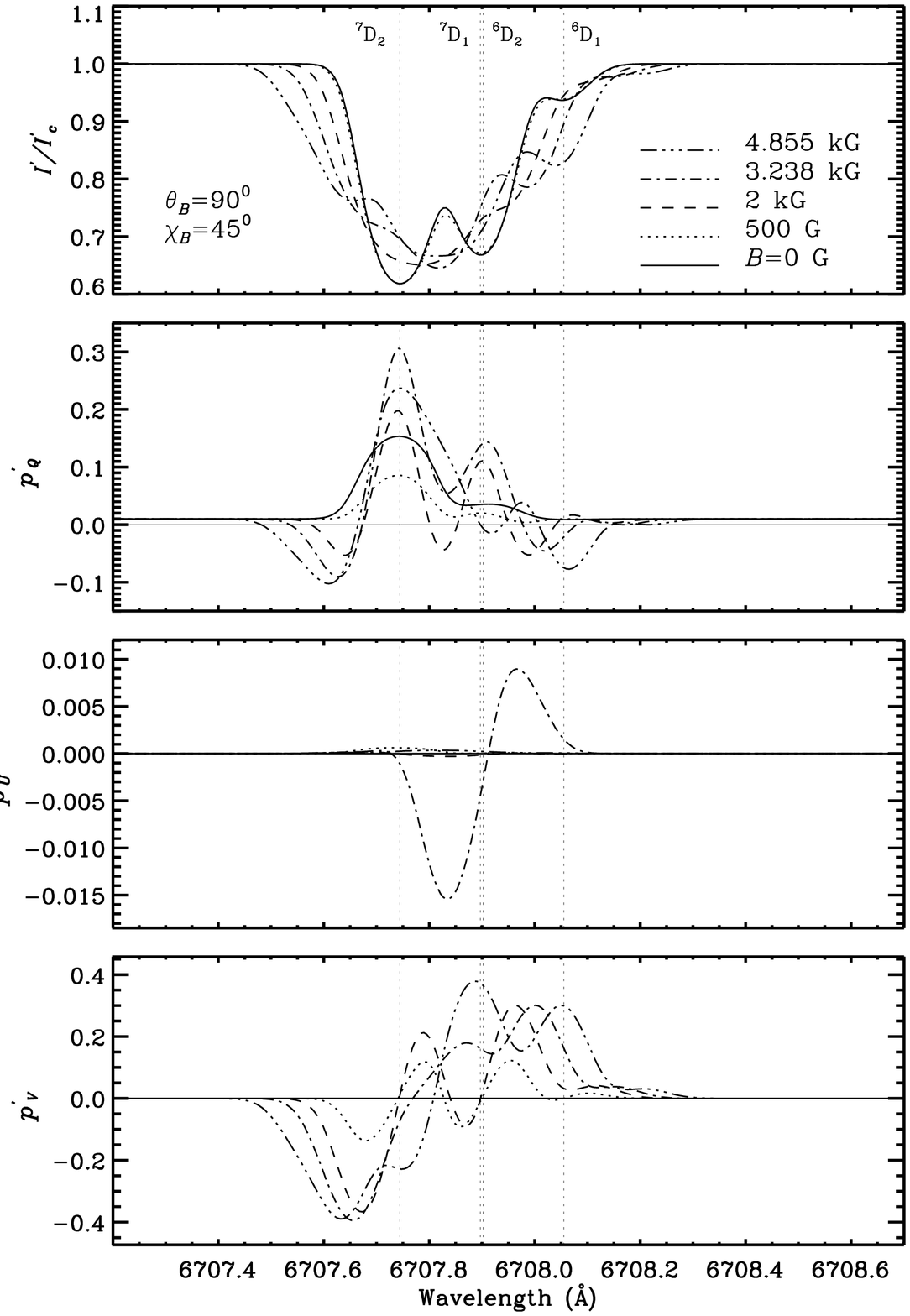}
\caption{Stokes profiles obtained by including the contribution from the continuum
for different values of $B$. Refer to Section~\ref{sec-3} for the scattering geometry.
The vertical dotted lines represent the positions of the D lines.
\label{stlc}
}
\end{center}
\end{figure}

In Figure~\ref{stlc}, we show the Stokes profiles obtained after including a weakly
polarized background continuum. We refer the reader to Section 4.3 of \citet{sow14a}
for details on how we add the continuum contribution and on the parameters
used for the continuum. We compare this figure with Figure 4 of \citet{sow14a} and
find that the HFS does not cause any change in the intensities. When $B=0$ the HFS
causes a depolarization in the core of $Q/I$ without affecting the shape of the profile.
For other field strengths, there is only a slight difference in the amplitude of the
profiles as compared to the case without HFS, although their shapes remain the same.
The $U/I$ profiles differ both in amplitude and shape for $B=3238$ G. This difference
is due to HFS. When HFS is neglected, there is only one level-crossing at this field
strength. On the other hand, when HFS is included, there are several level-crossings
around this field strength (see Figures~\ref{level-fig}(b) and (e)).
$V/I$ profiles have the same shapes and amplitudes as compared to the case without
hyperfine structure.

\subsection{Net Circular Polarization (NCP)}
\label{sec-5}
In this section, we present the plots of NCP defined as 
$\int V d\lambda$ as a function of the magnetic field strength $B$. Since the PB
effect causes nonlinear splitting of the magnetic components with respect to the
line center, the Stokes $V$ profiles become asymmetric. As a result of this asymmetry,
the integration of the Stokes $V$ over the full line profile yields a non-zero value. In
the linear Zeeman and complete PB regimes, the $V$ profiles show perfect antisymmetry
which causes the NCP to become zero. The A-O conversion
mechanism discussed in Section~\ref{sec-3} further enhances the asymmetry in Stokes $V$
profiles already caused by nonlinear MS, and thereby contributes to
the NCP. This mechanism is particularly efficient when the level-crossings satisfy
$\Delta\mu=\mu_{b^\prime}-\mu_b=1$.

\begin{figure}[ht]
\begin{center}
\includegraphics[scale=0.45]{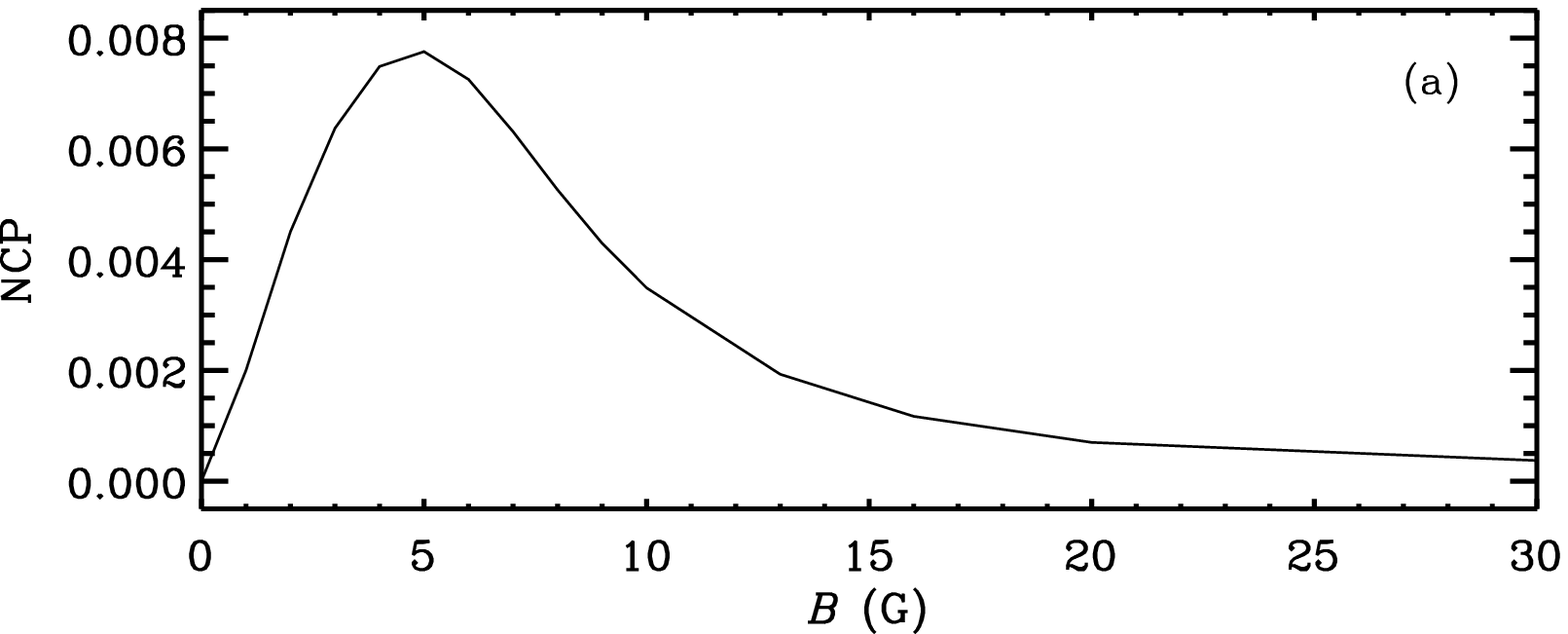}
\includegraphics[scale=0.45]{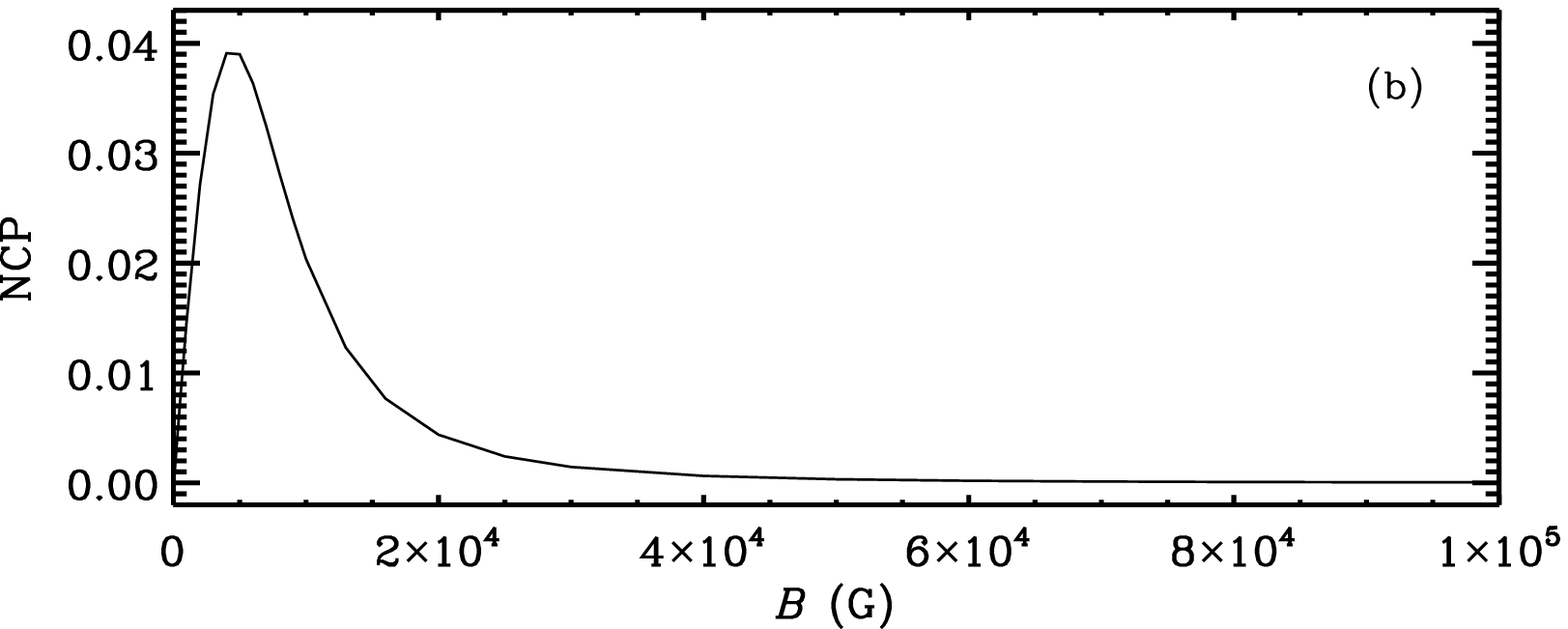}
\caption{Net circular polarization as a function of the magnetic field strength $B$.
The scattering geometry is characterized by: $\mu^\prime=0$, 
$\chi^\prime=0\degree$, $\mu=1$, $\chi=90^\degree$, $\theta_B=0\degree$, 
and $\chi_B=0\degree$. 
\label{ncp}
}
\end{center}
\end{figure}

In Figure~\ref{ncp}, we show the behavior of NCP in different field strength ranges
for the scattering geometry: $\mu^\prime=0$, $\chi^\prime=0\degree$, $\mu=1$, 
$\chi=90\degree$, $\theta_B=0\degree$, and $\chi_B=0\degree$. This choice of the field
geometry is made in order to obtain larger values for Stokes $V$. In panel (a), we 
show the weak field behavior of NCP. We attribute the non-zero NCP in this regime to the
PB effect in the $F$ states and the A-O conversion mechanism taking
place in the incomplete PB regime for the $F$ states. We find that the NCP increases with
increasing field strength, peaking around the level-crossing field strength
(see Tables~\ref{tab-2} and \ref{tab-3}), and decreases with further increase in $B$.
For fields of the order of kG we see a second peak in NCP whose magnitude is larger than
the first peak by an order. This is due to the PB effect in the $J$ states and the
A-O conversion mechanism occurring in the incomplete PB regime for
the $J$ states. With a further increase in the field strength, we enter the
complete PB regime for the $J$ states where the NCP becomes zero.

Detailed discussions on the various mechanisms producing NCP are presented in LL04.


\subsection{Polarization Diagrams}
\label{sec-6}
\begin{figure*}
\begin{center}
\includegraphics[scale=0.55]{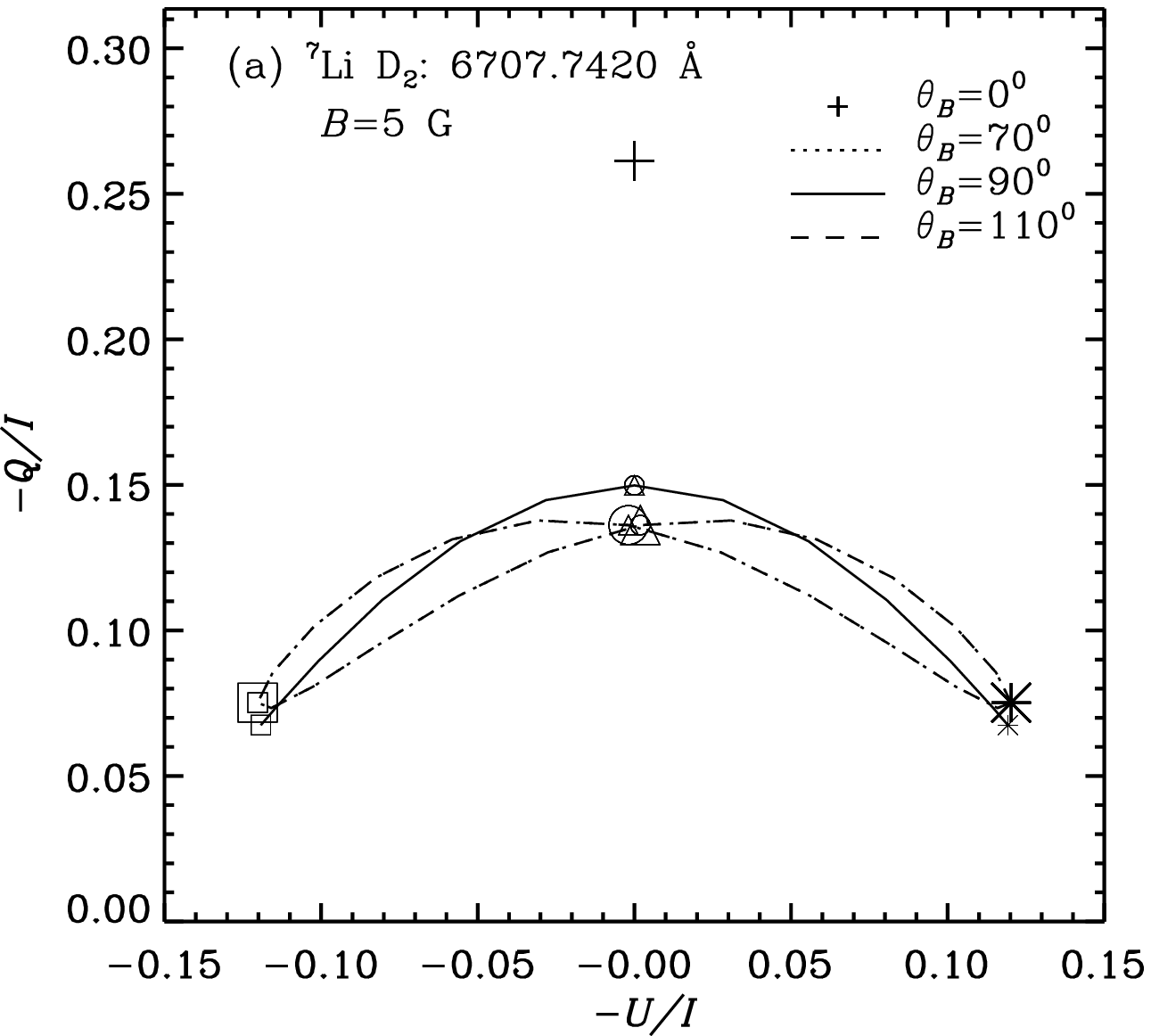}
\includegraphics[scale=0.55]{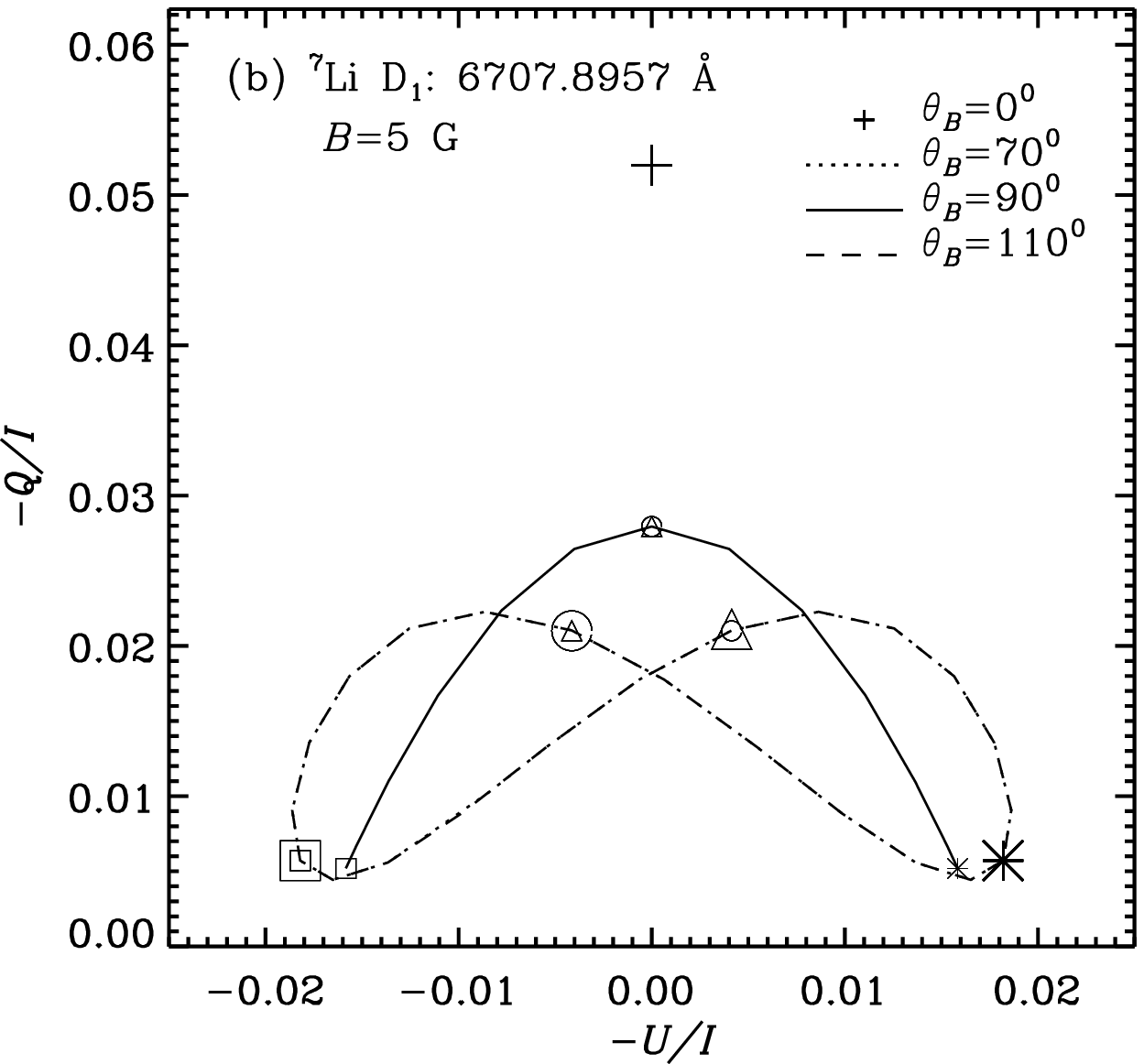}
\includegraphics[scale=0.55]{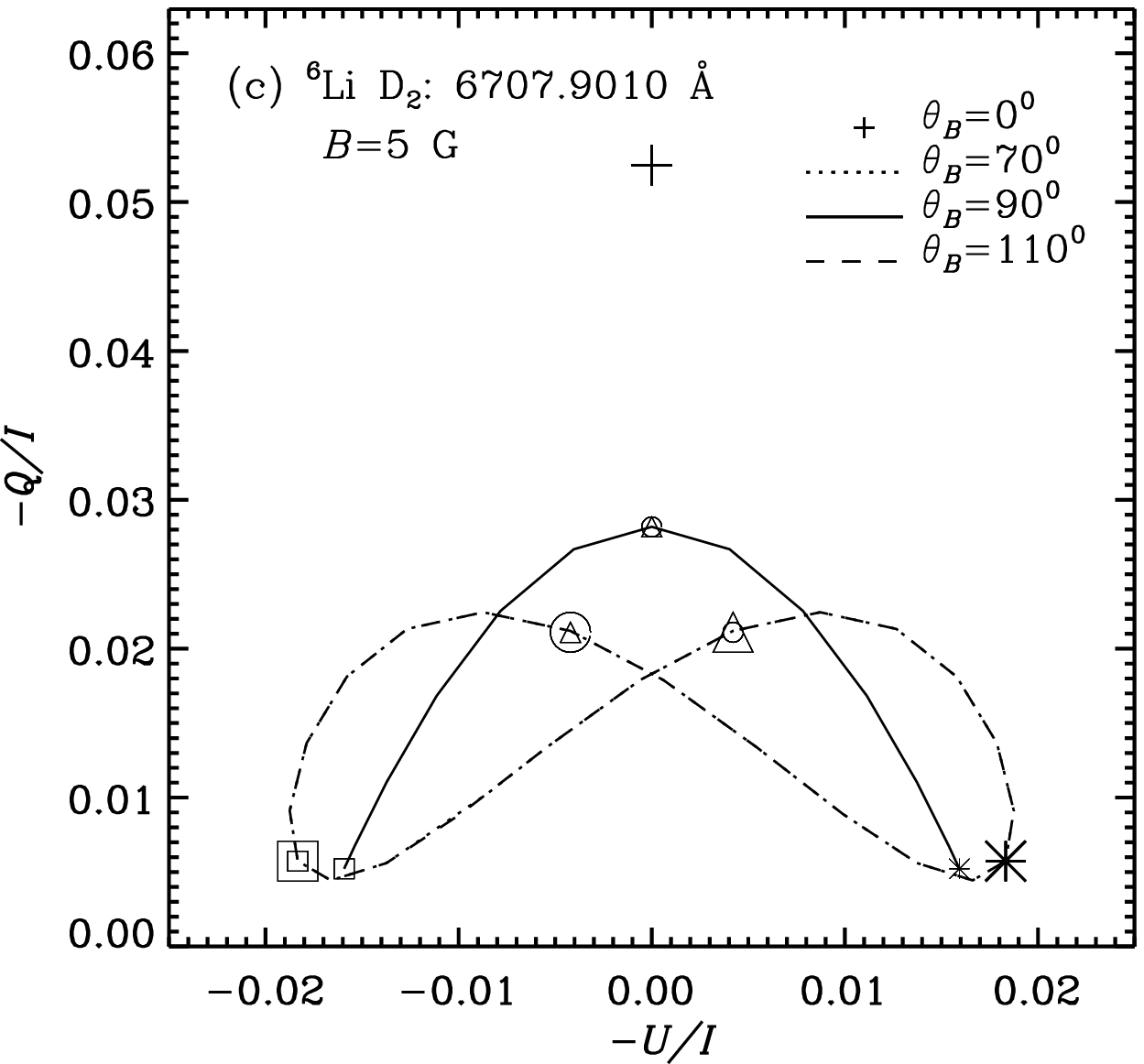}
\includegraphics[scale=0.55]{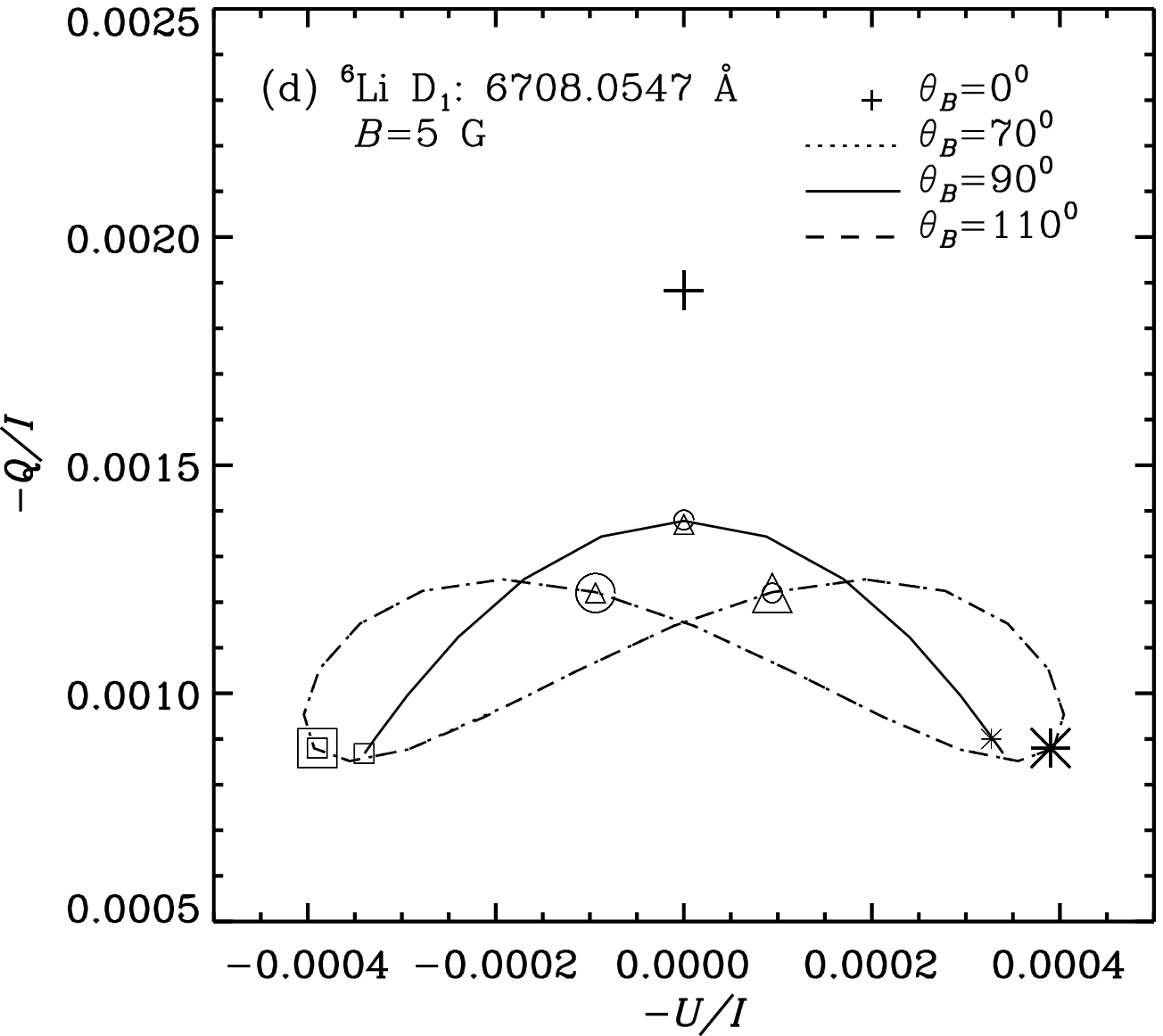}
\caption{Polarization diagrams obtained at the D line positions for $B=5$ G and 
different $\theta_B$ as indicated in the panels. The azimuth $\chi_B$ of the magnetic
field is varied from $0\degree$ to $360\degree$. The symbols on the curves mark the
$\chi_B$ values: $\ast-0\degree$, $\circ-70\degree$, $\square-180\degree$, and
$\vartriangle-270\degree$. Since the curves for the $\theta_B=70\degree$ and
$110\degree$ coincide, we use symbols that are bigger in size for the $\theta_B=110\degree$
case to distinguish it from the $\theta_B=70\degree$ curve. The geometry considered
is $\mu=0$, $\mu^\prime=1$, $\chi=0\degree$, and $\chi^\prime=0\degree$.
\label{pd-1}
}
\end{center}
\end{figure*}

In Figure~\ref{pd-1}, we present the plots of Stokes $Q/I$ versus Stokes $U/I$ 
(polarization diagrams) for a given $B$ and $\theta_B$ and for the full range 
of $\chi_B$. Refer to the figure caption for the incident and scattered ray directions.
$\theta_B$ takes values $0\degree$, $70\degree$, $90\degree$, and $110\degree$. We find
that the $\theta_B=70\degree$ and $110\degree$ curves perfectly coincide in all four
panels. They take same values for $Q/I$ and $U/I$ at $\chi_B=0\degree$ and
$\chi_B=180\degree$. However, we see that the dependence on $\chi_B$ of the
$\theta_B=70\degree$ curve is somewhat different from that of the $\theta_B=110\degree$
curve. By this, we mean that for the $\theta_B=70\degree$ case, the $Q/I$ value changes
in an anti-clockwise direction from the $\chi_B=0\degree$ point while it changes in a
clockwise direction from the $\chi_B=0\degree$ point for the $\theta_B=110\degree$ case.
The $Q/I$ value increases with increasing $\chi_B$, reaches a maximum and then decreases
till $\chi_B=180\degree$. $U/I$ makes a gradual transition from being positive to
negative. $Q/I$ again increases with an increase in $\chi_B$ and at $\chi_B=360\degree$
it resumes the same value it had at $\chi_B=0\degree$. $U/I$ now makes a transition
from being negative to positive. When $\theta_B=0\degree$ the magnetic field is along
the {\it z}-axis and exhibits azimuthal symmetry. Hence, $\theta_B=0\degree$ is just a
point in the polarization diagram. For $\theta_B=90\degree$ the diagram is symmetric
with respect to the $U/I=0$ line.

\begin{figure}
\begin{center}
\includegraphics[scale=0.55]{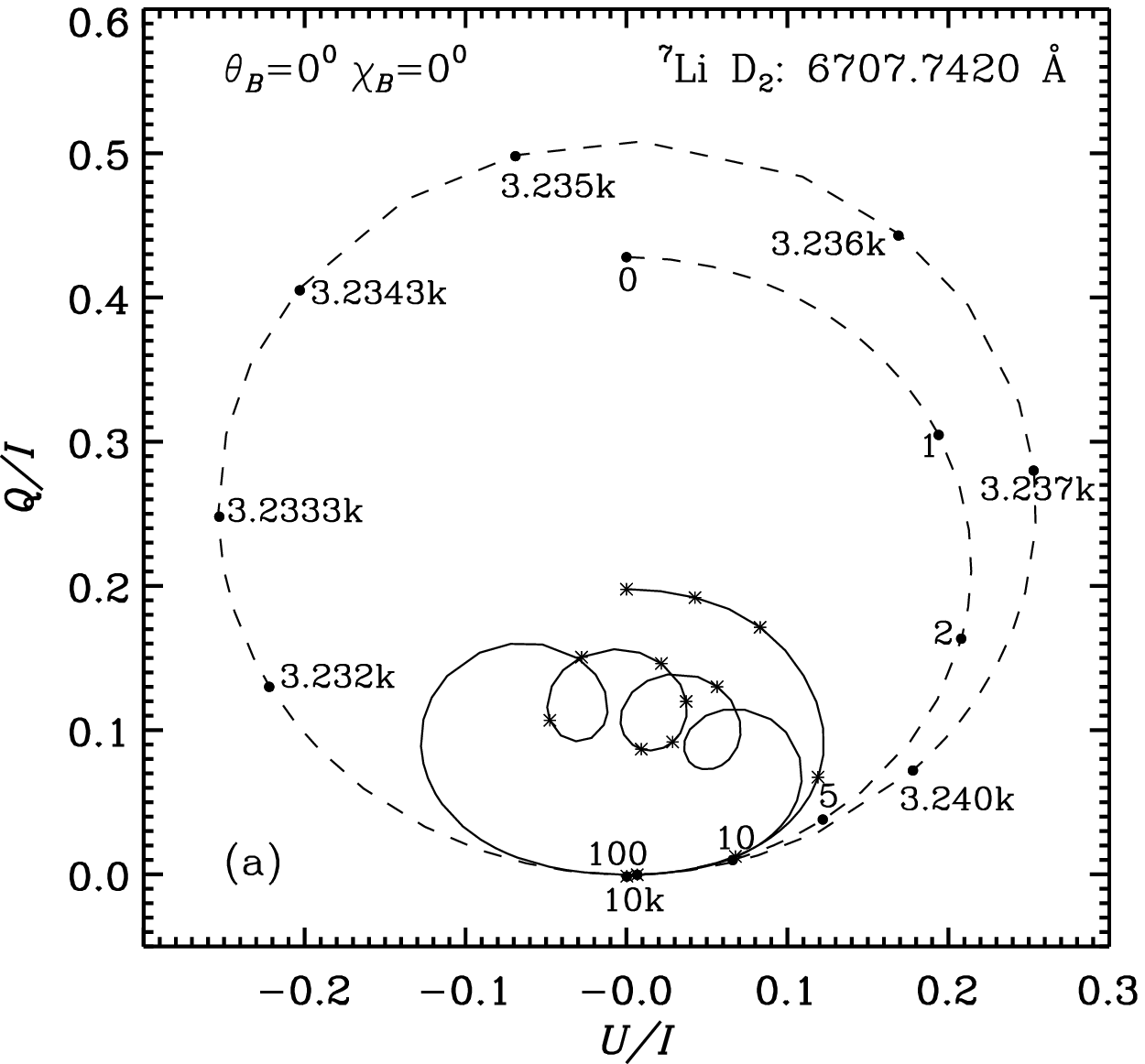}
\includegraphics[scale=0.55]{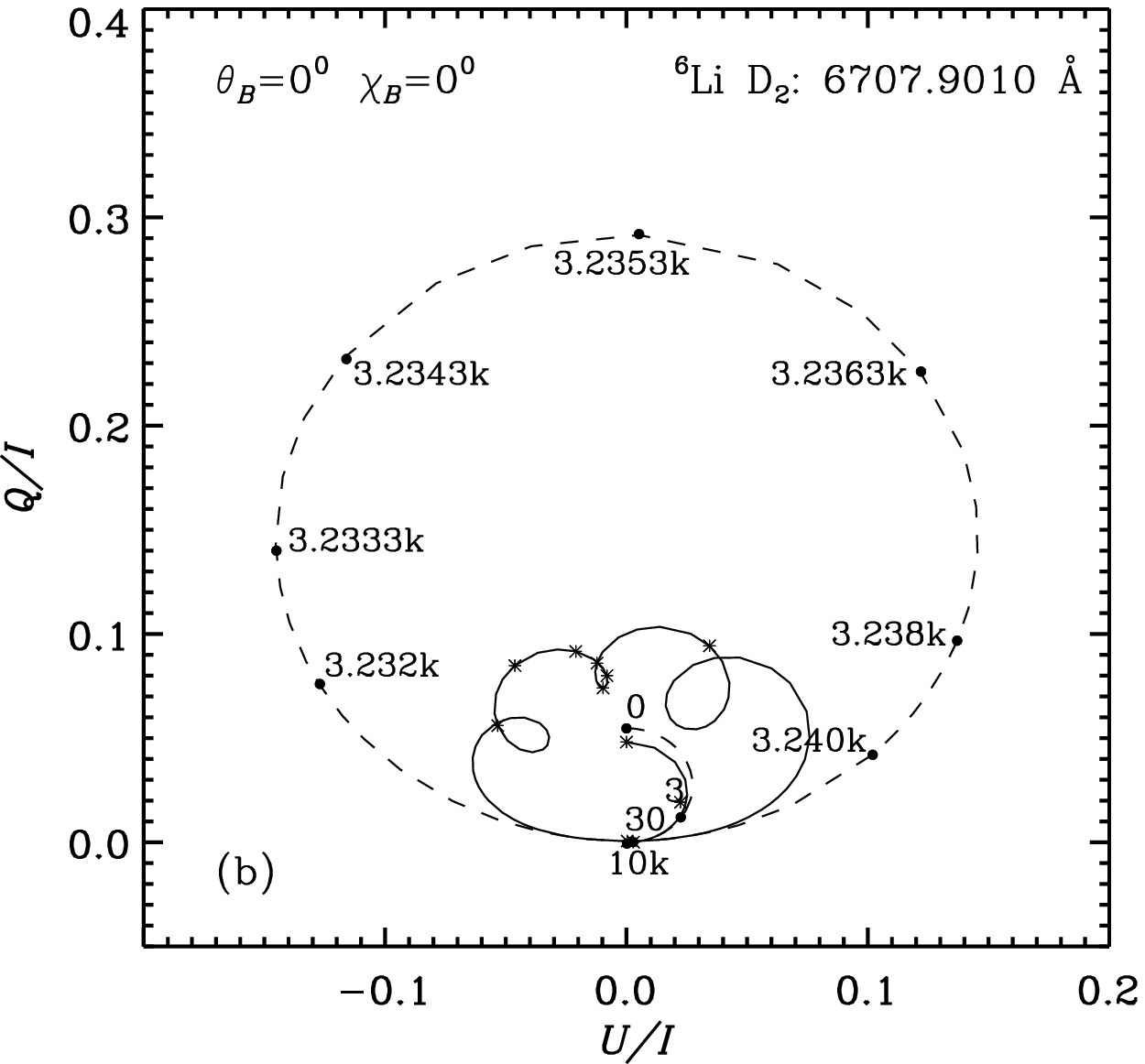}
\includegraphics[scale=0.55]{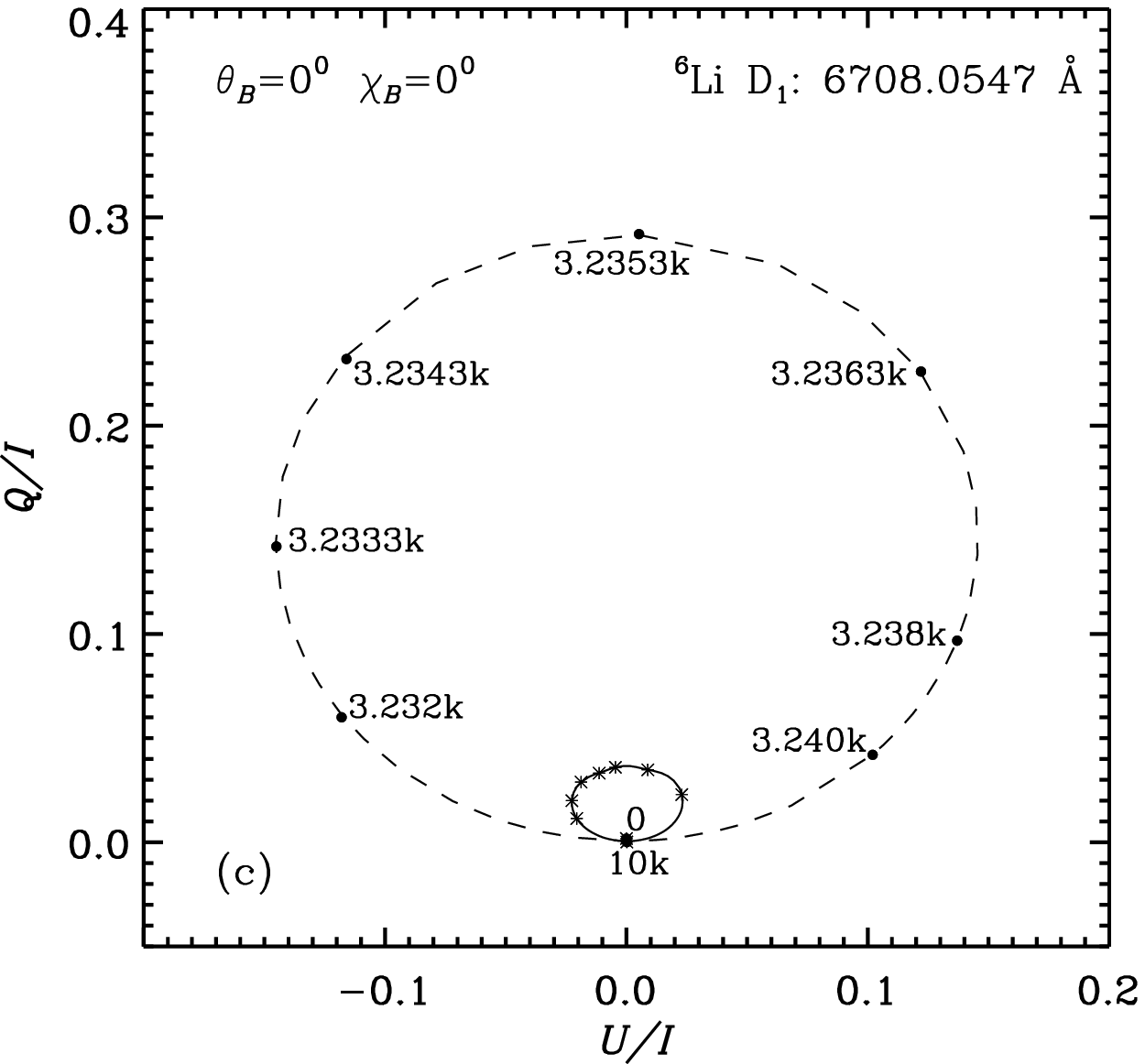}
\caption{Polarization diagrams obtained at the D line positions for a given 
orientation of the magnetic field. The dashed lines correspond to the pure $J$
state interference case without HFS while the solid lines correspond to the
combined theory case (including HFS). The magnetic field strength values are
marked along the dashed curves in Gauss, with ``k'' meaning a factor of 1000. The
asterisks on the solid curves represent the same field strength values as indicated
for the dashed curves. The scattering geometry considered is
$\mu=1$, $\mu^\prime=0$, $\chi=90\degree$, and $\chi^\prime=0\degree$.
\label{pd-3}
}
\end{center}
\end{figure}

In Figure~\ref{pd-3}, we compare the polarization diagrams obtained at different
wavelength points by varying the field strength $B$ for a two-term atom without HFS
(dashed curves) and a two-term atom with HFS (solid curves). The geometry considered
is described in the caption to the figure. In panel (a), we see a decrease in $Q/I$
with increasing field strength due to the Hanle effect. For fields greater than 100 G, we
enter the Hanle saturation regime. $Q/I$ starts to increase as we approach the
level-crossing field strength (around 3 kG). Loops (i.e., a single circular loop for
the dashed line and multiple small loops for the solid line) arise due to several
level-crossings (see Figure~\ref{level-fig}) where the coherence increases and $Q/I$
tends to approach its non-magnetic value. Comparing the solid and dashed curves
in Figure~\ref{pd-3}, the effects of HFS can be clearly seen. First, due to the
depolarization caused by HFS, the polarization diagram shrinks in size. Second,
multiple small loops are formed (see the solid lines in Figure~\ref{pd-3}). These
multiple loops arise due to several level-crossings that occur only when HFS is
included (see Figure~\ref{level-fig}(b), (c), (e), and (f)). For field strengths
larger than the level-crossing field strengths, the $Q/I$ value decreases
again and becomes zero around 10 kG. We see the effects due to
Rayleigh scattering in strong magnetic fields when we increase the field
strength beyond 10 kG \citep[similar to Figure 6(b) of][]{sow14a}. In panel (b),
we show the polarization diagram computed at the $^6$Li D$_2$ wavelength position. Since
the $^7$Li D$_1$ position nearly coincides with that of $^6$Li D$_2$, we see
the combined effect of both lines. However, due to the large abundance of $^7$Li,
the behavior of the polarization diagram is dominated by contribution from $^7$Li D$_1$.
Since $^7$Li D$_1$ is unpolarized, the small arcs seen for weak fields are due to the
$^6$Li D$_2$ line. After the Hanle saturation field strength (30 G), the polarization
diagrams essentially show behavior similar to the corresponding polarization diagrams
in panel (a). In panel (c), we show the polarization diagram for $^6$Li D$_1$ position.
The D$_1$ line remains unpolarized till the level-crossing field strength (around 3 kG)
is reached. Around the level-crossing field strength, we see a bigger loop for the case
without HFS (dashed line) and a smaller loop for the case with HFS (solid line).

\section{CONCLUSIONS}
\label{conclu}
We present a formalism to treat the combined interferences between the magnetic
substates of the hyperfine structure states pertaining to different fine structure
states of the same term including the effects of PRD
in scattering. Using the Kramers--Heisenberg approach, we calculate the
polarized scattering cross section (i.e., the redistribution matrix) for this process.
We also demonstrate the behavior of the redistribution matrix in a single scattering
of the incident unpolarized radiation by the lithium atoms. In the solar case, the
combined theory finds applications in modeling of spectral lines like lithium
6708 \AA{} for which the effects of both fine and hyperfine structure are significant. 

In the absence of magnetic fields, we recover the results already published by
\citet{bel09}. In the present paper, we illustrate the effects of a deterministic
magnetic field on the Stokes profiles of the lithium D line system. We cover the
entire field strength regime from a weak field Hanle regime to incomplete and complete
PB regimes. When the fields are weak, the Stokes profiles exhibit the well-known Hanle
signatures at the centers of the lithium D lines, namely, depolarization of $Q/I$ and 
rotation of polarization plane. We note that there are
Zeeman-like signatures for stronger fields.
We show the signatures of level-crossings and avoided crossings in Stokes profiles
and polarization diagrams. Unlike the pure $J$ state or $F$ state interferences, when
$J$ and $F$ state interferences are treated together, a multitude of level-crossings
and avoided crossings occur which produce multiple loops in the polarization diagrams
and interesting signatures in teh $U/I$ profiles. Non-zero NCP is seen for fields in the
incomplete PB regime which arises not only due to non-linear MS but also
due to the A-O conversion mechanism as already described in LL04. However, its diagnostic
potential needs to be explored. We perform the calculations including the effects of
PRD. However, its effect manifests itself only when one considers the transfer of the
line radiation in the solar atmospheric conditions. 

We thank the referee for very useful, detailed, constructive comments and suggestions
which helped us understand the results better and improve the paper substantially.
We acknowledge the use of the HYDRA cluster facility at the Indian Institute of
Astrophysics for the numerical computations related to the work presented in this paper. 

\appendix
\section{THE MAGNETIC REDISTRIBUTION FUNCTIONS FOR THE COMBINED $J$ AND $F$
STATE INTERFERENCES}
\label{a-a}
The magnetic redistribution functions of type-II in the case of combined $J$ and
$F$ state interferences have the same form as those in cases where only the interferences
between fine structure or hyperfine structure states are considered, except for
the increase in the dimension of the quantum number space. For our problem of
interest, they take the forms given by
\begin{eqnarray}
&& R^{\rm II,H}_{k_b\mu_bk_a\mu_ak_f\mu_f}(x_{ba},x_{ba}^\prime,\Theta)
=\frac{1}{\pi\sin\Theta}
{\rm exp}\bigg\{-\bigg[\frac{x_{ba}-x_{ba}^\prime+x_{k_a\mu_ak_f\mu_f}}
{2\sin(\Theta/2)}\bigg]^2\bigg\}
\nonumber \\ && \times
H\bigg(\frac{a}{\cos(\Theta/2)},\frac{x_{ba}+x_{ba}^\prime+x_{k_a\mu_ak_f\mu_f}}
{2\cos(\Theta/2)}\bigg)\ ,
\label{r-ii-h}
\end{eqnarray}
and
\begin{eqnarray}
&& R^{\rm II,F}_{k_b\mu_bk_a\mu_ak_f\mu_f}(x_{ba},x_{ba}^\prime,\Theta)
=\frac{1}{\pi\sin\Theta}
{\rm exp}\bigg\{-\bigg[\frac{x_{ba}-x_{ba}^\prime+x_{k_a\mu_ak_f\mu_f}}
{2\sin(\Theta/2)}\bigg]^2\bigg\}
\nonumber \\ && \times
2F\bigg(\frac{a}{\cos(\Theta/2)},\frac{x_{ba}+x_{ba}^\prime+x_{k_a\mu_ak_f\mu_f}}
{2\cos(\Theta/2)}\bigg)\ .
\label{r-ii-f}
\end{eqnarray}
Here, $\Theta$ is the scattering angle, and the functions $H$ and $F$ are the Voigt and
Faraday-Voigt functions \citep[see Equation (18) of][]{smi11a}.
The quantities appearing in the expressions for the
type-II redistribution functions have the following definitions:
\begin{eqnarray}
&&x_{ba}=\frac{\nu_{k_b\mu_bk_a\mu_a}-\nu}{\Delta\nu_{\rm D}}\ ;
\ \ x^\prime_{ba}=\frac{\nu_{k_b\mu_bk_a\mu_a}-\nu^\prime}{\Delta\nu_{\rm D}}\ ;
x_{k_a\mu_ak_f\mu_f}=\frac{\nu_{k_a\mu_ak_f\mu_f}}{\Delta\nu_{\rm D}}\ ;
\ \ a=\frac{\gamma}{4\pi\Delta\nu_{\rm D}}\ ,
\label{defns}
\end{eqnarray}
where $x_{ba}$ is the emission frequency, $x^\prime_{ba}$ is the absorption frequency,
$a$ is the damping parameter, and $\Delta\nu_{\rm D}$ is the Doppler width.

The auxiliary functions $h^{\rm II}$ and $f^{\rm II}$ appearing
in Equation~(\ref{final-rm}) can be constructed by making use of
Equations~(\ref{r-ii-h}) and (\ref{r-ii-f}) as
\begin{eqnarray}
&&(h^{\rm II}_{k_b\mu_b,k_{b^\prime}\mu_{b^\prime}})_{k_a\mu_ak_f\mu_f}
=\frac{1}{2}\bigg[R^{\rm II,H}_{k_b\mu_bk_a\mu_ak_f\mu_f}+
R^{\rm II,H}_{k_{b^\prime}\mu_{b^\prime}k_a\mu_ak_f\mu_f}\bigg]\ ,
\label{h-ii}
\end{eqnarray}
\begin{eqnarray}
&&(f^{\rm II}_{k_b\mu_b,k_{b^\prime}\mu_{b^\prime}})_{k_a\mu_ak_f\mu_f}
=\frac{1}{2}\bigg[R^{\rm II,F}_{k_{b^\prime}\mu_{b^\prime}k_a\mu_ak_f\mu_f}
-R^{\rm II,F}_{k_b\mu_bk_a\mu_ak_f\mu_f}\bigg]\ .
\label{f-ii}
\end{eqnarray}
These auxiliary functions contain the information regarding the Doppler redistribution
of photon frequencies.

\section{THE PSS}
\label{a-b}
The PSS is a basic test for checking the
correctness of the eigenvalues and eigenvectors obtained from a diagonalization
procedure, for a given problem. A detailed description of the PSS is given in LL04
(see pp. 321-325). In LL04, the manifestations of PSS are given separately for (a) a
two-level atom with hyperfine structure and (b) a two-term atom exhibiting only
FS. In this appendix, we formulate the PSS for the case of a two-term atom exhibiting
both FS and HFS. We basically follow the same procedure as described in LL04 to derive
the expression for the centers of gravity in frequency of the magnetic components.

The strengths of the magnetic components are given by
\begin{eqnarray}
&&\mathcal{S}_q^{k_a\mu_a,k_b\mu_b}=
|\langle L_aSI_s,k_a\mu_a|{\rm r}_q|L_bSI_s,k_b\mu_b\rangle|^2 \ ,
\label{str}
\end{eqnarray}
which are essentially the square of the complex amplitude of the transition between the
lower term (quantities with subscripts $a$) and the upper term (quantities with
subscripts $b$). ${\rm r}_q$ are the spherical components of the dipole moment operator.
Using the basis expansion defined in Equation~(\ref{basis-good}), the Wigner--Eckart
theorem and its corollary, we expand the above equation as
\begin{eqnarray}
&&\mathcal{S}_q^{k_a\mu_a,k_b\mu_b}=(2L_a+1)
\sum_{J_aJ_{a^\prime}J_bJ_{b^\prime}F_aF_{a^\prime}F_bF_{b^\prime}}
(-1)^{J_a+J_{a^\prime}+J_b+J_{b^\prime}} \nonumber \\ &&
\times
C^{k_a}_{J_aF_a}(L_aSI_s,\mu_a)
C^{k_a}_{J_{a^\prime}F_{a^\prime}}(L_aSI_s,\mu_a)
C^{k_b}_{J_bF_b}(L_bSI_s,\mu_b)
C^{k_b}_{J_{b^\prime}F_{b^\prime}}(L_bSI_s,\mu_b) \nonumber \\ &&
\times\sqrt{(2J_a+1)(2J_{a^\prime}+1)(2J_b+1)(2J_{b^\prime}+1)
(2F_a+1)(2F_{a^\prime}+1)(2F_b+1)(2F_{b^\prime}+1)}
\nonumber \\ &&
\times\left\lbrace
\begin{array}{ccc}
L_a & L_b & 1\\
J_b & J_a & S \\
\end{array}
\right\rbrace
\left\lbrace
\begin{array}{ccc}
L_a & L_b & 1\\
J_{b^\prime} & J_{a^\prime} & S \\
\end{array}
\right\rbrace
\left\lbrace
\begin{array}{ccc}
J_a & J_b & 1\\
F_b & F_a & I_s \\
\end{array}
\right\rbrace
\left\lbrace
\begin{array}{ccc}
J_{a^\prime} & J_{b^\prime} & 1\\
F_{b^\prime} & F_{a^\prime} & I_s \\
\end{array}
\right\rbrace \nonumber \\ &&
\times
\left (
\begin{array}{ccc}
F_b & F_a & 1\\
-\mu_b & \mu_a & -q \\
\end{array}
\right )
\left (
\begin{array}{ccc}
F_{b^\prime} & F_{a^\prime} & 1\\
-\mu_b & \mu_a & -q \\
\end{array}
\right )
|\langle L_a||{\bf r}||L_b\rangle|^2\ ,
\label{ustr}
\end{eqnarray}
with, $q=0$ for $\pi$, $+1$ for $\sigma_r$, and $-1$ for the $\sigma_b$
components. The transitions connecting the upper and the lower terms obey the
selection rules $\Delta L=0,\pm1$, $\Delta S=0$, $\Delta I_s=0$, and
$\Delta\mu=\mu_b-\mu_a=0,\pm1$. Summing the expression for the unnormalized strengths
over all the possible transitions, making use of the
orthogonality property of the $C$-coefficients given in Equation~(5a) of
\citet{cm05} and Equations~(2.23a) and (2.39) of LL04, we obtain
\begin{eqnarray}
 &&\sum_{k_ak_b\mu_a\mu_b}\mathcal{S}_q^{k_a\mu_a,k_b\mu_b}=
\frac{1}{3}(2L_a+1)(2S+1)(2I_s+1)|\langle L_a||{\bf r}||L_b\rangle|^2\ .
\label{csum}
\end{eqnarray}
Making use of the condition that
\begin{eqnarray}
 &&\sum_{k_ak_b\mu_a\mu_b}\mathcal{S}_q^{k_a\mu_a,k_b\mu_b}=1\ ,
\label{cnorm}
\end{eqnarray}
we write the expression for the normalized strengths as
\begin{eqnarray}
&&\mathcal{S}_q^{k_a\mu_a,k_b\mu_b}=
\frac{3}{(2S+1)(2I_s+1)}
\sum_{J_aJ_{a^\prime}J_bJ_{b^\prime}F_aF_{a^\prime}F_bF_{b^\prime}}
(-1)^{J_a+J_{a^\prime}+J_b+J_{b^\prime}} \nonumber \\ &&
\times
C^{k_a}_{J_aF_a}(L_aSI_s,\mu_a)
C^{k_a}_{J_{a^\prime}F_{a^\prime}}(L_aSI_s,\mu_a)
C^{k_b}_{J_bF_b}(L_bSI_s,\mu_b)
C^{k_b}_{J_{b^\prime}F_{b^\prime}}(L_bSI_s,\mu_b) \nonumber \\ &&
\times\sqrt{(2J_a+1)(2J_{a^\prime}+1)(2J_b+1)(2J_{b^\prime}+1)
(2F_a+1)(2F_{a^\prime}+1)(2F_b+1)(2F_{b^\prime}+1)}
\nonumber \\ &&
\times\left\lbrace
\begin{array}{ccc}
L_a & L_b & 1\\
J_b & J_a & S \\
\end{array}
\right\rbrace
\left\lbrace
\begin{array}{ccc}
L_a & L_b & 1\\
J_{b^\prime} & J_{a^\prime} & S \\
\end{array}
\right\rbrace
\left\lbrace
\begin{array}{ccc}
J_a & J_b & 1\\
F_b & F_a & I_s \\
\end{array}
\right\rbrace
\left\lbrace
\begin{array}{ccc}
J_{a^\prime} & J_{b^\prime} & 1\\
F_{b^\prime} & F_{a^\prime} & I_s \\
\end{array}
\right\rbrace \nonumber \\ &&
\times
\left (
\begin{array}{ccc}
F_b & F_a & 1\\
-\mu_b & \mu_a & -q \\
\end{array}
\right )
\left (
\begin{array}{ccc}
F_{b^\prime} & F_{a^\prime} & 1\\
-\mu_b & \mu_a & -q \\
\end{array}
\right ) \ .
\nonumber \\ &&
\label{nstr}
\end{eqnarray}
The centers of gravity in frequency of the magnetic components are defined as
\begin{eqnarray}
&&\Delta\nu_q= \sum_{k_ak_b\mu_a\mu_b}
\mathcal{S}_q^{k_a\mu_a,k_b\mu_b} \Delta\nu^{k_ak_b}_{\mu_a\mu_b}\ ,
\label{cog}
\end{eqnarray}
with
\begin{eqnarray}
 &&\Delta\nu^{k_ak_b}_{\mu_a\mu_b}=
\frac{E_{k_b}(L_bSI_s,\mu_b)-E_{k_a}(L_aSI_s,\mu_a)}{h}\ .
\label{shift}
\end{eqnarray}
Using Equations (\ref{nstr}) and (\ref{shift}) in Equation (\ref{cog}), and 
performing sums over $k_a$ and $k_b$ with the help of Equations
(5a) and (7) of \citet{cm05}, we obtain
\begin{eqnarray}
 &&\Delta\nu_q=\frac{1}{h}\frac{3}{(2S+1)(2I_s+1)}
\sum_{J_aJ_{a^\prime}J_bJ_{b^\prime}F_aF_{a^\prime}F_bF_{b^\prime}}
\sum_{\mu_a\mu_b} (-1)^{J_a+J_{a^\prime}+J_b+J_{b^\prime}}\nonumber \\ &&
\times\sqrt{(2J_a+1)(2J_{a^\prime}+1)(2J_b+1)(2J_{b^\prime}+1)
(2F_a+1)(2F_{a^\prime}+1)(2F_b+1)(2F_{b^\prime}+1)}
\nonumber \\ &&
\times\left\lbrace
\begin{array}{ccc}
L_a & L_b & 1\\
J_b & J_a & S \\
\end{array}
\right\rbrace
\left\lbrace
\begin{array}{ccc}
L_a & L_b & 1\\
J_{b^\prime} & J_{a^\prime} & S \\
\end{array}
\right\rbrace
\left\lbrace
\begin{array}{ccc}
J_a & J_b & 1\\
F_b & F_a & I_s \\
\end{array}
\right\rbrace
\left\lbrace
\begin{array}{ccc}
J_{a^\prime} & J_{b^\prime} & 1\\
F_{b^\prime} & F_{a^\prime} & I_s \\
\end{array}
\right\rbrace \nonumber \\ &&
\times
\left (
\begin{array}{ccc}
F_b & F_a & 1\\
-\mu_b & \mu_a & -q \\
\end{array}
\right )
\left (
\begin{array}{ccc}
F_{b^\prime} & F_{a^\prime} & 1\\
-\mu_b & \mu_a & -q \\
\end{array}
\right ) 
[\delta_{J_aJ_{a^\prime}}\delta_{F_aF_{a^\prime}}
\langle L_bSJ_bI_sF_b\mu_b|\mathcal{H}_T|L_bSJ_{b^\prime}I_sF_{b^\prime}\mu_b\rangle
\nonumber \\ &&
-\delta_{J_bJ_{b^\prime}}\delta_{F_bF_{b^\prime}}
\langle L_aSJ_aI_sF_a\mu_a|\mathcal{H}_T|L_aSJ_{a^\prime}I_sF_{a^\prime}\mu_a\rangle]\ .
\label{cog1}
\end{eqnarray}
We separate the atomic and magnetic Hamiltonians in the above expression.
It can be shown that the atomic part does not contribute to the centers of gravity.
Using Equations (2.42), (2.41), (2.36d), (2.26d) and (2.39) of LL04, we simplify
the magnetic Hamiltonian part and find that
\begin{eqnarray}
 &&\Delta\nu_q=-q\nu_{\rm L} \ ,
\label{cog2}
\end{eqnarray}
where $\nu_{\rm L}$ is the Larmor frequency associated with the applied magnetic field.
This result is the same as Equation (3.66) of LL04 which one would expect for a
two-term atom without any fine or hyperfine structure. This means that the centers of
gravity of the magnetic components in the PB regime have the same frequencies as
the individual components due to Zeeman effect that would arise from the
transitions between spinless lower and upper terms. In situations where the
electron and nuclear spins are negligible, this is expected from the PSS. 

We then verify that the eigenvalues and eigenvectors obtained by diagonalizing
$\mathcal{H}_T$, when used in Equation~(\ref{cog}), give the same value for
$\Delta\nu_q$ as that calculated from Equation~(\ref{cog2}).

\section{A-O CONVERSION MECHANISM}
\label{a-c}
The RM presented in Equation~(\ref{final-rm}) can be reduced to the phase matrix
by integrating the auxiliary functions over the incoming and outgoing frequencies.
The phase matrix will then take the form given by
\begin{eqnarray}
P_{ij}({\bm n},{\bm n^\prime};{\bm B})=\sum_{KK^\prime Q} W_{KK^\prime Q}({\bm B})
(-1)^Q \mathcal{T}^K_Q(i,{\bm n}) \mathcal{T}^{K^\prime}_{-Q}(j,{\bm n^\prime})\ ,
\label{p-mat}
\end{eqnarray}
where
\begin{eqnarray}
&& W_{KK^\prime Q}({\bm B})=\frac{3(2L_b+1)}{(2S+1)(2I_s+1)}
\left\lbrace
\begin{array}{ccc}
1 & 1 & K\\
L_b & L_b & L_a \\
\end{array}
\right\rbrace
\left\lbrace
\begin{array}{ccc}
1 & 1 & K^\prime\\
L_b & L_b & L_a \\
\end{array}
\right\rbrace
\nonumber \\ && \times
\sum_{J_bJ_{b^\prime}J_{b^{\prime\prime}}J_{b^{\prime\prime\prime}}}
\sum_{F_bF_{b^\prime}F_{b^{\prime\prime}}F_{b^{\prime\prime\prime}}}
\sum_{\mu_b\mu_{b^\prime}}
(-1)^{J_b+J_{b^\prime}+J_{b^{\prime\prime}}+J_{b^{\prime\prime\prime}}}
(-1)^{K+K^\prime}
\nonumber \\ && \times
\sqrt{(2J_b+1)(2J_{b^\prime}+1)(2J_{b^{\prime\prime}}+1)
(2J_{b^{\prime\prime\prime}}+1)(2F_b+1)(2F_{b^\prime}+1)(2F_{b^{\prime\prime}}+1)
(2F_{b^{\prime\prime\prime}}+1)}
\nonumber \\ && \times
\left\lbrace
\begin{array}{ccc}
L_b & L_b & K\\
J_b & J_{b^\prime} & S \\
\end{array}
\right\rbrace
\left\lbrace
\begin{array}{ccc}
L_b & L_b & {K^\prime}\\
J_{b^{\prime\prime}} & J_{b^{\prime\prime\prime}} & S \\
\end{array}
\right\rbrace
\left\lbrace
\begin{array}{ccc}
J_{b^\prime} & J_b & K\\
F_b & F_{b^\prime} & I_s \\
\end{array}
\right\rbrace
\left\lbrace
\begin{array}{ccc}
J_{b^{\prime\prime\prime}} & J_{b^{\prime\prime}} & {K^\prime}\\
F_{b^{\prime\prime}} & F_{b^{\prime\prime\prime}} & I_s \\
\end{array}
\right\rbrace
\nonumber \\ && \times
\left (
\begin{array}{ccc}
F_b & F_{b^\prime} & K\\
-\mu_b & \mu_{b^\prime} & -Q \\
\end{array}
\right )
\left (
\begin{array}{ccc}
F_{b^{\prime\prime}} & F_{b^{\prime\prime\prime}} & K^\prime\\
-\mu_b & \mu_{b^\prime} & -Q \\
\end{array}
\right )
\sqrt{(2K+1)(2K^\prime+1)}
\nonumber \\ && \times
\sum_{k_bk_{b^\prime}}
C^{k_b}_{J_bF_b}(L_bSI_s,\mu_b)
C^{k_b}_{J_{b^{\prime\prime}}F_{b^{\prime\prime}}}(L_bSI_s,\mu_b)
C^{k_{b^\prime}}_{J_{b^\prime}F_{b^\prime}}(L_bSI_s,\mu_{b^\prime})
C^{k_{b^\prime}}_{J_{b^{\prime\prime\prime}}F_{b^{\prime\prime\prime}}}
(L_bSI_s,\mu_{b^\prime})
\nonumber \\ && \times
\frac{1}{1+2\pi{\rm i}\nu(k_{b^\prime}\mu_{b^\prime},k_b\mu_b)
/A(L_aSI_s\rightarrow L_bSI_s)}\ .
\label{w-matrix}
\end{eqnarray}
Here, $A$ is the Einstein coefficient for the $L_a\rightarrow L_b$ transition and
$\nu(k_{b^\prime}\mu_{b^\prime},k_b\mu_b)=(E_{k_{b^\prime}\mu_{b^\prime}}-E_{k_b\mu_b})/h$.
We compute the $\mathcal{T}^K_Q$s for the geometry considered in
Section~\ref{sec-3} so that we can obtain an expression for the frequency
integrated fractional circular polarization, $\tilde p_V$, similar to the one
given in Section 10.20 of LL04. The explicit expressions for the
$\mathcal{T}^K_Q(i,{\bm n})$ in the atmospheric reference frame for a rotation of
the form $R\equiv(0,-\theta,-\chi)\times(\chi_B,\theta_B,0)$ are given by

\begin{eqnarray}
&& \mathcal{T}^0_0(0,{\bm n})=1\ , \nonumber \\ &&
\mathcal{T}^1_0(0,{\bm n})=0\ , \nonumber \\ &&
\mathcal{T}^1_1(0,{\bm n})=0\ , \nonumber \\ &&
\mathcal{T}^2_0(0,{\bm n})=\frac{1}{\sqrt{2}}\ \bigg[\frac{1}{4}\ (3{\rm cos}^2\theta-1)
\ (3{\rm cos}^2\theta_B-1)
+3\ {\rm sin}\theta\ {\rm cos}\theta\ {\rm sin}\theta_B
\ {\rm cos}\theta_B\ {\rm cos}(\chi-\chi_B)
\nonumber \\ && \ \ \ \ \ \ \ \ \ \
+\frac{3}{4}\ {\rm sin}^2\theta\ {\rm sin}^2\theta_B\ {\rm cos}2(\chi-\chi_B)\bigg]\ ,
\nonumber \\ &&
\mathcal{T}^2_1(0,{\bm n})=\frac{1}{\sqrt{2}}\ \bigg[\frac{\sqrt{3}}{2\sqrt{2}}
\ (3{\rm cos}^2\theta-1)\ {\rm sin}\theta_B\ {\rm cos}\theta_B
\nonumber \\ && \ \ \ \ \ \ \ \ \ \
-\frac{\sqrt{3}}{\sqrt{2}}\ {\rm sin}\theta\ {\rm cos}\theta
\bigg[{\rm e}^{{\rm i}(\chi-\chi_B)}\bigg({\rm cos}\theta_B-\frac{1}{2}\bigg)
\ ({\rm cos}\theta_B+1)-{\rm e}^{{\rm -i}(\chi-\chi_B)}
\bigg({\rm cos}\theta_B+\frac{1}{2}\bigg)\ (1-{\rm cos}\theta_B)\bigg]
\nonumber \\ && \ \ \ \ \ \ \ \ \ \
-\frac{\sqrt{3}}{2\sqrt{8}}\ {\rm sin}^2\theta\ {\rm sin}\theta_B
\big[{\rm e}^{{\rm 2i}(\chi-\chi_B)}(1+{\rm cos}\theta_B)
-{\rm e}^{{\rm -2i}(\chi-\chi_B)}(1-{\rm cos}\theta_B)\big]\bigg]\ ,
\nonumber \\ &&
\mathcal{T}^2_2(0,{\bm n})=\frac{1}{\sqrt{2}}\bigg[\frac{\sqrt{3}}{4\sqrt{2}}
\ (3{\rm cos}^2\theta-1)\ {\rm sin}^2\theta_B
\nonumber \\ && \ \ \ \ \ \ \ \ \ \
-\frac{\sqrt{3}}{2\sqrt{2}}\ {\rm sin}\theta \ {\rm cos}\theta\ {\rm sin}\theta_B
\big[{\rm e}^{{\rm i}(\chi-\chi_B)}(1+{\rm cos}\theta_B)
-{\rm e}^{{\rm -i}(\chi-\chi_B)}(1-{\rm cos}\theta_B)\big]
\nonumber \\ && \ \ \ \ \ \ \ \ \ \
+\frac{\sqrt{3}}{8\sqrt{2}}\ {\rm sin}^2\theta\big[{\rm e}^{{\rm 2i}(\chi-\chi_B)}
(1+{\rm cos}\theta_B)^2+{\rm e}^{{\rm -2i}(\chi-\chi_B)}
(1-{\rm cos}\theta_B)^2\big]\bigg]\ ,
\label{tkq0}
\end{eqnarray}
\begin{eqnarray}
&& \mathcal{T}^0_0(1,{\bm n})=0\ , \nonumber \\ &&
\mathcal{T}^1_0(1,{\bm n})=0\ , \nonumber \\ &&
\mathcal{T}^1_1(1,{\bm n})=0\ , \nonumber \\ &&
\mathcal{T}^2_0(1,{\bm n})=-\frac{\sqrt{3}}{2}\bigg[\frac{\sqrt{3}}{\sqrt{8}}
\ {\rm sin}^2\theta\ (3{\rm cos}^2\theta_B-1)
-\frac{2\sqrt{3}}{\sqrt{2}}\ {\rm sin}\theta\ {\rm cos}\theta\ {\rm sin}\theta_B
\ {\rm cos}\theta_B\ {\rm cos}(\chi-\chi_B)
\nonumber \\ && \ \ \ \ \ \ \ \ \ \
+\frac{\sqrt{3}}{\sqrt{8}}\ (1+{\rm cos}^2\theta)\ {\rm sin}^2\theta_B
\ {\rm cos}2(\chi-\chi_B)\bigg]\ ,
\nonumber \\ &&
\mathcal{T}^2_1(1,{\bm n})=-\frac{\sqrt{3}}{2}\bigg[\frac{3}{2}\ {\rm sin}^2\theta
\ {\rm sin}\theta_B\ {\rm cos}\theta_B
\nonumber \\ && \ \ \ \ \ \ \ \ \ \
+{\rm sin}\theta\ {\rm cos}\theta
\bigg[{\rm e}^{{\rm i}(\chi-\chi_B)}\bigg({\rm cos}\theta_B-\frac{1}{2}\bigg)
({\rm cos}\theta_B+1)-{\rm e}^{-{\rm i}(\chi-\chi_B)}
\bigg({\rm cos}\theta_B+\frac{1}{2}\bigg)
(1-{\rm cos}\theta_B)\bigg]
\nonumber \\ && \ \ \ \ \ \ \ \ \ \
-\frac{1}{4}\ (1+{\rm cos}^2\theta)\ {\rm sin}\theta_B
\big[{\rm e}^{2{\rm i}(\chi-\chi_B)}(1+{\rm cos}\theta_B)-
{\rm e}^{-2{\rm i}(\chi-\chi_B)}(1-{\rm cos}\theta_B)\big]\bigg]\ ,
\nonumber \\ &&
\mathcal{T}^2_2(1,{\bm n})=-\frac{\sqrt{3}}{2}\bigg[\frac{3}{4}\ {\rm sin}^2\theta
\ {\rm sin}^2\theta_B
\nonumber \\ && \ \ \ \ \ \ \ \ \ \
+\frac{1}{2}\ {\rm sin}\theta\ {\rm cos}\theta\ {\rm sin}\theta_B
\big[{\rm e}^{{\rm i}(\chi-\chi_B)}(1+{\rm cos}\theta_B)-
{\rm e}^{-{\rm i}(\chi-\chi_B)}(1-{\rm cos}\theta_B)\big]
\nonumber \\ && \ \ \ \ \ \ \ \ \ \
+\frac{1}{8}(1+{\rm cos}^2\theta)
\big[{\rm e}^{2{\rm i}(\chi-\chi_B)}(1+{\rm cos}\theta_B)^2+
{\rm e}^{-2{\rm i}(\chi-\chi_B)}(1-{\rm cos}\theta_B)^2\big]\bigg]\ ,
\label{tkq1}
\end{eqnarray}
\begin{eqnarray}
&& \mathcal{T}^0_0(2,{\bm n})=0\ , \nonumber \\ &&
\mathcal{T}^1_0(2,{\bm n})=0\ , \nonumber \\ &&
\mathcal{T}^1_1(2,{\bm n})=0\ , \nonumber \\ &&
\mathcal{T}^2_0(2,{\bm n})=\frac{\sqrt{3}}{2}\bigg[-\sqrt{6}
\ {\rm sin}\theta\ {\rm sin}\theta_B\ {\rm cos}\theta_B\ {\rm sin}(\chi-\chi_B)
+\frac{\sqrt{3}}{\sqrt{2}}\ {\rm cos}\theta\ {\rm sin}^2\theta_B
\ {\rm sin}2(\chi-\chi_B)\bigg]\ ,
\nonumber \\ &&
\mathcal{T}^2_1(2,{\bm n})=-{\rm i}\frac{\sqrt{3}}{2}\bigg[{\rm sin}\theta
\bigg[{\rm e}^{{\rm i}(\chi-\chi_B)}\bigg({\rm cos}\theta_B-\frac{1}{2}\bigg)
({\rm cos}\theta_B+1)+{\rm e}^{-{\rm i}(\chi-\chi_B)}
\bigg({\rm cos}\theta_B+\frac{1}{2}\bigg)(1-{\rm cos}\theta_B)\bigg]
\nonumber \\ && \ \ \ \ \ \ \ \ \ \
-\frac{1}{2}\ {\rm cos}\theta\ {\rm sin}\theta_B
\big[{\rm e}^{2{\rm i}(\chi-\chi_B)}
(1+{\rm cos}\theta_B)+{\rm e}^{-2{\rm i}(\chi-\chi_B)}(1-{\rm cos}\theta_B)\big]\bigg]\ ,
\nonumber \\ &&
\mathcal{T}^2_2(2,{\bm n})=-{\rm i}\frac{\sqrt{3}}{2}\bigg[\frac{1}{2}\ {\rm sin}\theta
\ {\rm sin}\theta_B\big[{\rm e}^{{\rm i}(\chi-\chi_B)}(1+{\rm cos}\theta_B)
+{\rm e}^{-{\rm i}(\chi-\chi_B)}(1-{\rm cos}\theta_B)\big]
\nonumber \\ && \ \ \ \ \ \ \ \ \ \
+\frac{1}{4}\ {\rm cos}\theta\big[{\rm e}^{2{\rm i}(\chi-\chi_B)}(1+{\rm cos}\theta_B)^2
-{\rm e}^{-2{\rm i}(\chi-\chi_B)}(1-{\rm cos}\theta_B)^2\big]\bigg]\ ,
\label{tkq2}
\end{eqnarray}
and
\begin{eqnarray}
&&  \mathcal{T}^0_0(3,{\bm n})=0\ ,
\nonumber \\ &&
\mathcal{T}^1_0(3,{\bm n})=\frac{\sqrt{3}}{\sqrt{2}}[{\rm cos}\theta
\ {\rm cos}\theta_B+{\rm sin}\theta\ {\rm sin}\theta_B\ {\rm cos}(\chi-\chi_B)]\ ,
\nonumber \\ &&
\mathcal{T}^1_1(3,{\bm n})=\frac{\sqrt{3}}{\sqrt{2}}\bigg[\frac{1}{\sqrt{2}}
\ {\rm cos}\theta\ {\rm sin}\theta_B -\frac{1}{2\sqrt{2}}\ {\rm sin}\theta
\big[{\rm e}^{{\rm i}(\chi-\chi_B)}(1+{\rm cos}\theta_B)
-{\rm e}^{{\rm -i}(\chi-\chi_B)}(1-{\rm cos}\theta_B)\big]\bigg]\ ,
\nonumber \\ &&
\mathcal{T}^2_0(3,{\bm n})=0\ ,
\nonumber \\ &&
\mathcal{T}^2_1(3,{\bm n})=0\ ,
\nonumber \\ &&
\mathcal{T}^2_2(3,{\bm n})=0\ .
\label{tkq3}
\end{eqnarray}
We then expand the summations over $K,K^\prime$, and $Q$ in Equation~(\ref{p-mat}) and
write down the expressions for the $P_{00}$ and $P_{33}$ elements. We substitute in the
expressions for $P_{00}$ and $P_{33}$ the $\mathcal{T}^K_Q$s evaluated for the incoming
and the outgoing rays by making use of Equations~(\ref{tkq0}) and (\ref{tkq3}) for
the geometry considered in Section~\ref{sec-3}. After elaborate algebra, we finally arrive
at an expression for the frequency integrated fractional circular polarization given by
\begin{equation}
\tilde p_V=\frac{-2\sqrt{6}W_{120}}{16-W_{220}-3{\rm Re}(W_{222})}\ .
\label{a-o-mech}
\end{equation}
As discussed in Section~\ref{sec-3} and in Section 10.20 of LL04, due to the double
summations over $K$ and $K^\prime$ in Equation~(\ref{p-mat}) and due to the fact that
the spherical tensor $\mathcal{T}^K_Q(3,{\bm n})$ are non-zero only when $K=1$,
orientation can be produced in the upper term even when the circular polarization is
not present in the incident radiation. This mechanism is therefore called the A-O
conversion mechanism. We identify that the term with $K=1$ in the numerator of
Equation~(\ref{a-o-mech}) is responsible for the A-O conversion mechanism. We have
discussed the signatures of this mechanism in the Stokes $V$ parameter in
Section~\ref{res}.

\end{document}